\documentclass[twocolumn,tighten]{aastex61}
\usepackage[]{txfonts}
\usepackage{natbib, twoopt}
\usepackage{float}
\usepackage{graphicx}
\usepackage{epstopdf}
\usepackage[caption=false]{subfig}
\usepackage{color}
\usepackage{amsfonts}
\usepackage{bm}
\usepackage{longtable}
\usepackage{pifont}
\newcommand{\cmark}{\ding{51}}%
\newcommand{\xmark}{\ding{55}}%

\usepackage{booktabs} 

\newcommand{\gskfont}{
  \fontfamily{pcr}
  \bfseries 
  \color{mediumorchid}
}
\newcommand{\mcfont}{
  \fontfamily{pcr}
  \bfseries 
  \color{applered}
}
\newcommand{\jcafont}{
  \fontfamily{pcr}
  \bfseries 
  \color{azure}
}
\newcommand{\pryfont}{
  \fontfamily{pcr}
  \bfseries 
  \color{chartreuse4}
}
\newcommand{\andfont}{
  \fontfamily{pcr}
  \bfseries 
  \color{DarkGoldenrod3}
}
\DeclareTextFontCommand{\gsk}{\gskfont}
\DeclareTextFontCommand{\mc}{\mcfont}
\DeclareTextFontCommand{\jca}{\jcafont}
\DeclareTextFontCommand{\pry}{\pryfont}
\DeclareTextFontCommand{\and}{\andfont}

\shorttitle{Si~\textsc{iv} flare simulations}
\shortauthors{Kerr, Carlsson, Allred, Young, Daw}

\begin{document}


	\title{{Si}~\textsc{iv} Resonance Line Emission During Solar Flares: \\Non-LTE, Non-equilibrium, Radiation Transfer Simulations}
	
	\author{Graham~S.~Kerr}
	\email{graham.s.kerr@nasa.gov}
	\altaffiliation{NPP Fellow, administered by USRA}
	\affil{NASA Goddard Space Flight Center, Heliophysics Sciences Division, Code 671, 8800 Greenbelt Rd., Greenbelt, MD 20771, USA}

	 \author{Mats Carlsson}
	 \affil{Rosseland Centre for Solar Physics, University of Oslo, P.O. Box 1029, Blindern, N-0315 Oslo, Norway}
	 \affil{Institute of Theoretical Astrophysics, University of Oslo, P.O. Box 1029, Blindern, N-0315 Oslo, Norway}
	 
	 \author{Joel~C.~Allred}
	 \affil{NASA Goddard Space Flight Center, Heliophysics Sciences Division, Code 671, 8800 Greenbelt Rd., Greenbelt, MD 20771, USA}

	  \author{Peter~R.~Young}
	  \affil{NASA Goddard Space Flight Center, Heliophysics Sciences Division, Code 671, 8800 Greenbelt Rd., Greenbelt, MD 20771, USA}
	  \affil{College of Science, George Mason University, Fairfax, VA 22030, U.S.A}
	  
	  \author{Adrian~N.~Daw}
	  \affil{NASA Goddard Space Flight Center, Heliophysics Sciences Division, Code 671, 8800 Greenbelt Rd., Greenbelt, MD 20771, USA}

	\date{Received / Accepted}
	
	\keywords{Sun: transition region - Sun: flares - Sun: UV radiation - radiative transfer - methods: numerical - line: formation}
	
	\begin{abstract}	
	The Interface Region Imaging Spectrograph (IRIS) routinely observes the Si~\textsc{iv} resonance lines. When analyzing observations of these lines it has typically been assumed they form under optically thin conditions. This is likely valid for the quiescent Sun, but this assumption has also been applied to the more extreme flaring scenario. We used 36 electron beam driven radiation hydrodynamic solar flare simulations, computed using the \texttt{RADYN} code, to probe the validity of this assumption. Using these simulated atmospheres we solved the radiation transfer equations to obtain the non-LTE, non-equilibrium populations, line profiles, and opacities for a model Silicon atom, including charge exchange processes. This was achieved using the `minority species' version of \texttt{RADYN}. The inclusion of charge exchange resulted in a substantial fraction of Si~\textsc{iv} at cooler temperatures than those predicted by ionisation equilibrium. All simulations with an injected energy flux $F>5\times10^{10}$~erg~cm$^{-2}$~s$^{-1}$ resulted in optical depth effects on the Si~\textsc{iv} emission, with differences in both intensity and line shape compared to the optically thin calculation. Weaker flares (down to $F\approx5\times10^{9}$~erg~cm$^{-2}$~s$^{-1}$) also resulted in Si~\textsc{iv} emission forming under optically thick conditions, depending on the other beam parameters. When opacity was significant, the atmospheres generally had column masses in excess of $5\times10^{-6}$ g~cm$^{-2}$ over the temperature range $40$ to $100$~kK, and the Si~\textsc{iv} formation temperatures were between $30$ and $60$~kK. We urge caution when analyzing Si~\textsc{iv} flare observations, or when computing synthetic emission without performing a full radiation transfer calculation.
	\end{abstract}


\section{Introduction}\label{sec:intro}
The solar transition region (TR) is the thin layer through which the atmosphere steeply transitions from chromospheric temperatures, densities, and ionisation fractions to coronal values. This means that the temperature climbs through several $\times10-100$~kK to the $>1$~MK corona, densities precipitously fall from $\approx10^{11}$~cm$^{-3}$ to $\approx10^{8-9}$~cm$^{-3}$, and the atmosphere changes from a partially ionised state to being almost entirely ionised. {The TR is optically thin to most radiation, with notable exceptions being the C~\textsc{ii} resonance lines that form at the base of the TR \citep{2015ApJ...811...80R}}.  

Solar flares release tremendous amounts of magnetic energy (up to $\approx10^{32-33}$~ergs in the span of several minutes), which is subsequently transported from the release site in the corona to the lower atmosphere. It is believed that flare energy is transported by large numbers of electrons that are accelerated out of the thermal background \citep{1971SoPh...18..489B,1976SoPh...48..197H}. These non-thermal electrons are ducted along magnetic field lines, precipitating through the TR into the chromosphere, where they lose their energy via Coulomb collisions \citep{1971SoPh...18..489B,1978ApJ...224..241E,2011SSRv..159..107H}. Flares are characterised by intense, broadband, enhancements to the solar radiative output, the bulk of which originates from the chromosphere \citep{2011SSRv..159...19F}, due to plasma heating and ionisation. Heating drives mass motions with upflows (`chromospheric evaporation') filling the flux tubes with denser chromospheric material, and downflows (`chromospheric condensations'), the properties of which are sensitive to the location and magnitude of flare heating \citep[e.g.][]{1980ApJ...239..725D,2006ApJ...638L.117M,2009ApJ...699..968M,2015ApJ...807L..22G,1985ApJ...289..414F,1985ApJ...289..425F,1985ApJ...289..434F,1989ApJ...346.1019F}. It is worth noting that additional energy transport mechanisms have recently been proposed, for example the dissipation of high-frequency Alfv\'enic waves \citep{2008ApJ...675.1645F, 2016ApJ...818L..20R, 2016ApJ...827..101K, 2018ApJ...853..101R}.

The extreme conditions through the TR make it an important interface for energy, mass, and radiation transport. The sensitivity of the numerous spectral lines produced by ions in the TR to local plasma conditions makes the TR a source of rich diagnostic potential of flare energy transport. For example, the Doppler motions of spectral lines resulting from chromospheric evaporations or condensations are a useful, and commonly used, observation with which to make a model-data comparison to critically attack models of flare energy transport. It is crucial to accurately forward model the radiation that is produced by a flare simulation. 

Two such lines that are now routinely observed by the Interface Region Imaging Spectrograph \citep[IRIS;][]{2014SoPh..289.2733D} are the Si~\textsc{iv} resonance lines at $\lambda = 1393.75$~\AA\ and $\lambda = 1402.77$~\AA. These lines have been exploited to study various solar phenomenon since the launch of IRIS, including flares, and are often analysed using the simplifying assumption that they are formed under optically thin conditions. This is likely a safe assumption for many situations. However, during impulsive heating events, such as solar flares, temperatures and densities can increase substantially over the quiet Sun values, potentially increasing the atmospheric opacity at these wavelengths. If the lines form under conditions where opacity effects are non-negligible then interpreting observations, particularly Doppler motions or other velocity effects, becomes more complex \citep[e.g.][]{2015ApJ...813..125K,2016ApJ...827..101K,2018arXiv180703373B}.

Flare observations of the Si~\textsc{iv} resonance lines have largely focused on investigating atmospheric flows, and while these lines have been analysed under the optically thin assumption there have been (albeit ambiguous) hints that opacity effects may be present. 

\cite{2015ApJ...811..139T} noted that the lines broadened and exhibited redshifts in flare ribbon sources of up to a few$\times10$~km~s$^{-1}$ to $>100$~km~s$^{-1}$ (which is surprisingly fast), predominately appearing as asymmetries in the red wing{, and not as a shift} of the whole line. Other studies noted instead that the lines were fully redshifted to few$\times10$~km~s$^{-1}$ \citep{2016ApJ...829...35W,2017ApJ...848..118L}. \cite{2017ApJ...848..118L} pointed out that the Si~\textsc{iv} can deviate strongly from a Gaussian shape in the flare ribbons, with some profiles suitable for fitting with a Gaussian function but others not. In a very small flare (B4 class) \cite{2016ApJ...829...35W} found that the Si~\textsc{iv} intensity increased by up to $1000\times${, and a line} width similar in magnitude to the optically thick C~\textsc{ii} lines. {They also noted that the Si \textsc{iv} emission showed persistent redshifts, in contrast to the Mg \textsc{ii} redshifts which were bursty in character.}

\cite{2015ApJ...810....4B} found oscillations in Si~\textsc{iv} observations from flare ribbons. They noted that while line ratios suggested an optically thin plasma (in the optically thin limit the intensity ratio of the $\lambda=1393.75$\AA\ to $\lambda=1402.77$\AA\ line should be equal to two, the ratio of their oscillator strengths), some pixels deviated from the optically thin value. Further, while not interpreted as such, some of the profiles shown in their Figure 7 could show signs of self-absorption features. 

High-Resolution Telescope Spectrometer \citep[HRTS;][]{1975JOSA...65...13B} observations of a light bridge spectrum from 1978 February 13 showed a Si~\textsc{iv} resonance line ratio of $10:1$, far from the optically thin limit \citep{1991ApJS...75.1337B}. It was noted by \cite{2018ApJ...855..134J} that an X1 flare was produced from the same region, that this flare had an unusually long decay phase, and that there were several smaller events in the region. They speculate that the light bridge spectrum was actually the result of ongoing flaring activity, and the anomalous intensity ratio was due to flares.  

To our knowledge no other solar flare measurements of the $\lambda=1393.75$\AA\ to $\lambda=1402.77$\AA\ ratio have been reported (for telemetry reasons not every observation from IRIS includes both resonance lines in the linelist). There have been some stellar flare observations of the Si~\textsc{iv} resonance line ratio (and other TR line ratios) that indicate the presence of opacity effects \citep{1999A&A...351L..23M,2002A&A...390..219B,2006A&A...454..889C}. \cite{1999A&A...351L..23M} relate this to an electron density enhancement during the flare up to $\approx10^{11.5}$~cm$^{-3}$, using an escape probabilities argument. If stellar flare observations indicate optical depth effects then it is certainly prudent to consider this in the solar case, though the atmospheric structure of M dwarf stars such as YZ CMi studied by \cite{1999A&A...351L..23M} is rather different than the solar atmosphere.

Opacity effects on the Si~\textsc{iv} lines  have been suggested in regard to other solar phenomenon, such as UV bursts. These are smaller scale impulsive heating events, thought to be caused by reconnection in the TR or chromosphere. There has recently been a healthy debate as to both their origin and their relation to Ellerman bombs \citep[e.g.][]{2014Sci...346C.315P,2015ApJ...808..116J,2018arXiv180505850Y}. Self-absorption effects during a UV burst were reported by \cite{2015ApJ...811...48Y}, alongside a resonance line intensity ratio of 1.7. Bi-directional flows were ruled out as the origin of the central reversal feature. Similar observations were reported by \cite{2017ApJ...845...16N}.

As well as the ratio of the resonance lines, the ratio of the Si~\textsc{iv} lines to O~\textsc{iv} lines (also observed by IRIS) offer a density diagnostic, though as discussed by \cite{2015ApJ...808..116J}, \cite{2016ApJ...832...77D}, and \cite{2018ApJ...857....5Y} it is important to understand certain caveats with regard to non-equilibrium ionisation and dependence on formation temperatures. We note this diagnostic here as it has become a fairly commonly used tool, but if opacity effects are non-negligible the utility of this ratio is ambiguous. 

It is common in both flare, burst, and more quiescent modelling to assume that the Si~\textsc{iv} emission is optically thin, so that atomic databases such as CHIANTI \citep{1997A&AS..125..149D,2013ApJ...763...86L} are used, along with a modelled atmospheric structure or differential emission measure (DEM) model, to compute the emissivity without performing full radiation transfer. Examples include \cite{2015ApJ...802....5O}, \cite{2016ApJ...817...46M}, and \cite{2017ApJ...850..153N} who used radiation magnetohydrodynamic (RMHD) modelling to study various solar features. They found reasonable agreement with IRIS observations, particularly when non-equilibrium ionisation effects were included.

The effects of photoexcitation, non-equilibrium ionisation, kappa-distributions (rather than Maxwellian), and the effects of high energy tails of electron distributions on the formation of Si~\textsc{iv} and O~\textsc{iv} have been investigated in detail by \cite{2014ApJ...780L..12D,2017ApJ...842...19D}, \cite{2016A&A...589A..68D,2017A&A...603A..14D,2018A&A...610A..67D}. They noted that to be able to reproduce observed ratios of Si~\textsc{iv} $1402.77$~\AA\ to O~\textsc{iv} $1401.2$~\AA, inclusion of non-equilibrium effects were crucial. Low-temperature shoulders were present in the ionisation fractions of Si~\textsc{iv} ions compared to the assumption of ionisation equilibrium.

In flares the Si~\textsc{iv} emission has been computed in a similar manner to investigate the flows resulting from multi-threaded loop simulations versus single loop simulations \citep{2016ApJ...827..145R,2018ApJ...856..149R}, where it was found that multi-threaded loops with heating durations on the order $50-100$~s, are required to reproduce the observed persistent redshifts. Also, \cite{2018ApJ...856..178P} synthesised Si~\textsc{iv} from electron beam driven nanoflare simulations or from simulations with \textsl{in-situ} energy deposition, noting that observed Doppler motions were only consistent with the electron beam scenario (and then only for certain non-thermal electron distribution parameters). 

We could find only one flare model in which the Si~\textsc{iv} emission was computed in an alternative way, where optical depths were included in a simplified manner \citep{1985ApJ...299L.103D}. Using a semi-empirical flare atmosphere \cite{1985ApJ...299L.103D} modelled the Si~\textsc{iv} lines, finding them to be formed in optically thick conditions, significantly more intense than the typical quiet Sun profiles, and with a deep central reversal feature. 

Since the formation properties of the Si~\textsc{iv} resonance lines in flare models that take account of potential opacity effects have been explored relatively little in comparison to other IRIS lines, such as the Mg~\textsc{ii} h \& k lines \citep[e.g.][]{2016ApJ...827..101K,GSK_Thesis,2017ApJ...842...82R}, we present the initial results from a survey of electron beam driven solar flares in which the Si~\textsc{iv} resonance lines were computed in detail. This is a prudent investigation given hints of opacity effects reported in flares or other heating events, and given the importance of the Si~\textsc{iv} lines as a means to understand energy transport and mass motions. Additionally, it is important to understand if our flare models are able to reproduce not just one observation but many, so combining Si~\textsc{iv} with the other spectral lines observed by IRIS can assess the validity of flare models. 

The numerical approach, model atom and flare simulations are introduced in Section~\ref{sec:codes_descript}. Si~\textsc{iv} resonance line emission is discussed in Section~\ref{sec:siiv_form} where the line profiles, lightcurves, and ratios are presented, along with details regarding opacity and the atmospheric properties in the line formation region. In Section~\ref{sec:iris_comparison} the Si~\textsc{iv} lines are placed in context with other IRIS observables (Mg~\textsc{ii} k line, Mg~\textsc{ii} 2791~\AA\ line, C~\textsc{ii} 1334~\AA\ line, and the O~\textsc{i} 1356~\AA\ line). We summarise our findings in Section~\ref{sec:summary}.


\section{Numerical Simulations}\label{sec:codes_descript}

\subsection{The \texttt{RADYN} code}\label{sec:radyn_descript}
The radiation hydrodynamics code \texttt{RADYN} \citep{1992ApJ...397L..59C,1997ApJ...481..500C} is a well established resource with which to study the dynamics of the flaring atmosphere. It has the facility to model flares via an injection of non-thermal particles \citep{1999ApJ...521..906A,2005ApJ...630..573A,2015ApJ...809..104A}, and more recently via dissipation of Alfv\'enic waves \citep{2016ApJ...827..101K}. We present a brief introduction here but see \cite{2015ApJ...809..104A} for a more detailed description. \texttt{RADYN} solves the coupled, non-linear equations of hydrodynamics, radiation transport, and time-dependent (non-equilibrium) atomic level populations in a 1D plane-parallel atmosphere, on an adaptive grid \citep{1987JCoPh..69..175D}, that represents one leg of a symmetric flux tube. 

Typically, the detailed NLTE, radiation transport is solved for three elements that are considered important for {the} energy balance in the chromosphere (H, He, and Ca), with continua from other species treated in LTE as background metal opacities using the Uppsala opacity package \citep{Gustaffson_1973}. Optically thin losses are included by summing all transitions from the CHIANTI atomic database, apart from those transitions solved in detail. Additional backwarming and photoionisations by soft X-ray, extreme ultraviolet, and ultraviolet radiation is included as described in \cite{2015ApJ...809..104A}. 

Radiation transport is computed assuming complete redistribution (CRD), which may overestimate radiation losses from certain transitions \citep{2002ApJ...565.1312U}. We mitigate the effects of not employing the more physically realistic (but computationally expensive) partial redistribution (PRD) by truncating the Lyman lines at 10 doppler widths, and by omitting the Mg~\textsc{ii} h \& k lines (which in part counters the overestimation of losses from the Ca~\textsc{ii} H \& K lines). Thermal conduction is a modified form of Spitzer conductivity such that it saturates to avoid exceeding the electron free streaming limit \citep{1980ApJ...238.1126S}. 

When injecting flare energy via non-thermal electron beams, the electron distribution is computed by solving the Fokker-Planck equations (note that we do not include return current effects in these simulations). The non-thermal electron distribution is described by a power law energy spectrum with total energy flux $F$~erg~cm$^{-2}$~s$^{-1}$ above a low energy cutoff $E_{c}$~keV, with spectral index $\delta$. Direct non-thermal collisional excitations and ionisations from the ground state of hydrogen and helium by the electron beams are included. We use the expressions from \cite{1993A&A...274..917F}. 

The solution from \texttt{RADYN} (where H, He, and Ca are solved in detail, and heating and cooling by their radiation affects the thermodynamic response), can then be used to solve the time-dependent, NLTE, radiation transport and non-equilibrium atomic level populations for a `minority', or `trace', species not considered in detail during the initial simulation. This approach uses the hydrodynamic variables at each timestep, stored from the initial full solution, performing only the radiation transport calculations with the assumption that the thermodynamics are unaffected by the resulting radiation. The minority species is removed from the background opacity package so as to avoid double counting. Such an approach was used by \cite{2003ApJ...597.1158J} to investigate basal emission using C~\textsc{ii}. We employ this `minority species' version of \texttt{RADYN} here to investigate Si~\textsc{iv} emission during solar flares, and refer to this as \texttt{MS\_RADYN} hereafter so as to avoid confusion with the full RHD \texttt{RADYN} flare simulations. {A point to note here is that we save the hydrodynamic variables required for use in \texttt{MS\_RADYN} at every timestep used internally during the original RHD solution, so the Si \textsc{iv} computation is not constrained by the time interval selected by the user for the main \texttt{RADYN} output}.

\subsection{Model Silicon Atom}\label{sec:model_atom}
A 30 level-with-continuum silicon atom that included the ground terms of Si~\textsc{i}, Si~\textsc{ii},  Si~\textsc{iii},  Si~\textsc{iv}, \&  Si~\textsc{v}, along with several excited levels of each charge state, was used to model 31 bound-bound transitions and 33 bound-free transitions.  The major differences between our model atom and the CHIANTI approach is the inclusion of non-equilibrium ionisation, photoionisations plus detailed radiative recombinations, charge exchange processes, and the potential effects of opacity. Details of the construction of this model atom are presented in Appendix~\ref{sec:detailedatom}.

\begin{figure}
	\centering 
	\hbox{
		\subfloat{\includegraphics[width = 0.5\textwidth, clip = true, trim = 0cm 0cm 0cm 0cm]{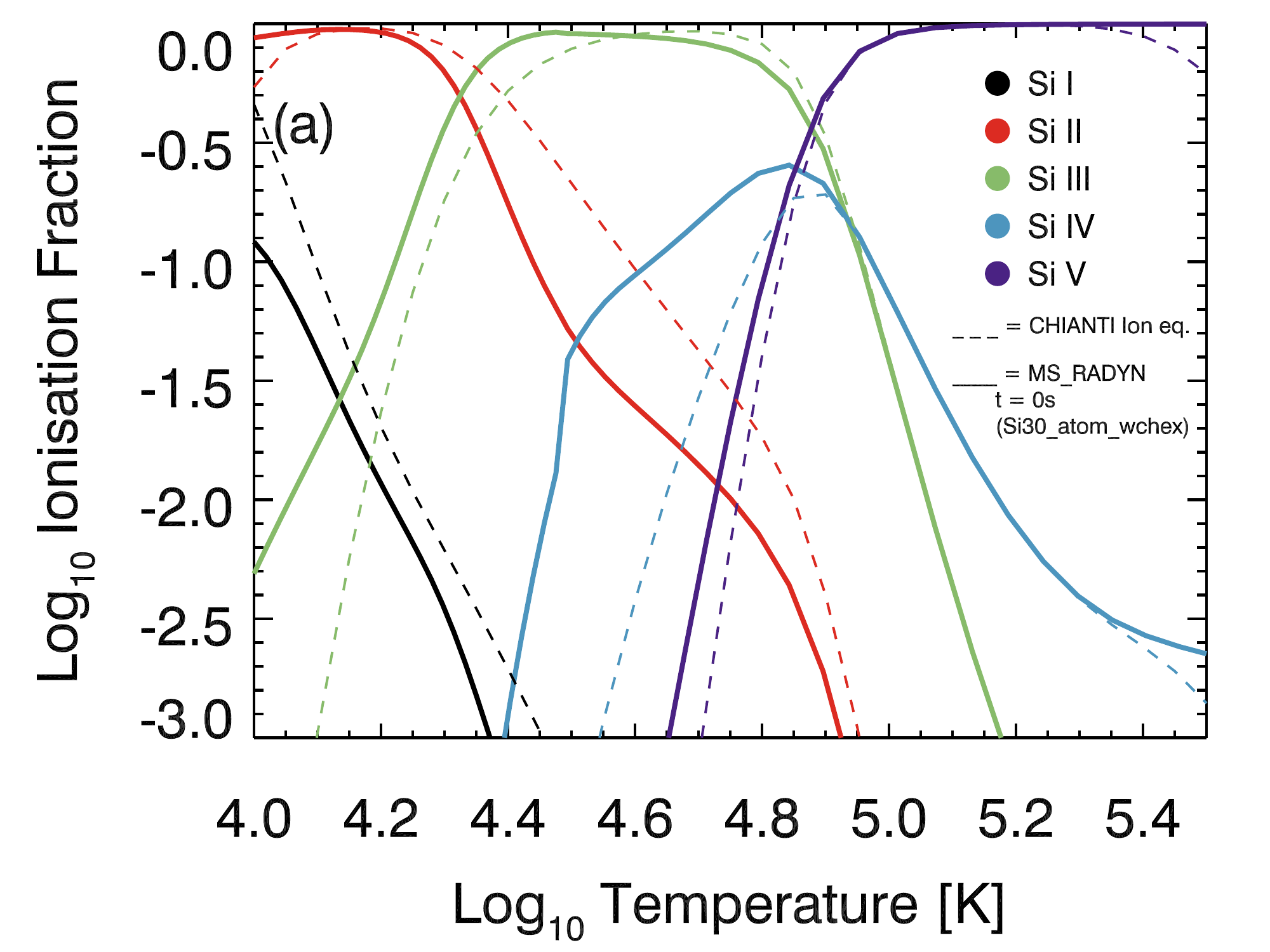}}	
		 }
	\hbox{
	        \subfloat{\includegraphics[width = 0.5\textwidth, clip = true, trim = 0cm 0cm 0cm 0cm]{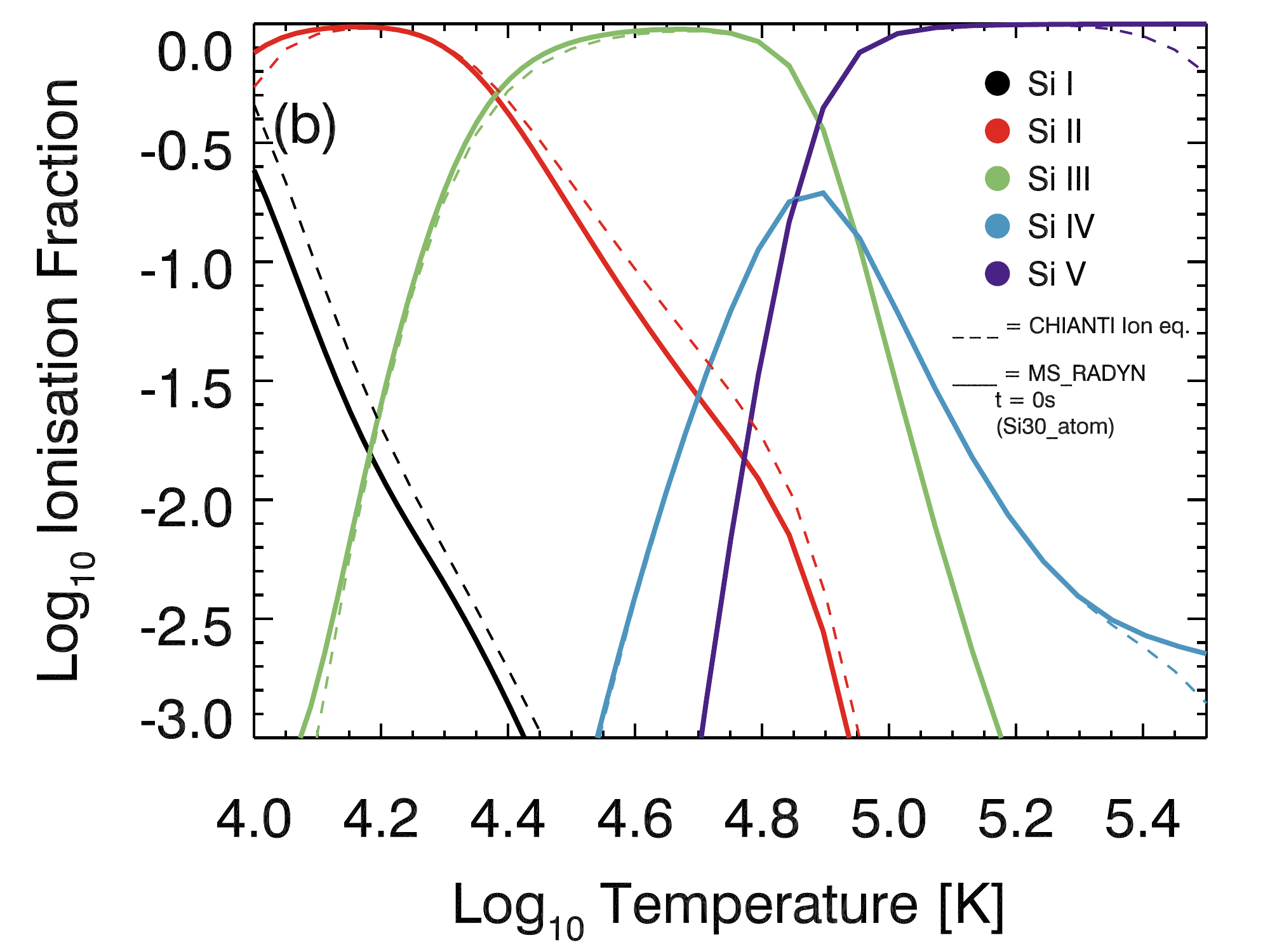}}	
                   }
	\caption{\textsl{Si ionisation fractions as functions of temperature from a \texttt{MS\_RADYN} calculation (solid lines) and from CHIANTI (dashed lines). Si~\textsc{i} (black), Si~\textsc{ii} (red), Si~\textsc{iii} (green), Si~\textsc{iv} (blue) and Si~\textsc{v} (purple) are shown. Panel (a) shows the case where charge exchange is included, where Si~\textsc{iv} peaks at $\mathrm{log}~T = 4.84$ in \texttt{MS\_RADYN}, with more emission from cooler layers, and at $\mathrm{log}~T = 4.90$ from CHIANTI. Panel (b) is the case without charge exchange, illustrating the importance of this process.}}
	\label{fig:si_ionfracs}
\end{figure}

Comparing the ionisation fractions at $t=0$~s in a \texttt{MS\_RADYN} simulation to those predicted by the ionisation equilibrium in CHIANTI reveals that the inclusion of photoionisations plus detailed radiative recombinations, and charge exchange in our model atom, results in an increased fraction of Si~\textsc{iv} at lower temperatures, a higher and broader peak, and a shift to a somewhat lower peak temperature of $\mathrm{log}~T = 4.84$ ($T = 69.6$~kK) compared to the CHIANTI equilibrium ionisation value of  $\mathrm{log}~T = 4.9$ ($T = 79.4$~kK). Other differences are present in Si~\textsc{i}, \textsc{ii} \& \textsc{iii} also. This is shown in Figure~\ref{fig:si_ionfracs}(a) where the dashed lines are the CHIANTI ionisation equilibrium fractions, and the solid lines are the \texttt{MS\_RADYN} fractions. Removing charge exchange yields ionisation fractions that are very similar to CHIANTI (Figure~\ref{fig:si_ionfracs}(b)), highlighting the importance of this process. If we do not include charge exchange or photoionisations, treating recombinations and ionisations as in CHIANTI, then we obtain ionisation fractions that match CHIANTI. Hereafter we refer to the 30 level-with-continuum model atom that includes charge exchange as \texttt{Si30\_atom\_wchex}, and the model atom that excludes charge exchange (but which is otherwise the same) as \texttt{Si30\_atom}. Unless specified all results use \texttt{Si30\_atom\_wchex}, though we at times show \texttt{Si30\_atom} results for illustration of the differences. Since charge exchange is a process dependent on the density of hydrogen or helium, and their ions, the ionisation fractions will vary somewhat with different pre-flare atmospheres.

\subsection{Parameter Survey}\label{sec:param_survey}
\begin{figure*}
	\centering
		\includegraphics[width = \textwidth, clip = true, trim = .8cm .1cm 0.25cm .5cm ]{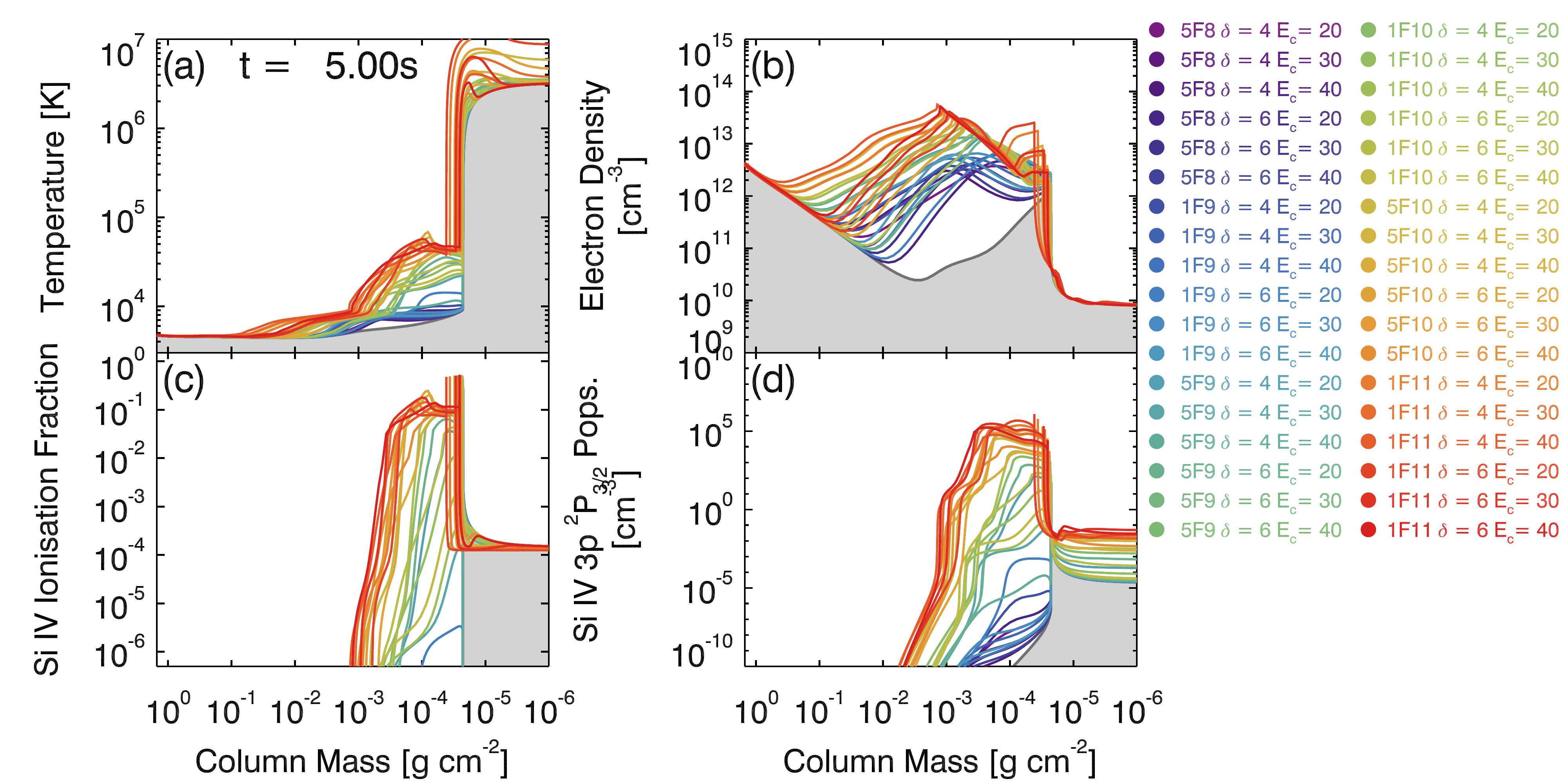}	
          	\caption{\textsl{The flare atmospheres after $t=5$~s of energy injection. Colour represents the simulation, with the beam parameters indicated. Panel (a) shows temperature, (b) shows electron density, (c) shows the Si~\textsc{iv} ionisation fraction, and (d) the population density of the Si~\textsc{iv} 1402.77\AA\ upper level. All properties are functions of column mass, where the leftmost side is the photosphere and the rightmost the corona. The grey shaded areas show $t=0$~s. The Si~\textsc{iv} data were computed using \texttt{MS\_RADYN} with \texttt{Si30\_atom\_wchex}}}
	\label{fig:atmos_response}
\end{figure*}

Thirty six flare simulations were produced, covering a wide range of energy fluxes, and several typical values of $E_{c}$ and $\delta$, in order to assess the effect of both the strength of energy injection, as well as location of peak deposition, on the formation of the {Si}~\textsc{iv} resonance lines. Flare energy was injected into a $10$~Mm loop spanning the sub-photosphere through to the corona, which was initially in radiative equilibrium. The temperature at the loop apex was $T = 3.2$~MK and the hydrogen number density was $n_{H} = 6.6\times10^{9}$~cm$^{-3}$. The photospheric temperature was $T=5800$~K, and the transition region was initially located at an altitude of $z\approx1.156$~Mm (column mass, $c_{\mathrm{mass}} \approx 2.25\times10^{-5}$~g~cm$^{-2}$). To maintain coronal and photospheric energy balance, additional non-radiative heating was applied to grid cells with column mass $c_{\mathrm{mass}} < 1\times10^{-6}$~g~cm$^{-2}$ and $c_{\mathrm{mass}} >7.6$~g~cm$^{-2}$. To mimic the effect of incoming disturbances from the other leg of the flaring loop we employ a reflecting boundary condition at the loop apex while a transmitting boundary was used at the base. The pre-flare atmosphere is shown as the grey shaded portions of Figure~\ref{fig:atmos_response}.

Using the notation $X\mathrm{F}Y$ to mean $X\times10^{Y}$~erg~cm$^{-2}$~s$^{-1}$, the electron beam parameters simulated were: \\$F = [5\mathrm{F}8,~1\mathrm{F}9,~5\mathrm{F}9,~1\mathrm{F}10,~5\mathrm{F}10,~1\mathrm{F}11]$, $\delta = [4,~6]$ and $E_{c} = [20,~30,~40]$~keV. Energy was injected at a constant rate for $t=10$~s, and the atmospheres allowed to cool for a subsequent $40$~s. A moderate-to-strong flare might have an energy flux on the order $>5\mathrm{F}10$, weaker flares on the order $\approx1\mathrm{F}9-1\mathrm{F}10$, and microflares (or smaller heating events) on the order $<1\mathrm{F}9$. Our simulations therefore cover a wide range of heating events (of course, if the typical flare footpoint sizes have been underestimated due to the spatial resolution of observations then our stronger flare may actually represent only modest flares). Hereafter we refer to simulations as, for example, $1\mathrm{F}11\delta4\mathrm{E_{c}}20$ to mean a simulation with parameters $F = 1\times10^{11}$~erg~cm$^{-2}$~s$^{-1}$, $E_{c} = 20$~keV, and $\delta = 4$.

Each of the flare simulations was then used as input along with the Si atom \texttt{Si30\_atom\_wchex} described in Section~\ref{sec:model_atom} in the \texttt{MS\_RADYN} code.

\subsection{Atmospheric Response}\label{sec:hydro}

In each simulation the injection of flare energy resulted in temperature and electron density enhancements. This resulted in an increased population of {Si}~\textsc{iv}, which occurred more rapidly in the stronger flares. Figure~\ref{fig:atmos_response} summarises the atmospheric response, showing the temperature, electron density, {Si}~\textsc{iv} ionisation fraction, and the population of the {Si}~\textsc{iv} 1402.77\AA\ upper level, as functions of column mass at $t=5$~s.

The weaker flares produced only modest enhancements where chromospheric temperatures barely exceed $T\approx6-10$~kK and electron densities reached a few~$\times10^{12}$~cm$^{-3}$ in the upper chromosphere. {In these simulations the temperature enhancement is not sufficient to produce a substantial increase of the Si~\textsc{iv} fraction.} It is clear that the stronger flares produced a much more dramatic response, with temperatures exceeding $T = 20$~kK to $T >50$~kK in the mid-to-upper flare chromosphere. Note that here we are defining the flare chromospheres as the regions below which the temperatures climb, and electron densities drop precipitously through the transition region to the corona (i.e. below $T\gtrapprox100$~kK and $n_{e}\lessapprox10^{10}$~cm$^{-3}$). The enhanced temperature over a wider swathe of the chromosphere and TR results in a substantial increase of the Si~\textsc{iv} populations, such that the formation location of the Si~\textsc{iv} resonance lines shifts lower in altitude.

The thickness of the atmosphere over which Si~\textsc{iv} populations are enhanced changes over time. Simulations which initially have a high Si~\textsc{iv} fraction over a large geometric region, may end with only a narrow layer of enhanced Si~\textsc{iv} if the atmosphere compresses. The opposite can also be true so that some simulations build more slowly to a sufficient temperature and density structure to allow the fraction of Si~\textsc{iv} to increase. Generally speaking, the {larger} the energy injected is, and the softer the electron beam is (that is, the higher proportion of lower energy electrons that deposit their energy in the upper atmosphere), then the larger the enhancement to the Si~\textsc{iv} density.


\section{Formation of {Si}~\textsc{iv} in Flares}\label{sec:siiv_form}
	\subsection{Line Profiles and Ratios}\label{sec:profiels_ratio}

To compare the results of solving the full NLTE, non-equilibrium radiation transfer problem, rather than producing synthetic line profiles assuming $\textsl{a priori}$ that they are optically thin, we show the resonance line profiles, ratios and lightcurves computed via each method. 

To compute the optically thin case, that we refer to as \texttt{Model B} hereafter, we used the following procedure. The contribution function, $G(\lambda,n_e,T)$~[erg~cm$^{-3}$~s$^{-1}$~sr$^{-1}$], allows the calculation of the intensity emitted at each depth point $z$ in our flare atmospheres, $I_{\lambda z}$. These were constructed using standard CHIANTI software, and contain information about various atomic properties (ionisation and recombination rates etc.,). Ionisation equilibrium was assumed here. $G(\lambda,n_e,T)$'s were obtained from the atomic data in CHIANTI V8.0.7, and were tabulated with a resolution of  $\delta \mathrm{log}~T = 0.05$ and  $\delta \log n_{e} = 0.5$. At each grid cell and each time in the \texttt{RADYN} atmospheres we interpolate $G(\lambda,n_e,T)$ to the electron density and temperature in that grid cell, then compute $I_{\lambda z}$ by:
	
	\begin{equation}\label{eq:emiss_chianti}
		I_{\lambda z} =A_{b}G(\lambda,n_e,T) n_{e}(z)n_{H}(z)\delta z,
	\end{equation}

\noindent where $A_{b}$ is the abundance (we used photospheric abundance from \citealt{2009ARA&A..47..481A}, see Appendix~\ref{sec:coronalabundance}), $\delta z$ is the size of the grid cell, and $n_{H}$ is the hydrogen density. Within each grid cell the line is thermally broadened using the local temperature, and Doppler shifted by any velocity gradients. The intensity is then the sum of the contribution from each grid cell along the whole loop, $I_{\lambda} = \sum_{z} I_{\lambda z}$. In the wavelength range selected, only the $G(\lambda,n_e,T)$'s associated with the resonance line transitions were considered. Other species/transitions that contribute to the solar spectrum within that wavelength range were omitted. This will not affect the intensity of Si~\textsc{iv} produced, but permits a simpler analysis of line ratios, and comparison to \texttt{MS\_RADYN} results, which also do not include blends from other species. 

For both methods we show results from disk centre. We assume no geometric effects on Doppler shifts since Si~\textsc{iv} forms relatively low in the atmosphere in relation to the length of the loop, and in a narrow region (i.e. we assume the portion of the loop in which the resonance lines originate is vertical).

\begin{figure}
	\centering 
	\hbox{
		\subfloat{\includegraphics[width = 0.5\textwidth, clip = true, trim = 0cm 0cm 0cm 0cm]{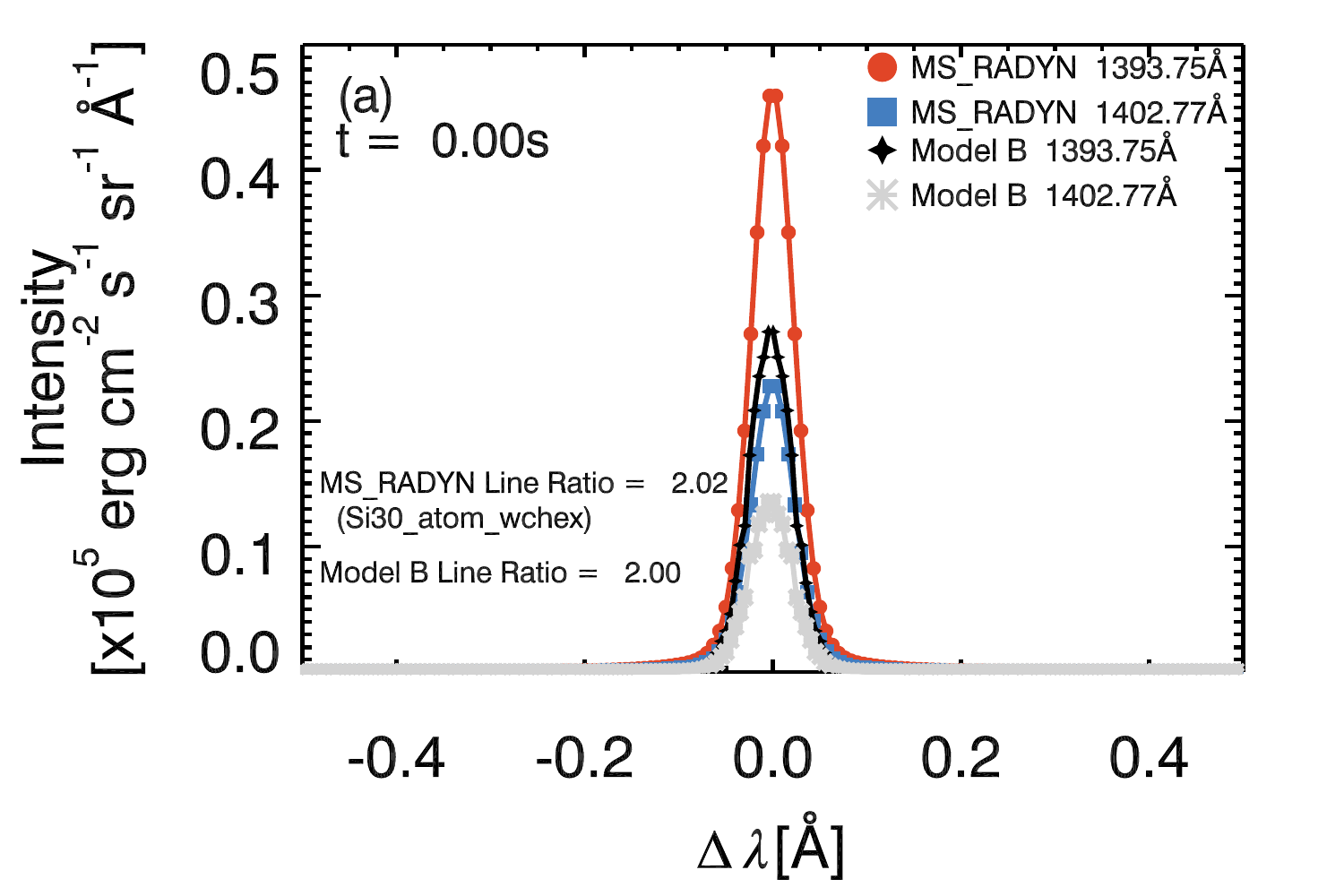}}	
		 }
	\hbox{
	        \subfloat{\includegraphics[width = 0.5\textwidth, clip = true, trim = 0cm 0cm 0cm 0cm]{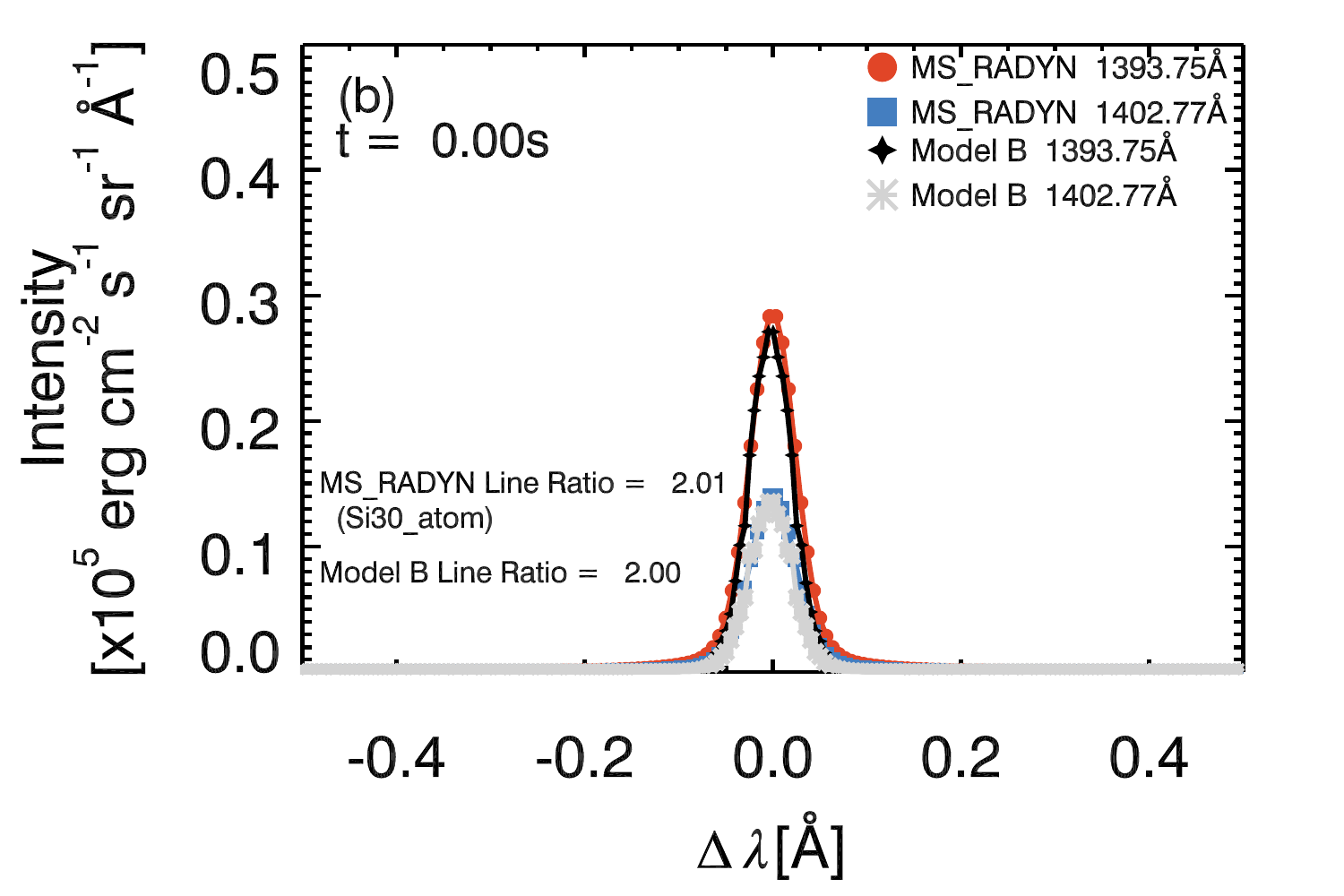}}	
                   }
	\caption{\textsl{Comparing resonance line profiles from \texttt{MS\_RADYN} ($1393$\AA\ red circles, and $1402$\AA\ blue squares), and \texttt{CHIANTI} ($1393$\AA\ black diamonds, and $1402$\AA\ grey stars), at $t=0$~s in our simulations. Panel (a) shows \texttt{MS\_RADYN} results using model atom \texttt{Si30\_atom\_wchex} that includes charge exchange. Panel (b) shows model atom \texttt{Si30\_atom} that excludes charge exchange.}}
	\label{fig:compare_t0}
\end{figure}

Resonance lines profiles at $t=0$~s from each method are shown in Figure~\ref{fig:compare_t0}. Panel (a) shows the case with charge exchange, whereas panel (b) shows the case without. While there are intensity differences between the two models when charge exchange is included, the line shapes are comparable. In both cases the resonance line ratio $R_{\mathrm{res}} = I_{1393}/I_{1402}\approx2$, as would be expected in the optically thin scenario. Excluding charge exchange results in a very close match between \texttt{MS\_RADYN} and \texttt{Model B}.

\begin{figure*}[h]
	\centering
	\hbox{
	\hspace{0.2in}
		\subfloat{\includegraphics[width = 0.225\textwidth, clip = true, trim = 0cm 0cm 0cm 0cm]{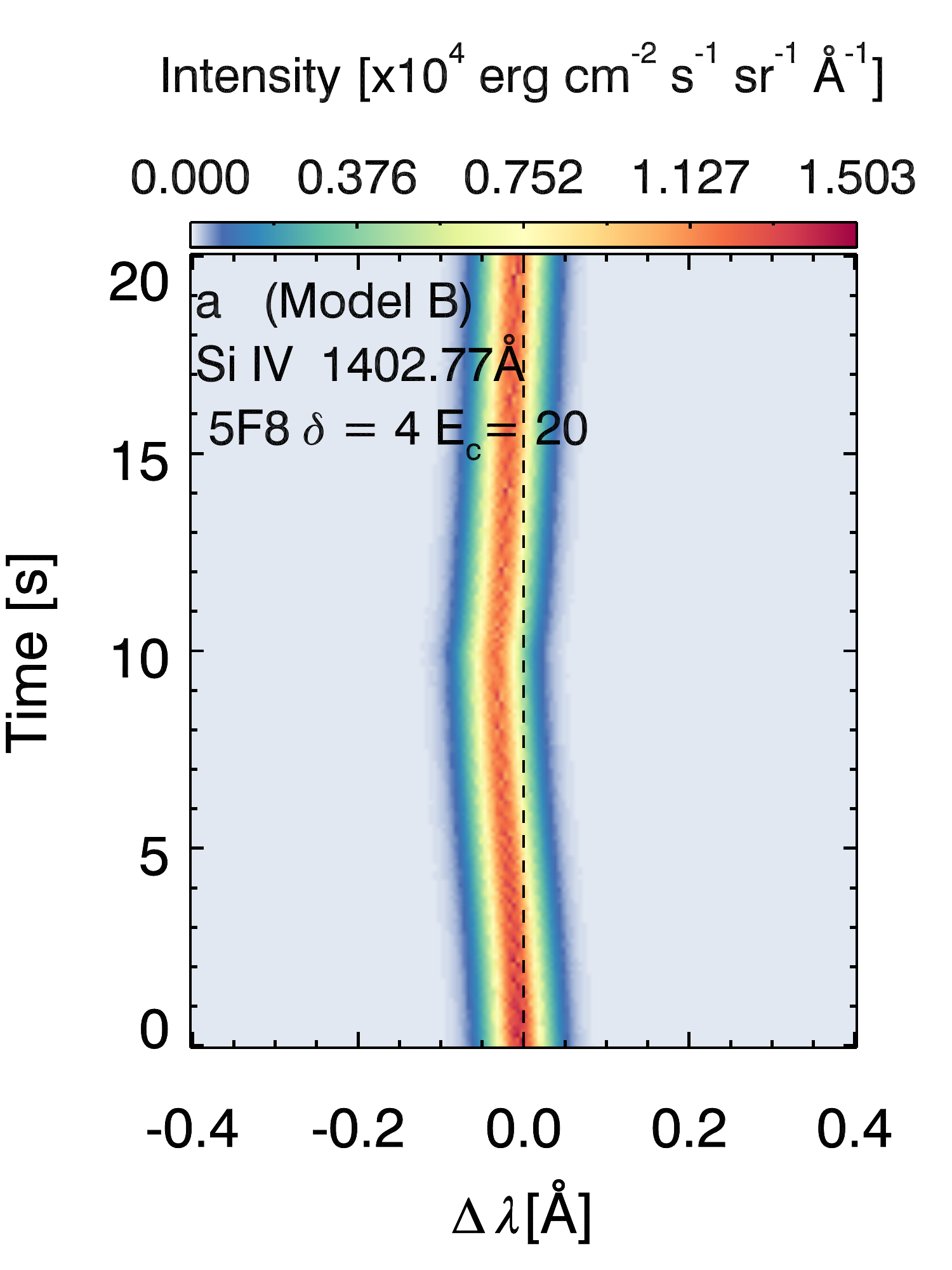}}	
	\hspace{-0.175in}
		\subfloat{\includegraphics[width = 0.225\textwidth, clip = true, trim = 0cm 0cm 0cm 0cm]{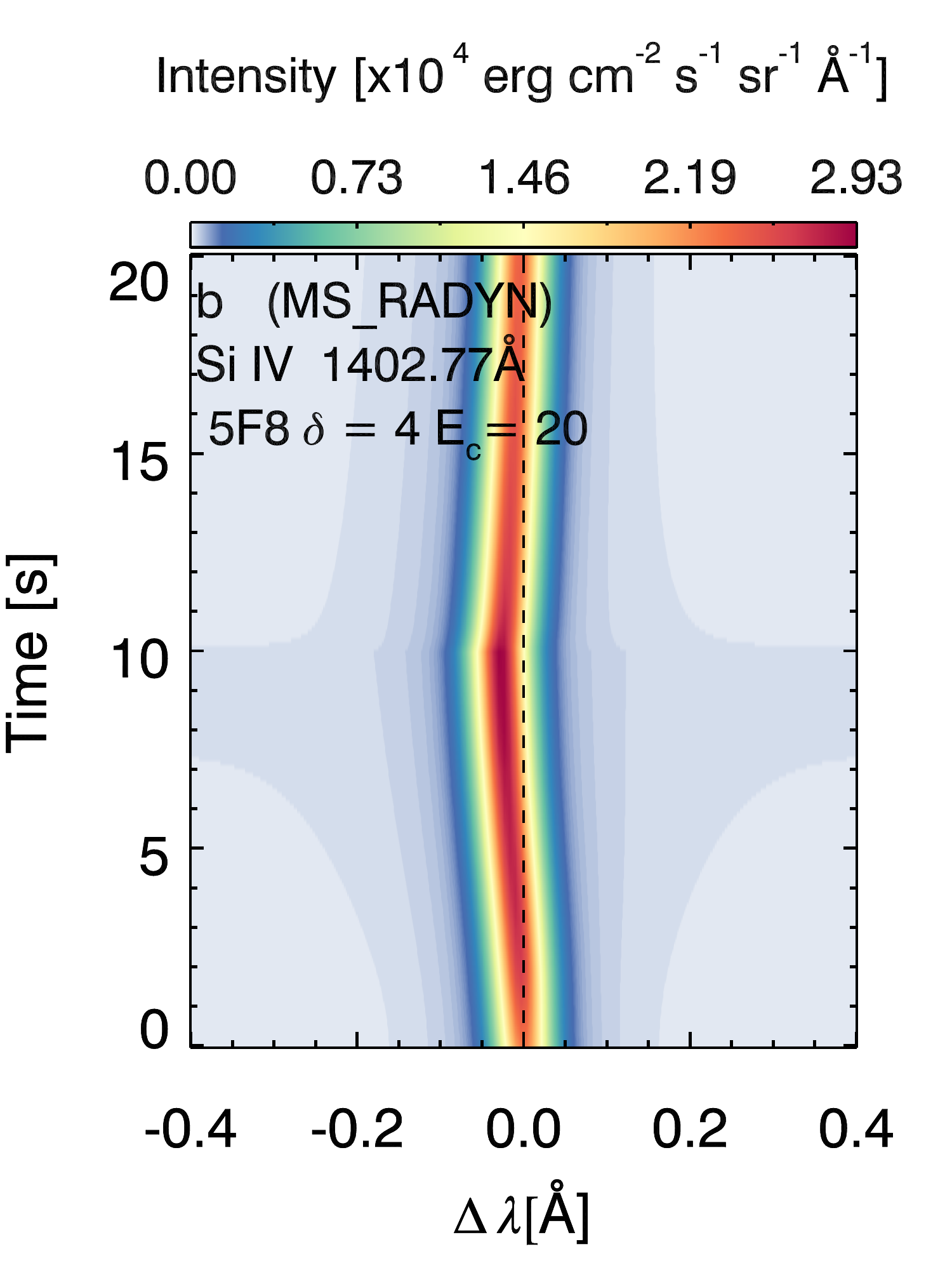}}
	\hspace{0.35in}
		\subfloat{\includegraphics[width = 0.225\textwidth, clip = true, trim = 0cm 0cm 0cm 0cm]{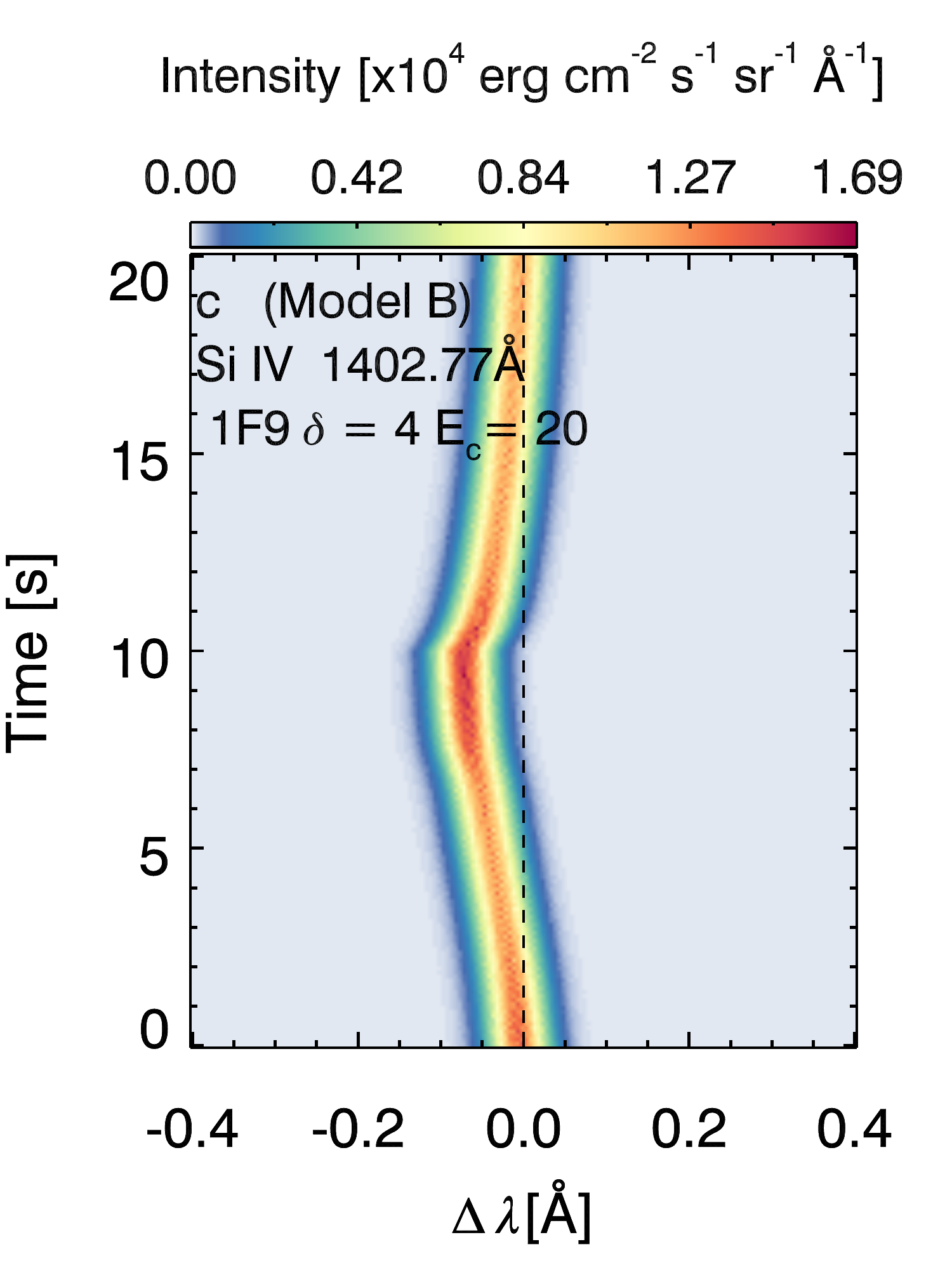}}	
	\hspace{-0.175in}
		\subfloat{\includegraphics[width = 0.225\textwidth, clip = true, trim = 0cm 0cm 0cm 0cm]{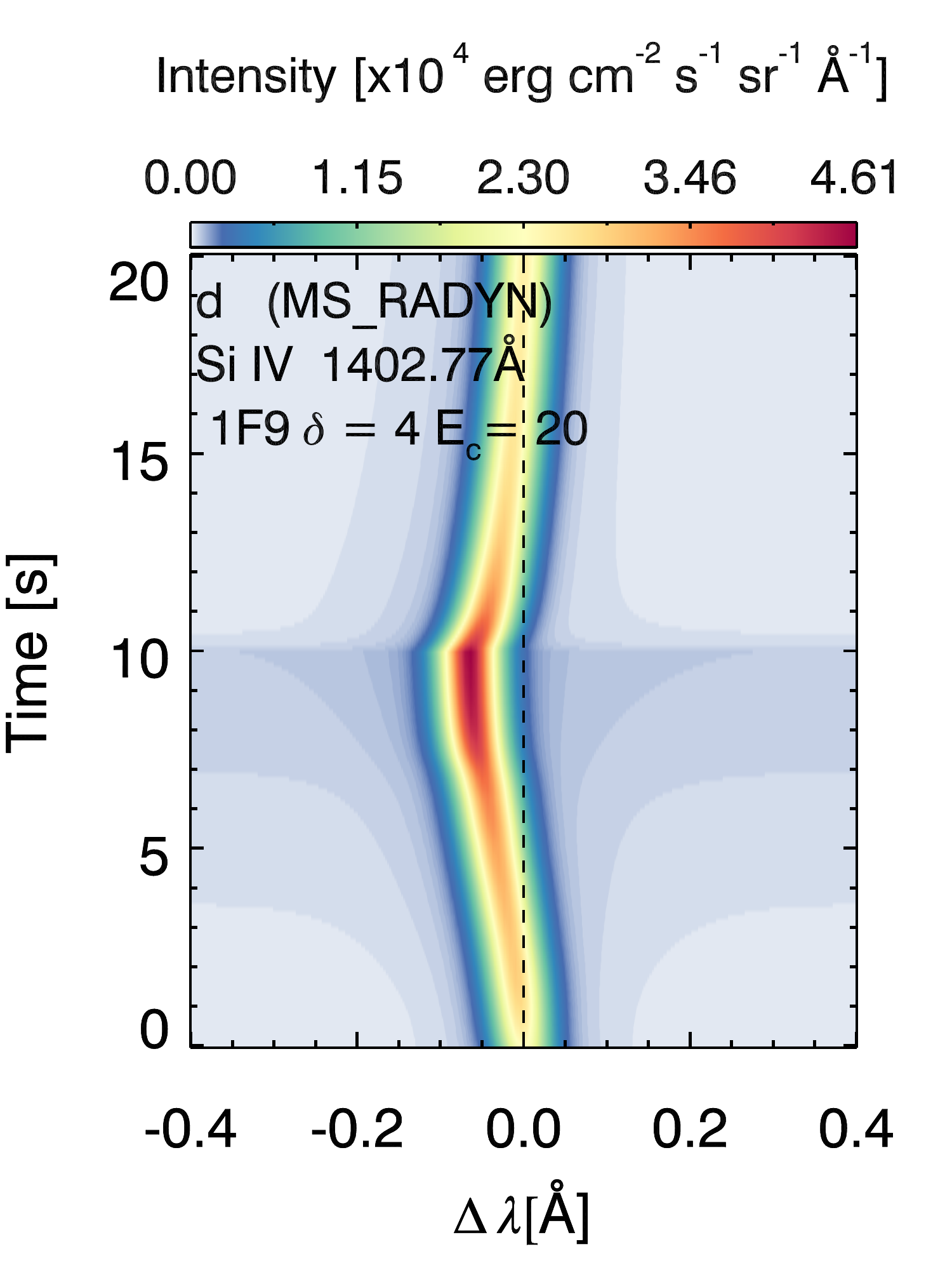}}
	         }
	 \vspace{-0.25in}
	  \hbox{
	  \hspace{0.2in}
		\subfloat{\includegraphics[width = 0.225\textwidth, clip = true, trim = 0cm 0cm 0cm 0cm]{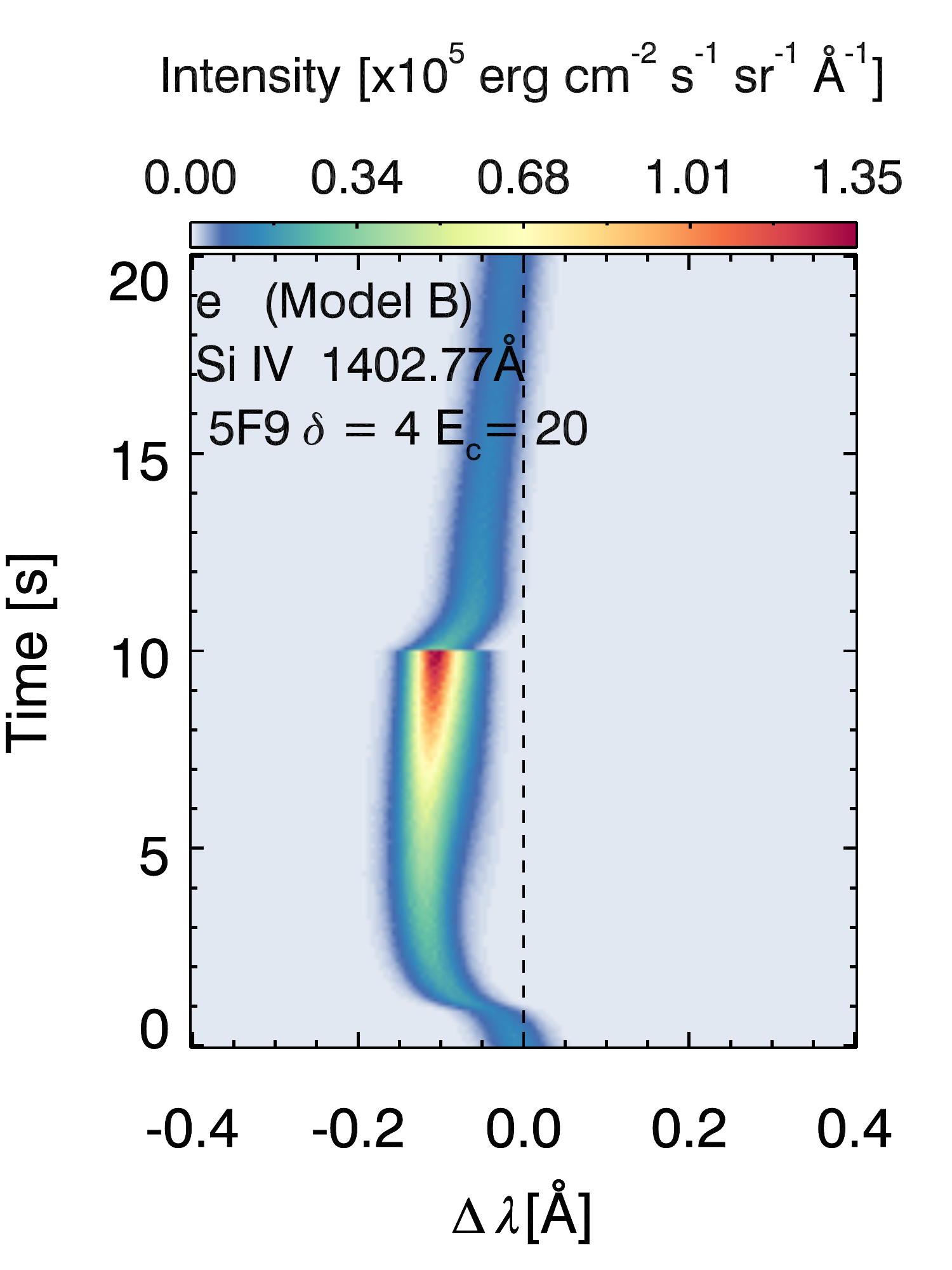}}	
	\hspace{-0.175in}
		\subfloat{\includegraphics[width = 0.225\textwidth, clip = true, trim = 0cm 0cm 0cm 0cm]{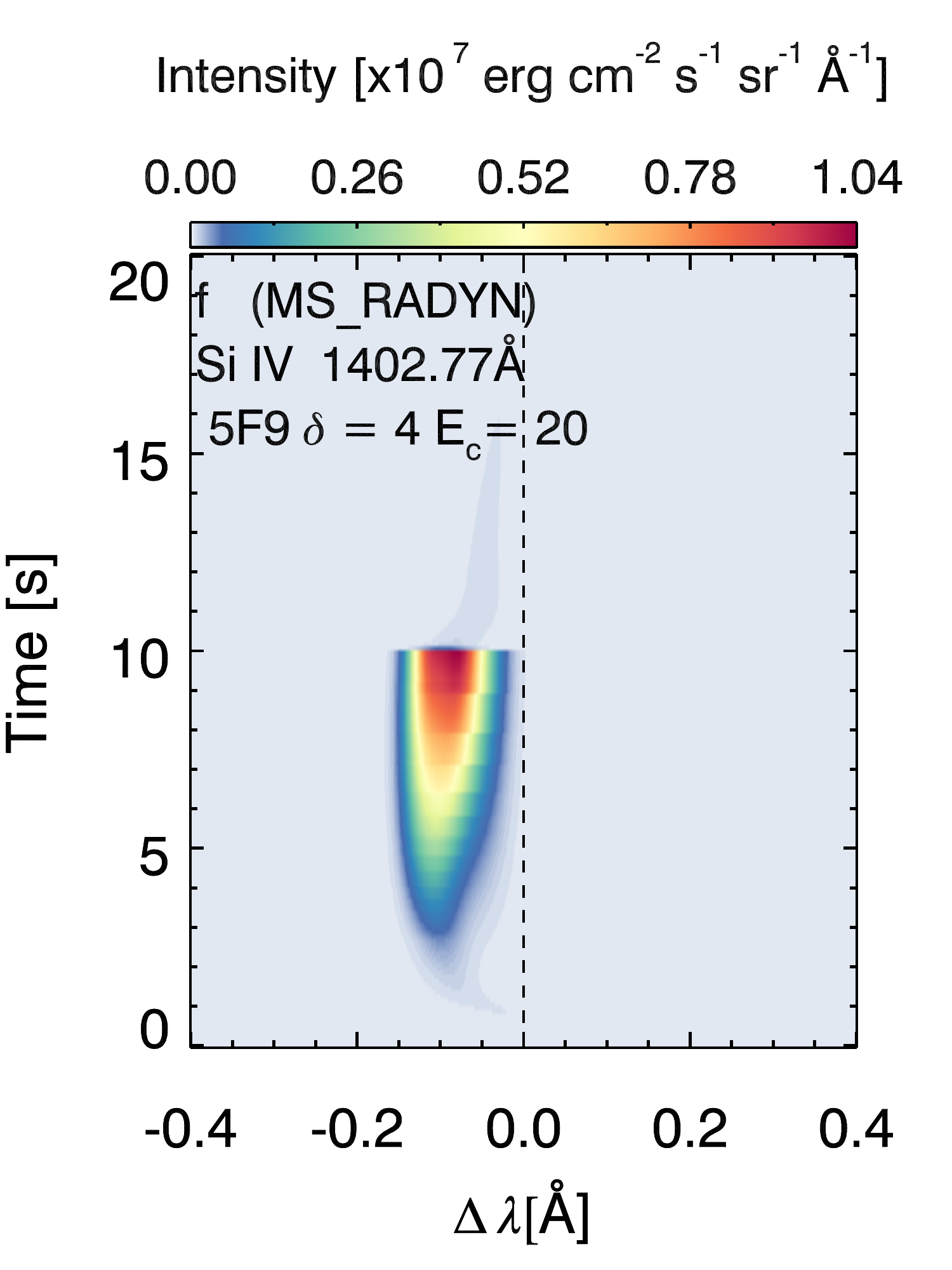}}
	\hspace{0.35in}
		\subfloat{\includegraphics[width = 0.225\textwidth, clip = true, trim = 0cm 0cm 0cm 0cm]{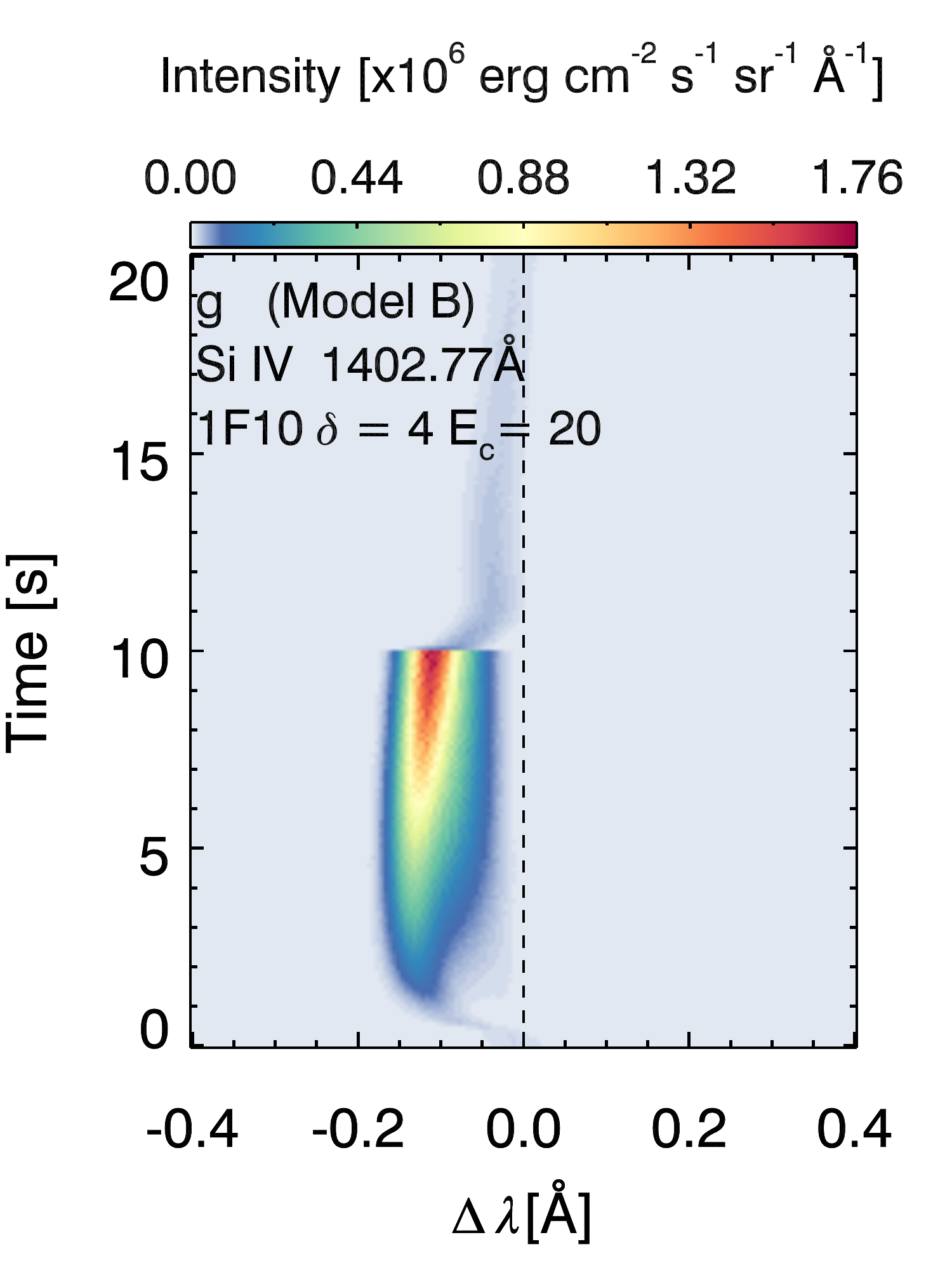}}	
	\hspace{-0.175in}
		\subfloat{\includegraphics[width = 0.225\textwidth, clip = true, trim = 0cm 0cm 0cm 0cm]{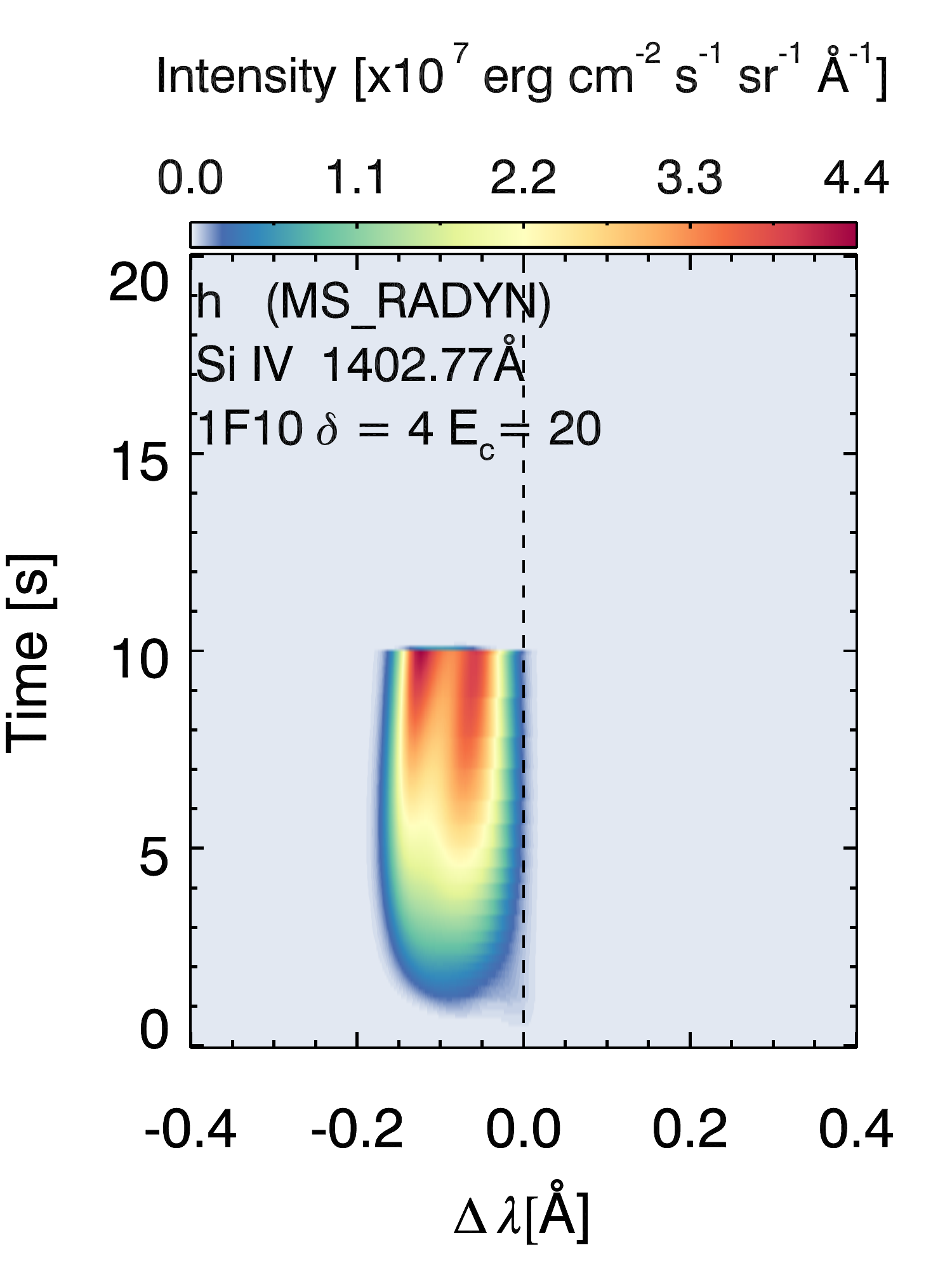}}
	         }
	 \vspace{-0.25in}
	  \hbox{
	  \hspace{0.2in}
		\subfloat{\includegraphics[width = 0.225\textwidth, clip = true, trim = 0cm 0cm 0cm 0cm]{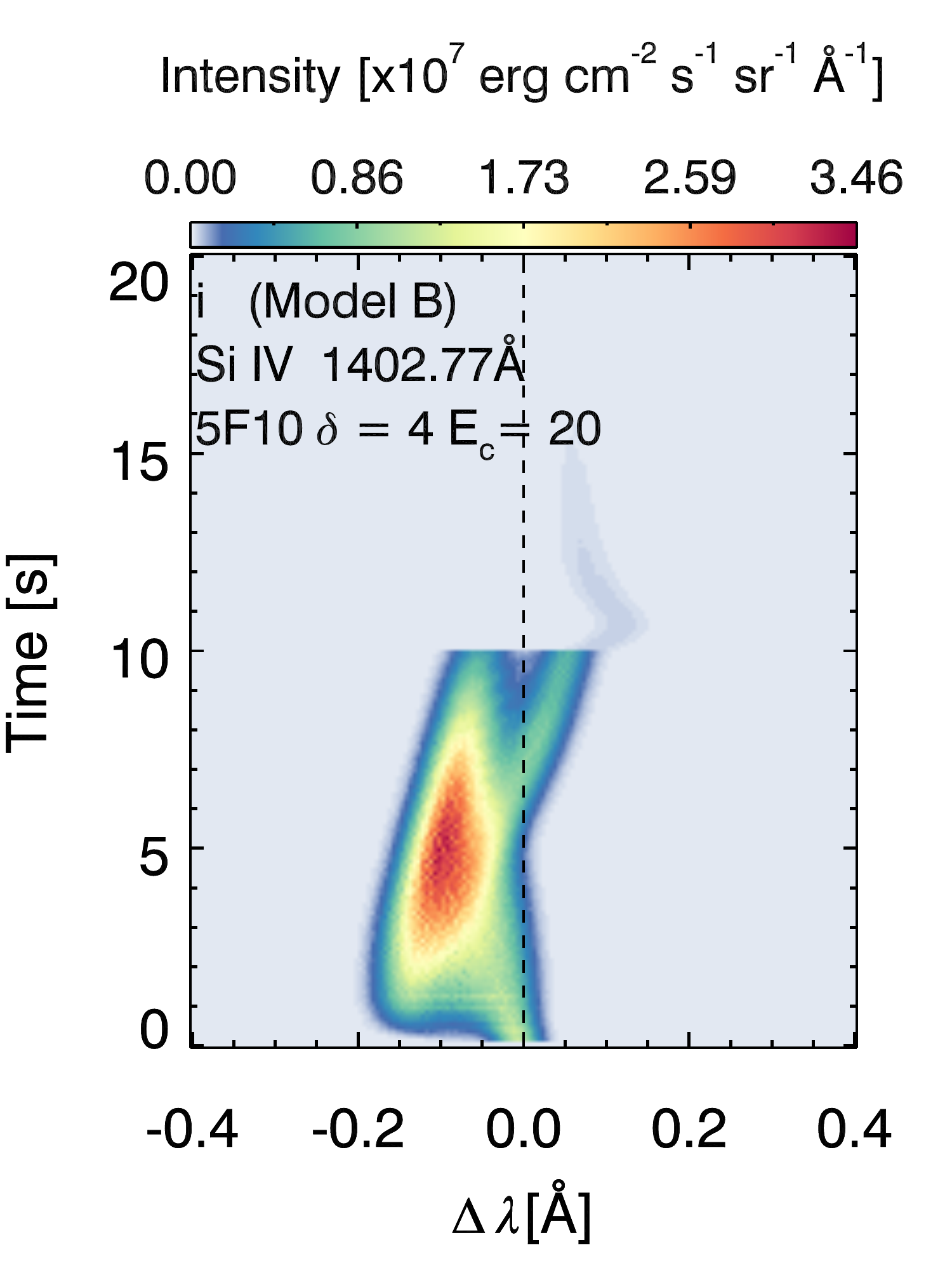}}	
	\hspace{-0.175in}
		\subfloat{\includegraphics[width = 0.225\textwidth, clip = true, trim = 0cm 0cm 0cm 0cm]{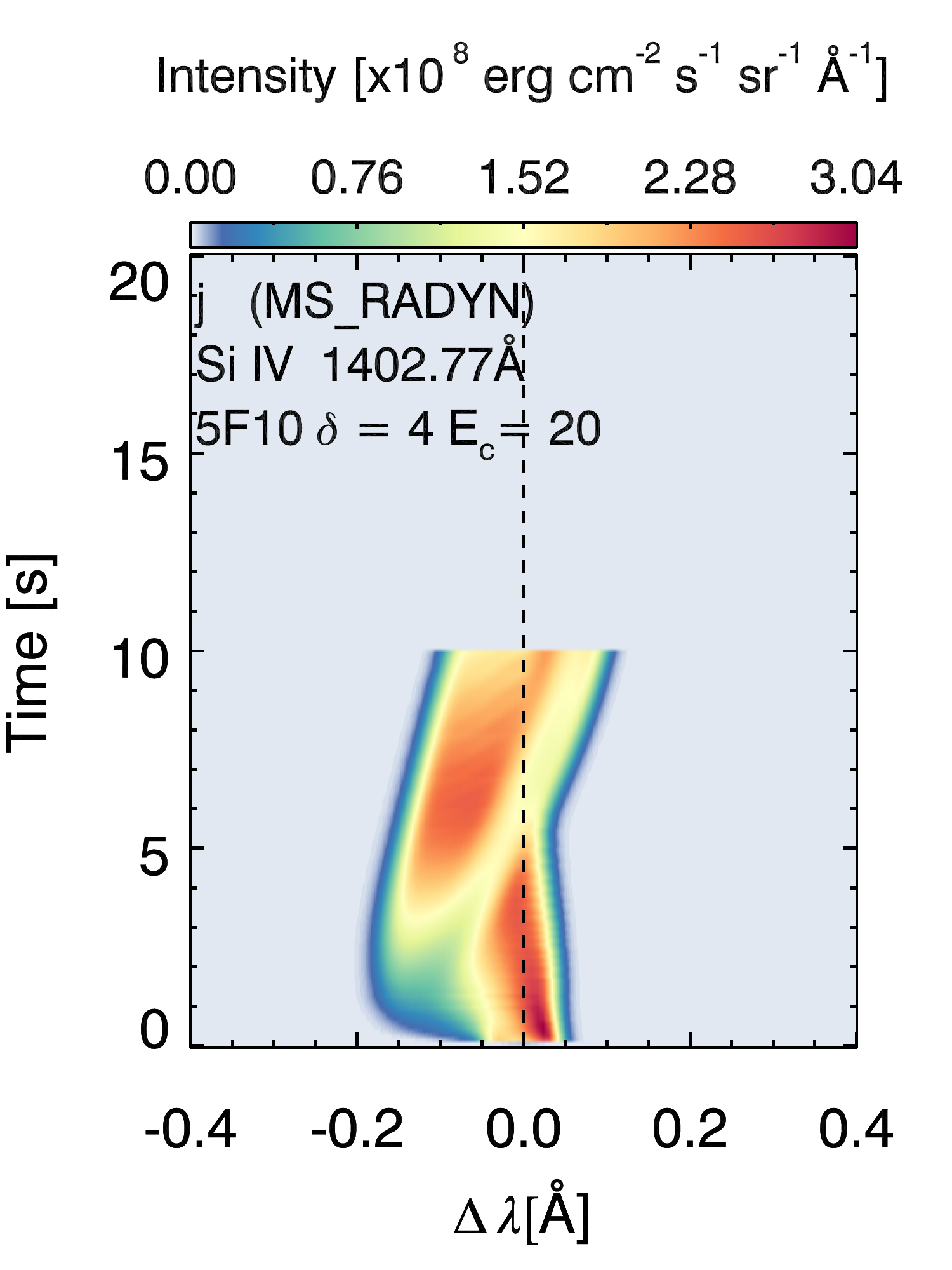}}
	\hspace{0.35in}
		\subfloat{\includegraphics[width = 0.225\textwidth, clip = true, trim = 0cm 0cm 0cm 0cm]{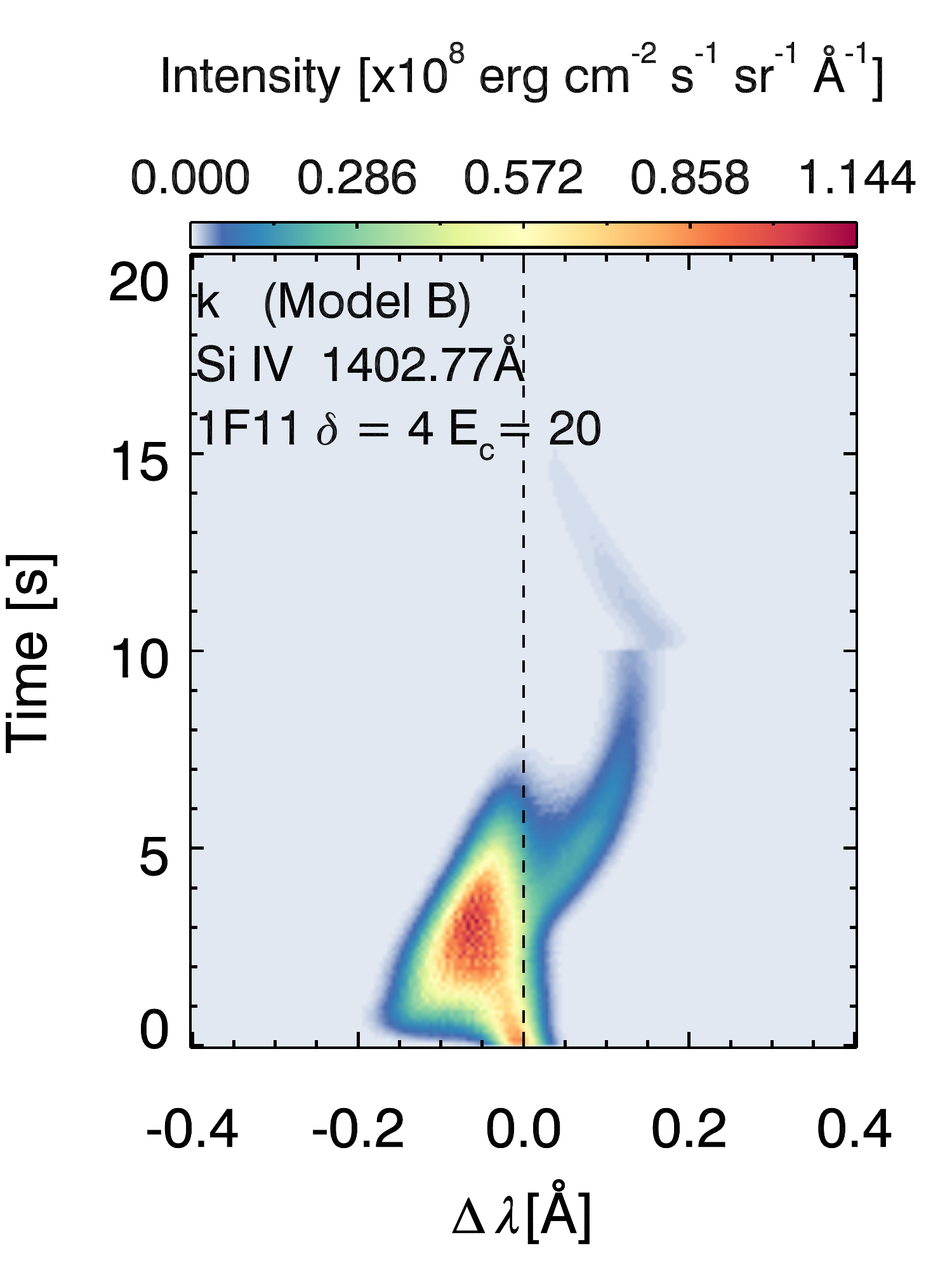}}	
	\hspace{-0.175in}
		\subfloat{\includegraphics[width = 0.225\textwidth, clip = true, trim = 0cm 0cm 0cm 0cm]{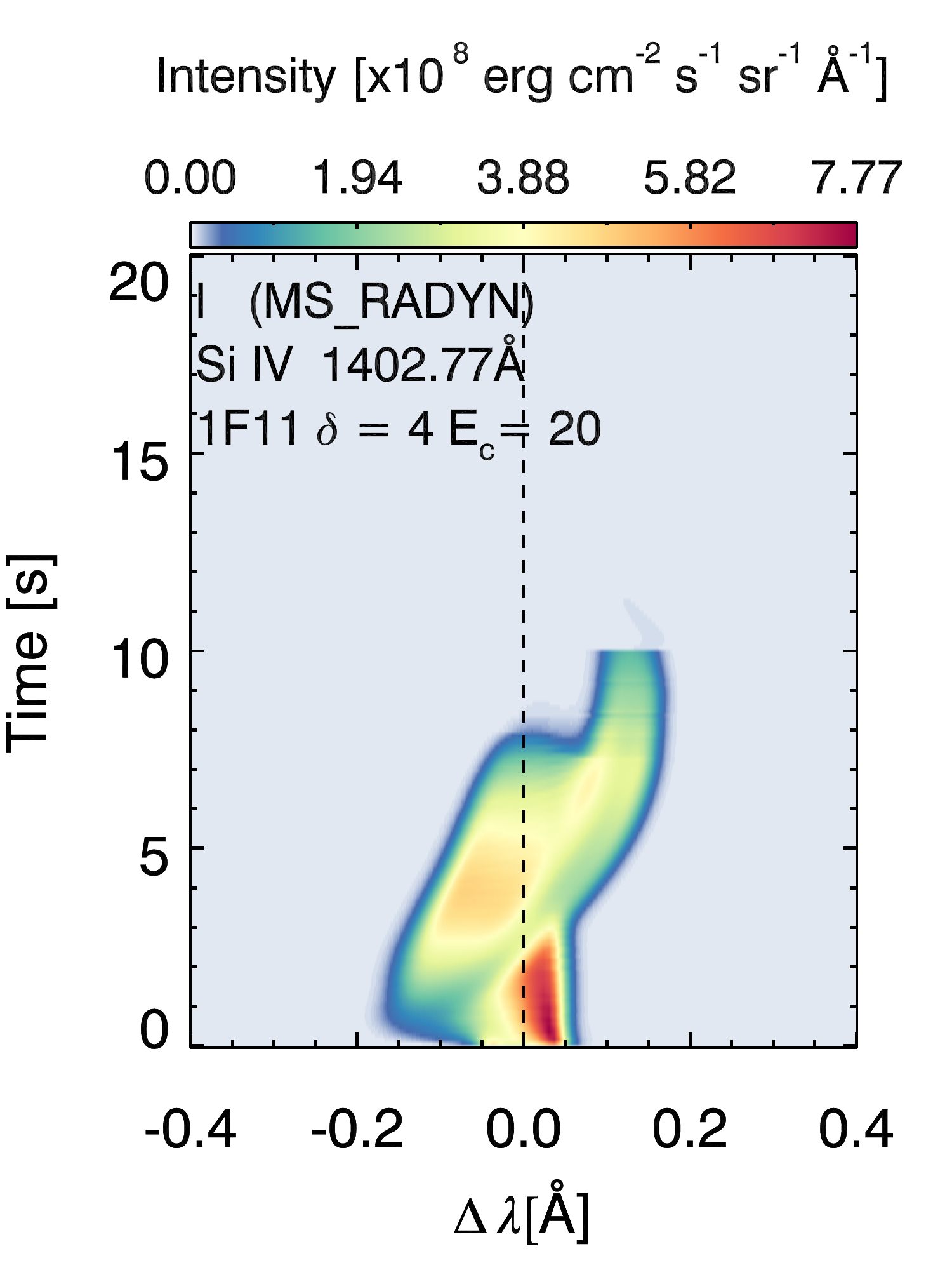}}
	         }
	\caption{\textsl{Si~\textsc{iv} 1402.77\AA\ line profiles in the first 20~s of the flare simulations computed using \texttt{Model B} (a,c,e,g,i,k), and computed using \texttt{MS\_RADYN} (b,d,f,h,j,l). Six flare simulations are shown. Each has non-thermal electron distribution parameters of $\delta=4$ and $E_{c} = 20$~keV, with the injected energy flux varying for each pair of images: 5F8 (a,b), 1F9 (c,d), 5F9 (e,f), 1F10 (g,h), 5F10 (i,j) 1F11 (k,l). Within each image the rows are different times in the simulation, so that the profiles are shown as wavelength vs time. Note that the scale is different for each panel, indicated in the relevant colour bars. For simulations with very intense emission in the heating phase, the cooling phase is not visible due to scaling.}}
	\label{fig:stackplots_1}
\end{figure*}

During the flares the Si~\textsc{iv} resonance lines exhibit changes in line intensity, shape and Doppler motions. These changes range from modest in the lower energy simulations to dramatic in the higher energy flares. Figure~\ref{fig:stackplots_1} illustrates the effect of varying the injected energy flux for simulations with fixed spectral index and low energy cutoff ($\delta = 4$ and $E_{c} = 20$~keV), on the $\lambda=1402.77$~\AA\ line. This figure also shows the differences that result from using either \texttt{Model B} or \texttt{MS\_RADYN} to synthesise the line. For each of the six values of energy flux two pairs of images are presented showing a stackplot of line intensity as functions of wavelength and time. The lefthand image of each pair shows the line computed via \texttt{Model B} and the righthand image shows the line computed by \texttt{MS\_RADYN}. The colour scale varies from panel-to-panel so as to illustrate the evolution of each simulation over the large intensity range present between the weaker and stronger flares. The sharp discontinuity in some panels at $t\sim10$~s is due to the rapid cooling of the atmosphere following cessation of the electron beam. In those cases the flare enhancements were so strong that the scaling on the figures makes it difficult to see the cooling phase. 
 
For 5F8 the response is very similar between the two techniques, with intensity enhancements, slight blueshifts, but overall a fairly symmetric line profile. The \texttt{MS\_RADYN} profile is somewhat more intense. Similarly, the 1F9 simulation produces fairly comparable profiles, albeit with an increased intensity difference. As discussed later, in these simulations the \texttt{MS\_RADYN} computed emission is optically thin, with differences resulting from using a full radiation transfer approach.

Further increasing the energy flux does eventually lead to marked differences between the two techniques. In the 5F9, 1F10, 5F10 and 1F11 simulations \texttt{Model B} produced Si~\textsc{iv} emission that is much weaker than the equivalent \texttt{MS\_RADYN} profiles (by an order of magnitude in some cases), is more symmetric, and is, generally, single peaked. In contrast the \texttt{MS\_RADYN} Si~\textsc{iv} emission is broader, with multiple line components, self-absorption features, strong asymmetries, and do not always share the same character of Doppler shift as the \texttt{Model B} profiles. The profiles are generally flatter, with wider cores. These difference arise because the Si~\textsc{iv} emission is affected by optical depth effects in these more energetic simulations. For these electron beam parameters, there is a transition from blueshift to redshift that occurs for injected fluxes $>1\mathrm{F}10$. 

\begin{figure*}
	\centering
	\hbox{
	\hspace{1.125in}
	\subfloat{\includegraphics[width = 0.225\textwidth, clip = true, trim = 0cm 0cm 0cm 0cm]{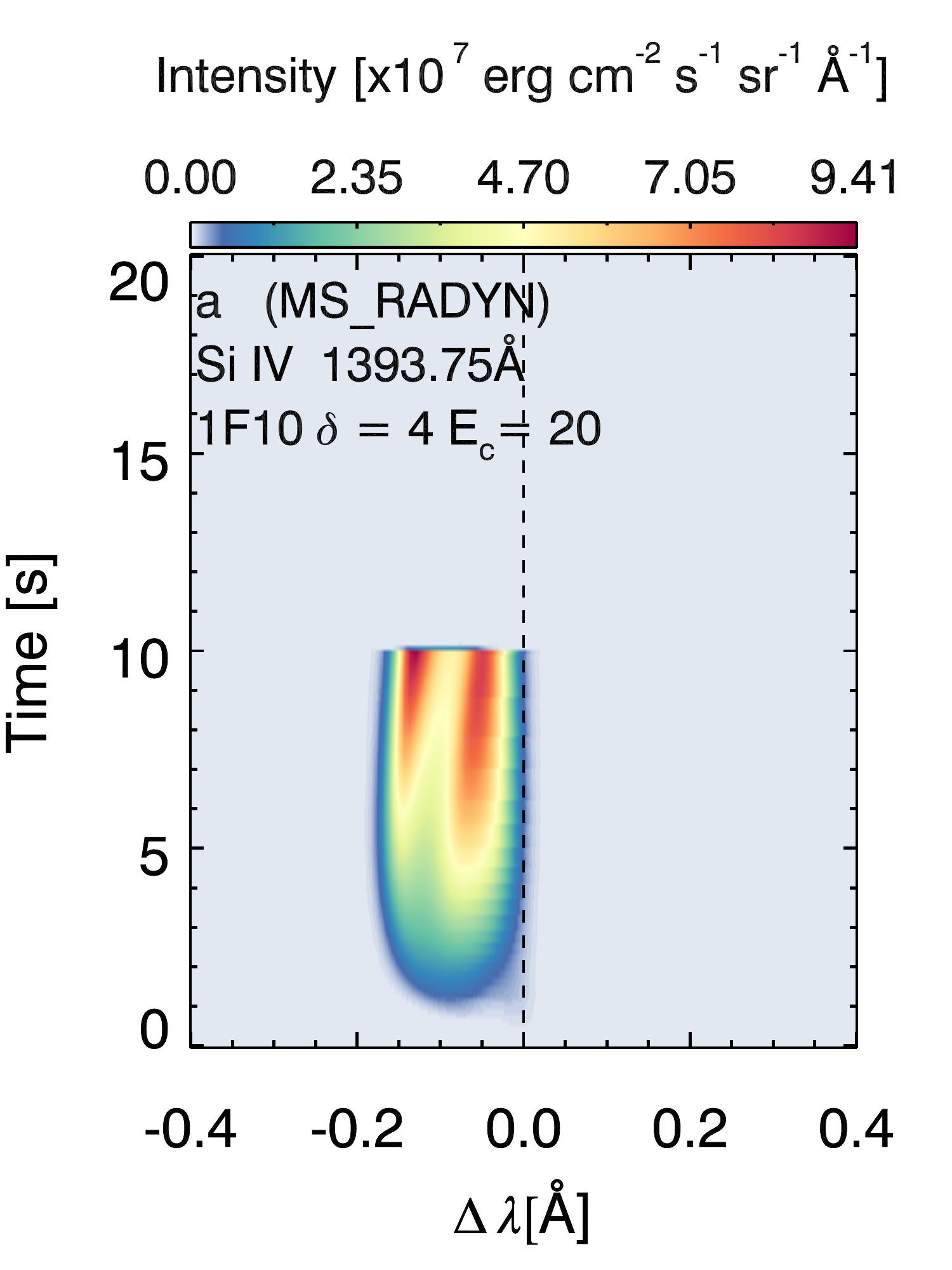}}	
	\subfloat{\includegraphics[width = 0.225\textwidth, clip = true, trim = 0cm 0cm 0cm 0cm]{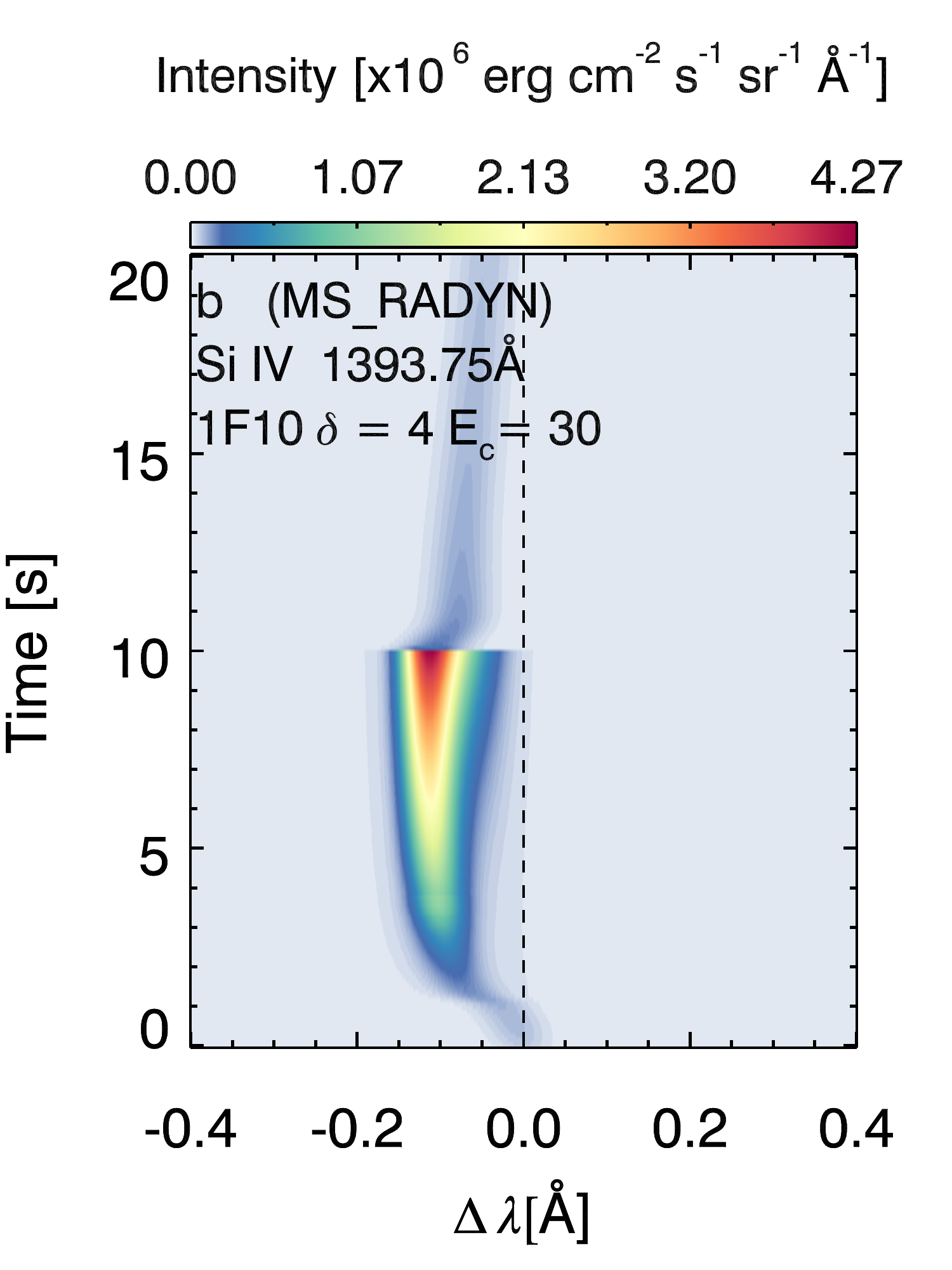}}
	\subfloat{\includegraphics[width = 0.225\textwidth, clip = true, trim = 0cm 0cm 0cm 0cm]{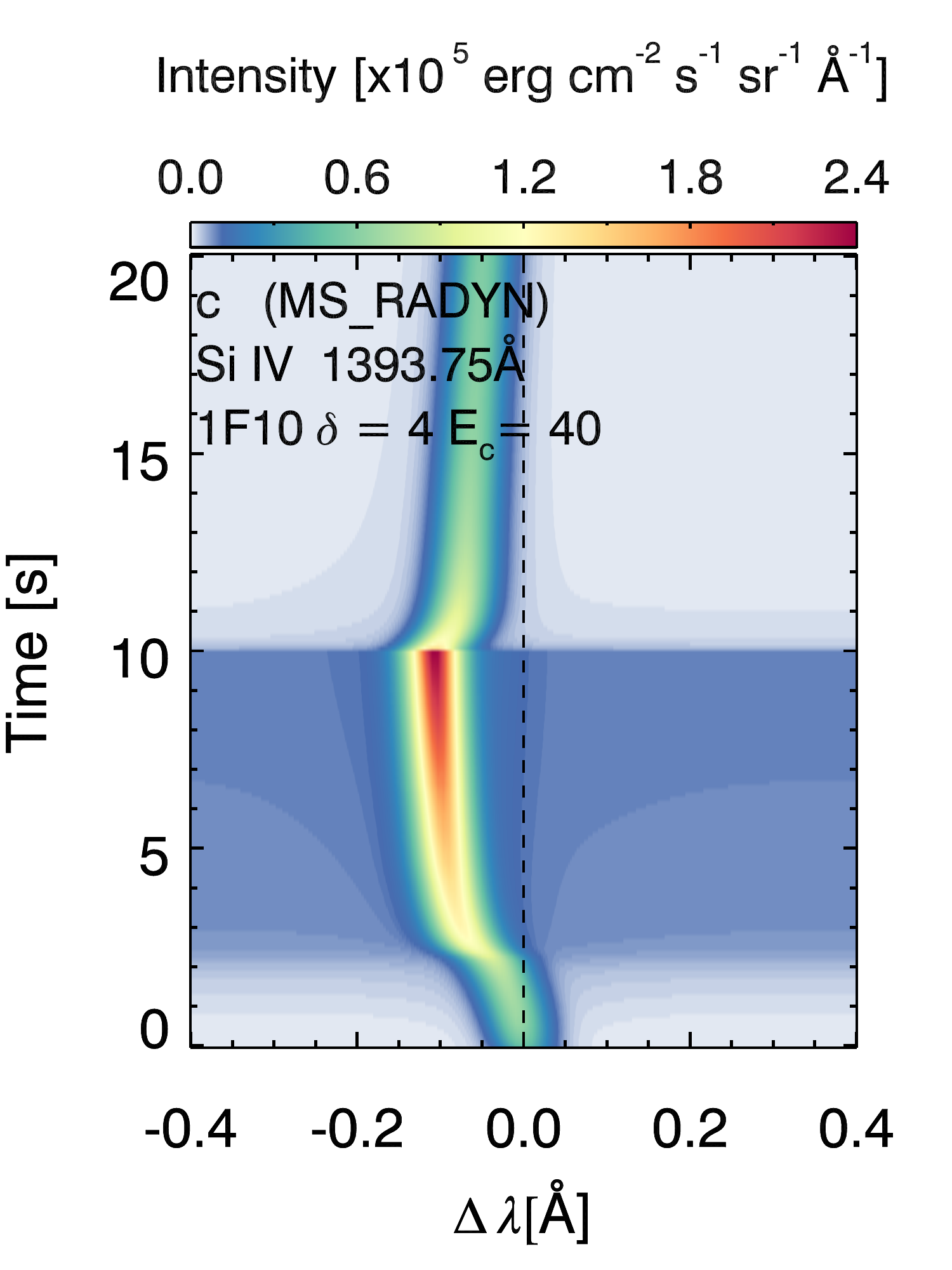}}	
	         }
	 \vspace{-0.2in}
	 \hbox{
	 \hspace{1.125in}
	\subfloat{\includegraphics[width = 0.225\textwidth, clip = true, trim = 0cm 0cm 0cm 0cm]{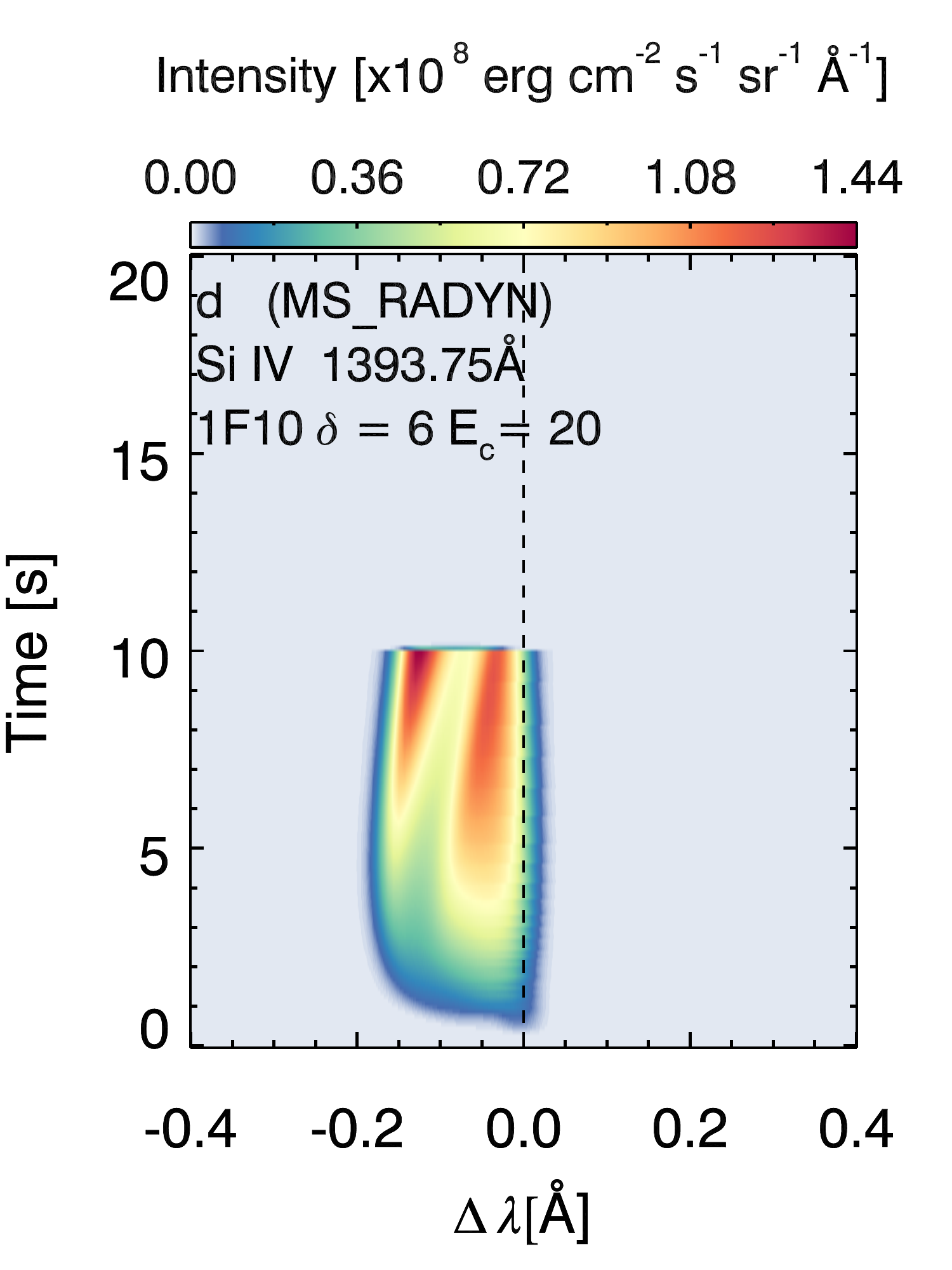}}	
	\subfloat{\includegraphics[width = 0.225\textwidth, clip = true, trim = 0cm 0cm 0cm 0cm]{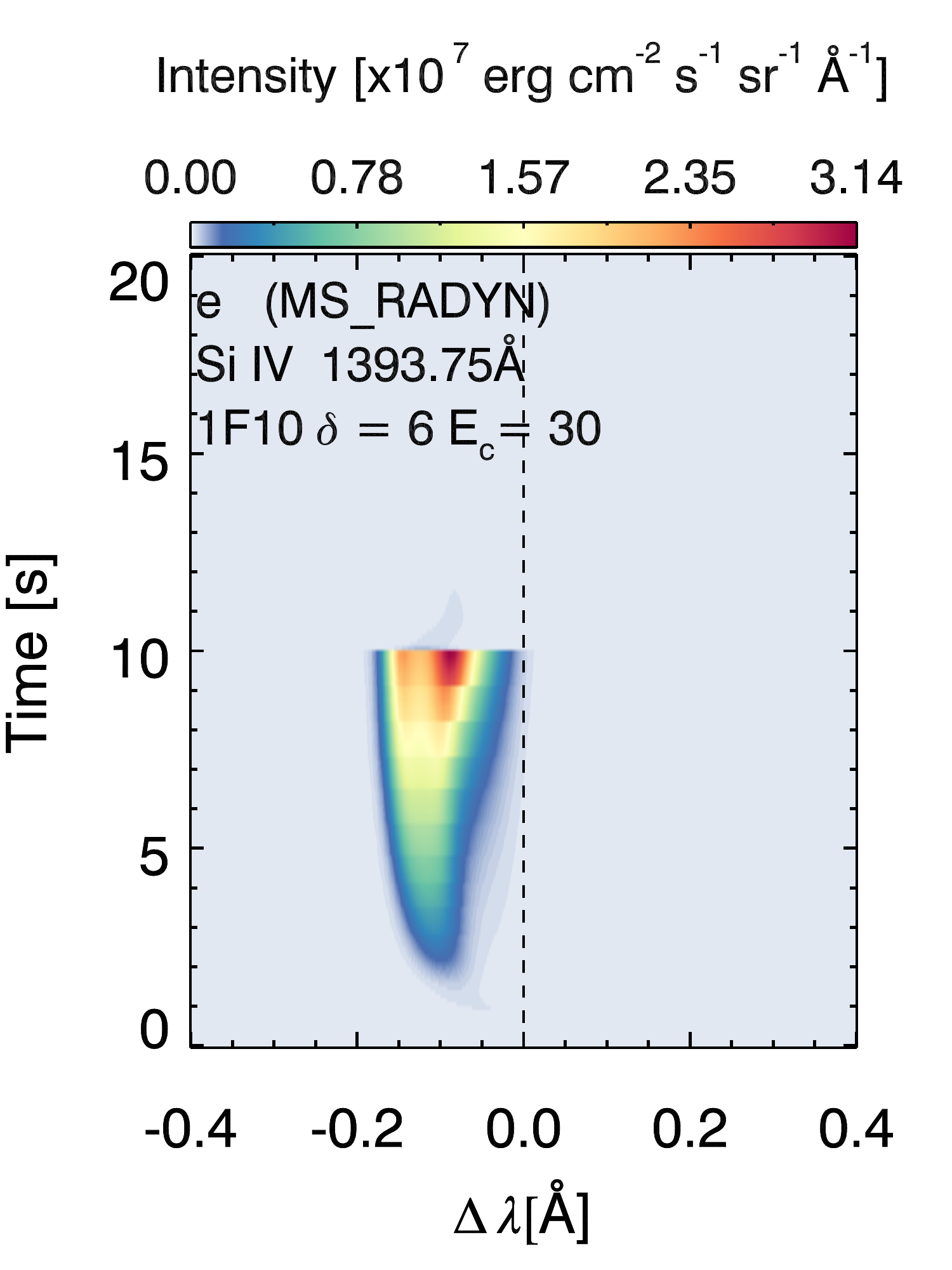}}
	\subfloat{\includegraphics[width = 0.225\textwidth, clip = true, trim = 0cm 0cm 0cm 0cm]{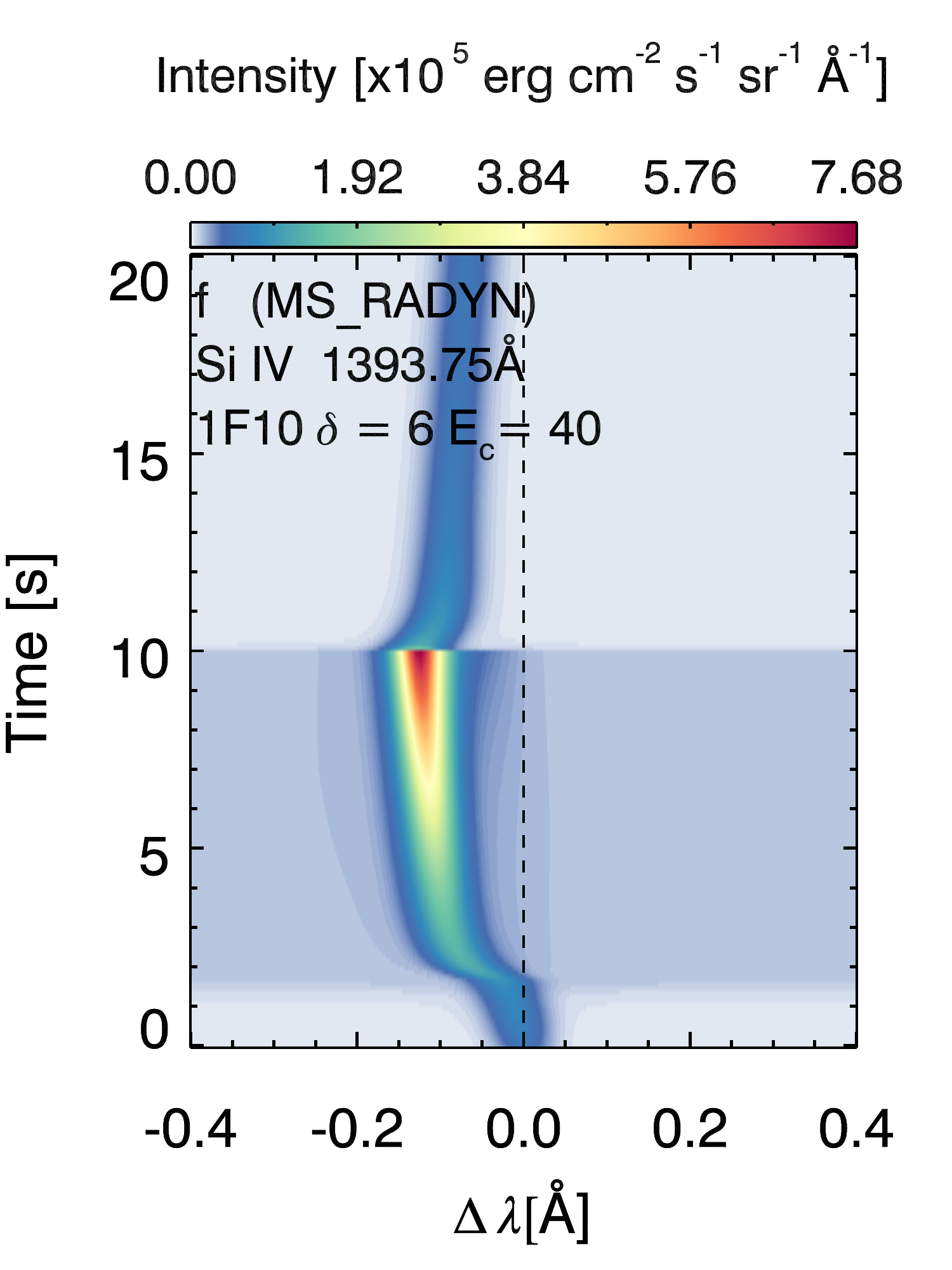}}	
	         }
	\caption{\textsl{Similar to Figure~\ref{fig:stackplots_2} but now showing the effect of varying the other non-thermal electron beam parameters on the \texttt{MS\_RADYN} results. Injected energy is fixed at 1F10 for all simulations, but the spectral indices vary so that panels (a,b,c) have $\delta=4$ and panels (d,e,f) have $\delta=6$. The low energy cutoff also varies, from $E_{c}=20,30,40$~keV  left to right in each row.}}
	\label{fig:stackplots_2}
\end{figure*}

The effect of varying the other non-thermal beam parameters are presented in Figure~\ref{fig:stackplots_2}, focusing only on the \texttt{MS\_RADYN} results. The top row of which shows $\delta=4$ and the bottom row $\delta=6$, with $E_{c} = 20,30,40$~keV from left to right, and the injected energy fixed at 1F10. Here we show the Si~\textsc{iv} $1393.75$~\AA\ line. At this value of injected energy flux the atmosphere at the heights at which Si~\textsc{iv} forms is upflowing, so that all of the emission here is blueshifted to some degree. For $\delta=4$ and $E_{c} = 20$~keV the emission shows self-absorption and asymmetries. As the low energy cutoff increases (i.e. energy is carried by more energetic, deeply penetrating electrons) the main atmospheric response occurs somewhat deeper in the atmosphere. The lines become narrower, weaker and single peaked. Similarly, the bottom row of Figure~\ref{fig:stackplots_2} shows that the beam parameters that would heat the upper atmosphere ($\delta=6$ with $E_{c} = [20,30]$) exhibit high intensity, self-absorbed Si~\textsc{iv} emission with asymmetries. Those profiles are wider than the $\delta=6$ with $E_{c} = 40$~keV case. As we discuss in \S~\ref{sec:lineformation} the opacity effects are dependent on the location of heating. For some profiles shown in Figure~\ref{fig:stackplots_2}, such as $\delta=6$ with $E_{c}=30$~keV, it takes a few seconds for the signatures that are indicative of optical depth effects to appear.  

\begin{figure*}
	\centering
		\includegraphics[width = 0.85\textwidth, clip = true, trim = 0.cm 0.cm 0.0cm 0.cm]{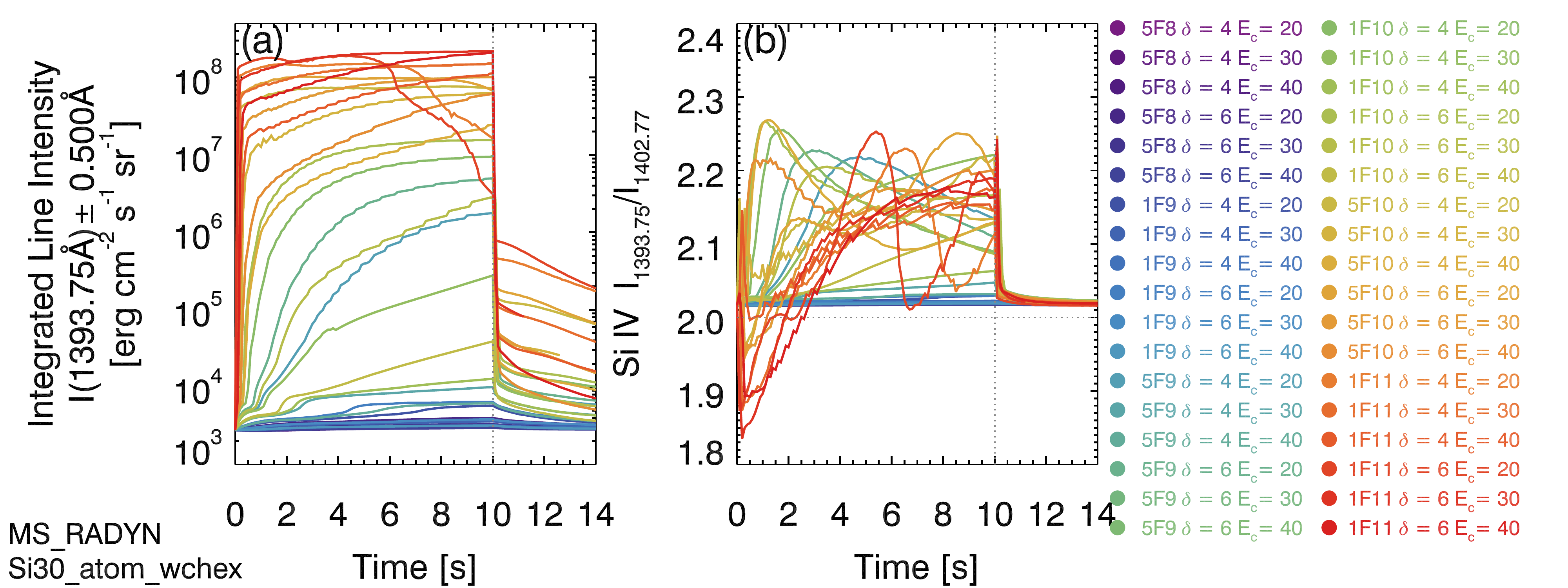}	
          	\caption{\textsl{Panel (a) shows lightcurves of the integrated intensity ($\lambda_\mathrm{rest}\pm0.5$\AA) of the Si~\textsc{iv} 1393.75\AA, as computed by \texttt{MS\_RADYN}, for the first $t=14$~s of the flare simulations (recall that the heating phase lasts $t=10$~s, indicated by the dotted line). Panel (b) shows the ratio $R_{\mathrm{res}} = I_{1393}/I_{1402}$ over the same duration. In both panels colour represents the different simulations.}}
	\label{fig:lcurves_radyn}
\end{figure*}

\begin{figure*}
	\centering
		\includegraphics[width = 0.85\textwidth, clip = true, trim = 0.cm 0.cm 0.0cm 0.cm]{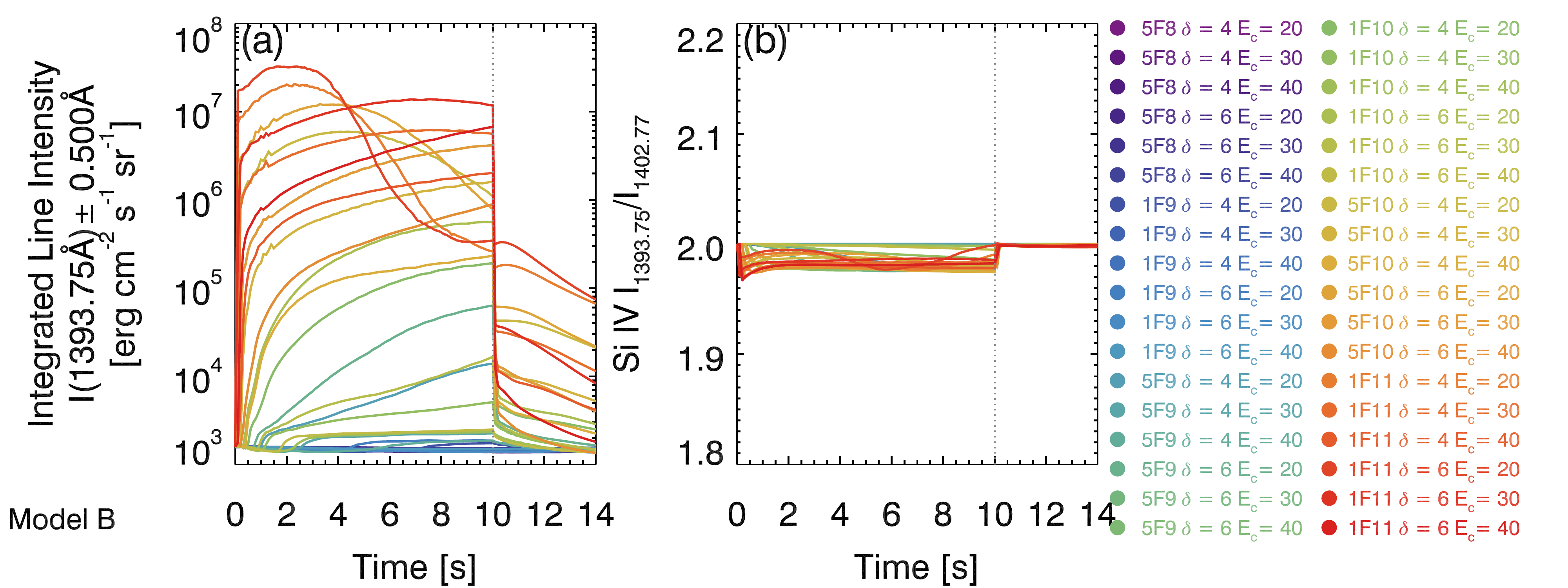}	
          	\caption{\textsl{As in Figure~\ref{fig:lcurves_radyn} but showing results as computed using the optically thin assumption via \texttt{Model B}.}}
	\label{fig:lcurves_chianti}
\end{figure*}

To compare all of the simulations we show lightcurves of integrated line intensity, as well as line intensity ratios in Figure~\ref{fig:lcurves_radyn} (which shows the \texttt{MS\_RADYN} results) and Figure~\ref{fig:lcurves_chianti} (which shows the \texttt{Model B} results), for all 36 simulations. Panel (a) in each figure shows the $1393.75\pm0.5$~\AA\ integrated intensities on a log scale to show all of the simulations. Panel (b) in each figure shows the ratio $R_{\mathrm{res}} = I_{1393}/I_{1402}$. As before colour represents the individual simulations. 

It is clear that in both models the intensities scale considerably in magnitude with the level of injected flare energy flux. Generally the temporal behaviour is shared among both the \texttt{MS\_RADYN} and \texttt{Model B} derived profiles, where profiles typically peak at the end of the heating phase, and rapidly decline after the cessation of the electron beams.  For those simulations in which the Si~\textsc{iv} lightcurves peak earlier, or which show more structure, then there is some disagreement between \texttt{Model B} and \texttt{MS\_RADYN}. For example there are peaks near $t\approx5.5$~s in Figures~\ref{fig:lcurves_radyn}, and at $t\approx2$~s in Figures~\ref{fig:lcurves_chianti}. 

The resonance line ratios differ significantly between the two models, acutely in the stronger flares. When computed via \texttt{Model B} the ratio is very close to two, as expected in the optically thin case, though with some deviations during the flare. This is likely caused by a temperature dependence of the line ratio that is present when the lines form in cooler plasma. When computed by \texttt{MS\_RADYN} there is significant deviation, with values as low as $R_{\mathrm{res}} \approx 1.84$ and as high as $R_{\mathrm{res}} \approx 2.27$. This implies that optical depth effects are substantial at these times. As discussed by \cite{2015ApJ...811...80R} the intensity ratio can in effect take any value, and represents the ratio of the source functions of the line. The magnitudes of the source functions depend on where they thermalise in the atmosphere. Since Si~\textsc{iv} 1393.75\AA\ has twice the opacity of the Si~\textsc{iv} 1402.77\AA\ (proportional to the ratio of oscillator strengths of the lines, since they share a common lower level, and $\chi_{\nu} \propto n_{\mathrm{l}}f_{\mathrm{lu}}$) it thermalises higher in altitude where temperatures are larger. While the Mg~\textsc{ii} h \& k lines typically have a ratio much lower than their optically thin limit both in the quiet Sun and in flares \citep{2013ApJ...772...89L,2015A&A...582A..50K}, the Si~\textsc{iv} resonance line intensity ratio is more variable and can exceed the optically thin limit. We speculate that this is due to the Si~\textsc{iv} lines being less optically thick than Mg~\textsc{ii} h \& k, and that the 1393.75~\AA\ line appears to become affected by opacity more than the 1402.77~\AA. A detailed investigation of the formation properties will elucidate, and will be the focus of a subsequent investigation. 

	\subsection{Line Formation}\label{sec:lineformation}

	\subsubsection{Contribution Functions}\label{sec:contrfns}
	A useful way to understand the complex spectral line formation under optically thick conditions, is to study the contribution functions to the emergent intensity, $C_{\nu\mu}$ \citep{1986A&A...163..135M,1998LNP...507..163C}. This effectively shows where in the atmosphere emission at frequency $\nu$ is formed. Integrating $C_{\nu\mu}$ through a depth scale yields the emergent intensity $I_{\nu\mu}$. Using the height $z$ as the depth scale, then the emergent intensity is:

\begin{equation}\label{eq:contrfn}
	I_{\nu\mu} = \int C_{\nu\mu}(z)~\mathrm{d}z =  \int \frac{1}{\mu}~\chi_{\nu}(z)~S(z)~e^{-\tau_{\nu}(z)/\mu}~\mathrm{d}z,
\end{equation}

\noindent where $\chi_{\nu}(z)$ is the monochromatic opacity, $S(z)$ is the frequency independent source function (since we use the assumption of CRD), $\tau_{\nu}(z) = \int \chi_\nu \mathrm{d}z$ is the optical depth, and $\mu = \cos\theta$ is the viewing angle relative to the normal. We use $\mu\approx0.953$ throughout this work, which is near disk centre, and drop the $\mu$ subscript. 
                  
For each \texttt{MS\_RADYN} simulation $C_{\nu}(z)$ was computed for the resonance lines at $\lambda = 1393.75$~\AA\ and $\lambda = 1402.77$~\AA, and the height of optical depth unity ($\tau_{\lambda} = 1$) compared to the locations of strong emission. If $C_{\nu}(z)$ originates near the height of $\tau_{\lambda} = 1$ then optical depth effects are important for line formation. If emission originates sufficiently far from the height of $\tau_{\lambda} = 1$ then the line can be considered optically thin. In some cases the emergent line profile can be a combination of emission produced where opacity effects are considerable and optically thin emission. Opacity effects were time dependent in some simulations, with the importance varying over time, as a function of the atmospheric structure. In some cases the optical depth at the formation height did not reach unity, but was still non-negligible ($\tau_{\lambda} \ge 0.1$). Finally, the optical depth effects were wavelength dependent, such that the width of the line that was affected by opacity varied. In the atmospheres where the Si~\textsc{iv} fraction significantly increased a large part of the line was affected by opacity, while in atmospheres with more modest increases only a narrow region around the line core became optically thick (where the absorption profile is maximum).

\begin{table*}
\centering
\begin{tabular}{| c | c c c | c c c |} \toprule
    \textbf{\texttt{Si30\_atom\_wchex}}  & \textbf{} & \textbf{Si~\textsc{iv} Opacity Effects?} & \textbf{} & \textbf{} & \textbf{Si~\textsc{iv} Opacity Effects?} & \textbf{}  \\
     \midrule
    \textbf{Energy Flux} & \textbf{} & \textbf{$\boldsymbol{\delta = 4}$} & \textbf{} & \textbf{} & \textbf{$\boldsymbol{\delta = 6}$} & \textbf{}  \\ 
    \textbf{[erg~cm$^{-2}$~s$^{-1}$]} & \textbf{$\boldsymbol{E_{c} = 20}$~keV} & \textbf{$\boldsymbol{E_{c} = 30}$~keV} & \textbf{$\boldsymbol{E_{c} = 40}$~keV} & \textbf{$\boldsymbol{E_{c} = 20}$~keV} & \textbf{$\boldsymbol{E_{c} = 30}$~keV} & \textbf{$\boldsymbol{E_{c} = 40}$~keV} \\ \toprule
    \textbf{1F11} & \cmark & \cmark & \cmark & \cmark & \cmark & \cmark \\
    \textbf{5F10} & \cmark & \cmark & \cmark & \cmark & \cmark & \cmark \\
    \textbf{1F10} & \cmark & \xmark & \xmark & \cmark & \cmark & \xmark \\
    \textbf{5F9}   & \cmark & \xmark & \xmark & \cmark & \cmark & \cmark  \\
    \textbf{1F9}   & \xmark & \xmark & \xmark & \xmark & \xmark & \xmark \\
    \textbf{5F8}   & \xmark & \xmark & \xmark & \xmark & \xmark & \xmark  \\
   \bottomrule
    \end{tabular}
    \caption{{{Presence of Si~\textsc{iv} resonance line optical depth effects in the \texttt{MS\_RADYN} flare simulations using \texttt{Si30\_atom\_wchex}. The duration and strength of optical depth effects varies in each simulation, we are just indicating here if they are non-negligible at some point during the flare.}}}
    \label{tab:opticallythickorthin}
\end{table*}

\begin{table*}
\centering
\begin{tabular}{| c | c c c | c c c |} \toprule
    \textbf{\texttt{Si30\_atom}}  & \textbf{} & \textbf{Si~\textsc{iv} Opacity Effects?} & \textbf{} & \textbf{} & \textbf{Si~\textsc{iv} Opacity Effects?} & \textbf{}  \\
     \midrule
    \textbf{Energy Flux} & \textbf{} & \textbf{$\boldsymbol{\delta = 4}$} & \textbf{} & \textbf{} & \textbf{$\boldsymbol{\delta = 6}$} & \textbf{}  \\ 
    \textbf{[erg~cm$^{-2}$~s$^{-1}$]} & \textbf{$\boldsymbol{E_{c} = 20}$~keV} & \textbf{$\boldsymbol{E_{c} = 30}$~keV} & \textbf{$\boldsymbol{E_{c} = 40}$~keV} & \textbf{$\boldsymbol{E_{c} = 20}$~keV} & \textbf{$\boldsymbol{E_{c} = 30}$~keV} & \textbf{$\boldsymbol{E_{c} = 40}$~keV} \\ \toprule
    \textbf{1F11} & \cmark & \cmark & \cmark & \cmark & \cmark & \cmark \\
    \textbf{5F10} & \cmark & \cmark & \xmark & \cmark & \cmark & \cmark \\
    \textbf{1F10} & \xmark & \xmark & \xmark & \cmark & \xmark & \xmark \\
    \textbf{5F9}   & \xmark & \xmark & \xmark & \xmark & \cmark & \cmark  \\
    \textbf{1F9}   & \xmark & \xmark & \xmark & \xmark & \xmark & \xmark \\
    \textbf{5F8}   & \xmark & \xmark & \xmark & \xmark & \xmark & \xmark  \\
   \bottomrule
    \end{tabular}
    \caption{{{Presence of Si~\textsc{iv} resonance line optical depth effects in the \texttt{MS\_RADYN} flare simulations using \texttt{Si30\_atom}.}}}
    \label{tab:opticallythickorthin_nochex}
\end{table*}

\begin{figure*}
	\centering
	\hbox{
		\subfloat{\includegraphics[width = 0.33\textwidth, clip = true, trim = 0cm 0cm 0cm 0cm]{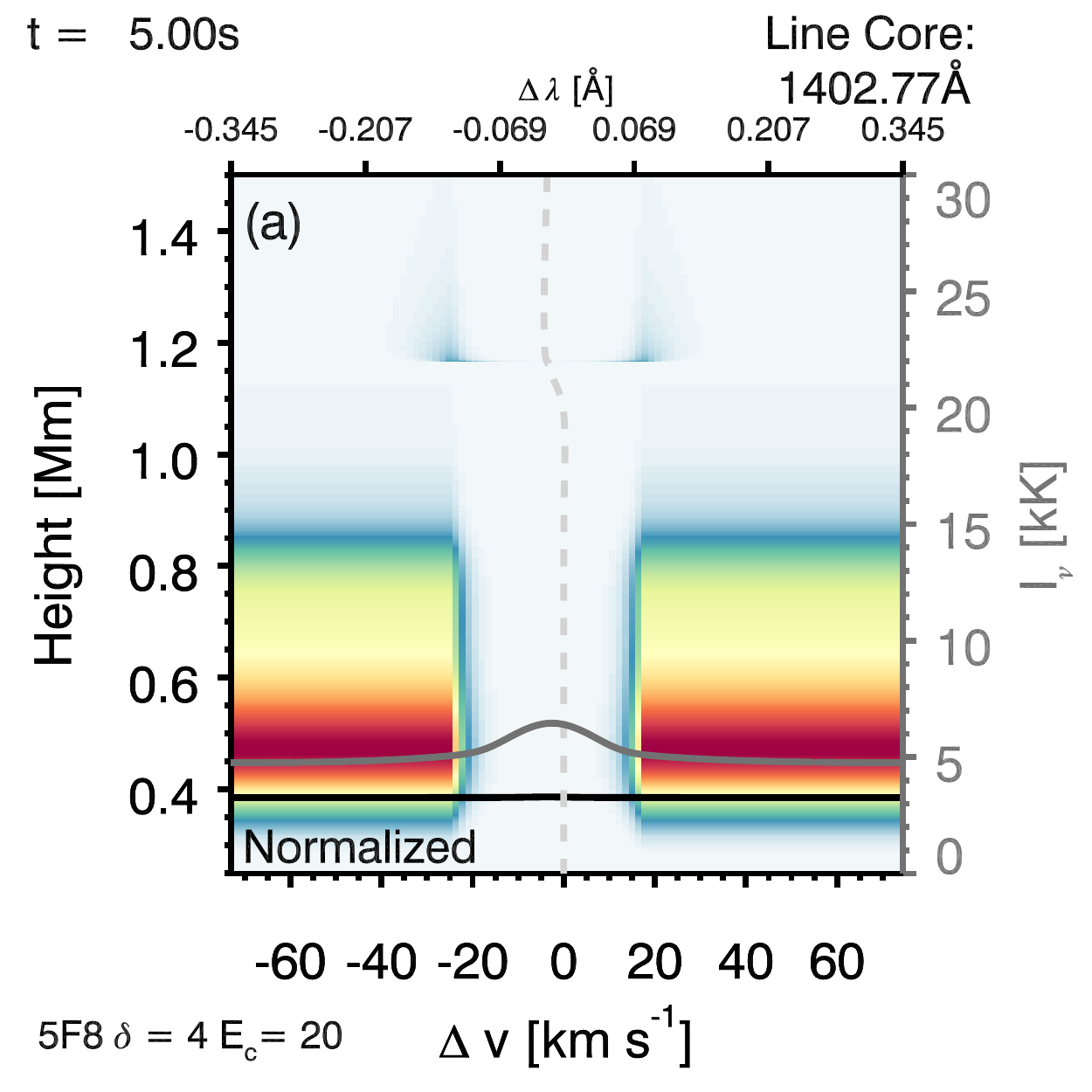}}	
		\subfloat{\includegraphics[width = 0.33\textwidth, clip = true, trim = 0cm 0cm 0cm 0cm]{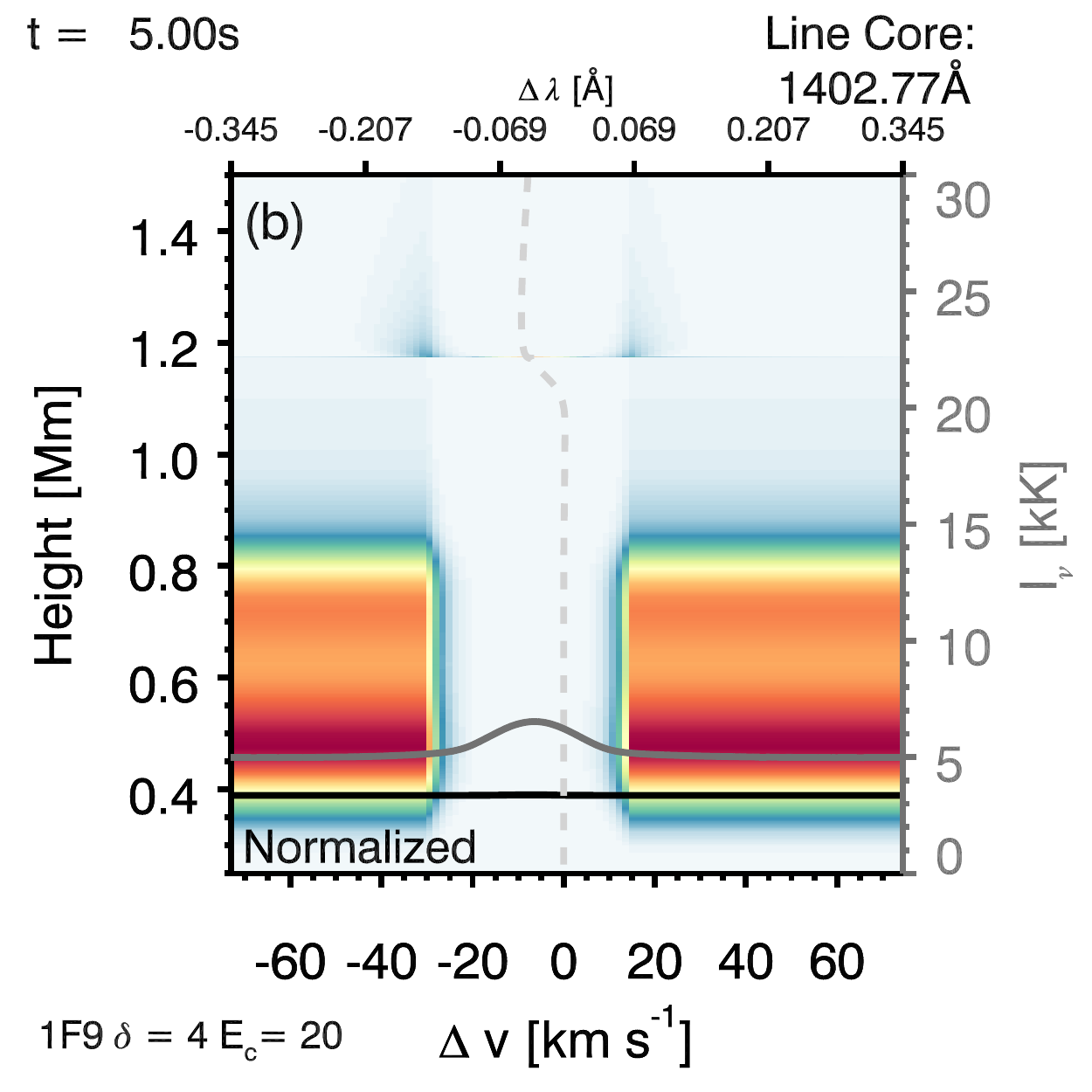}}
		\subfloat{\includegraphics[width = 0.33\textwidth, clip = true, trim = 0cm 0cm 0cm 0cm]{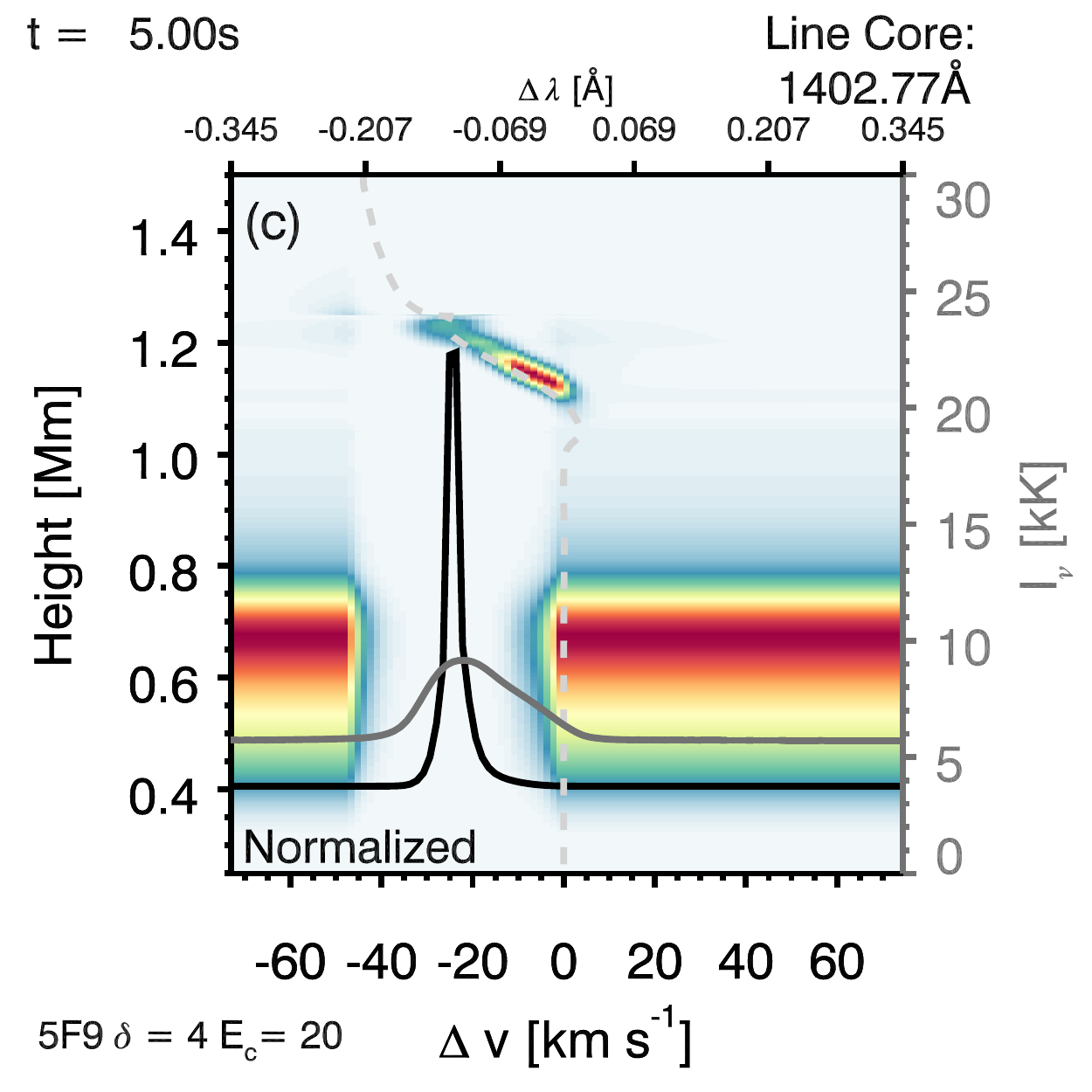}}	
	         }
	\hbox{
		\subfloat{\includegraphics[width = 0.33\textwidth, clip = true, trim = 0cm 0cm 0cm 0cm]{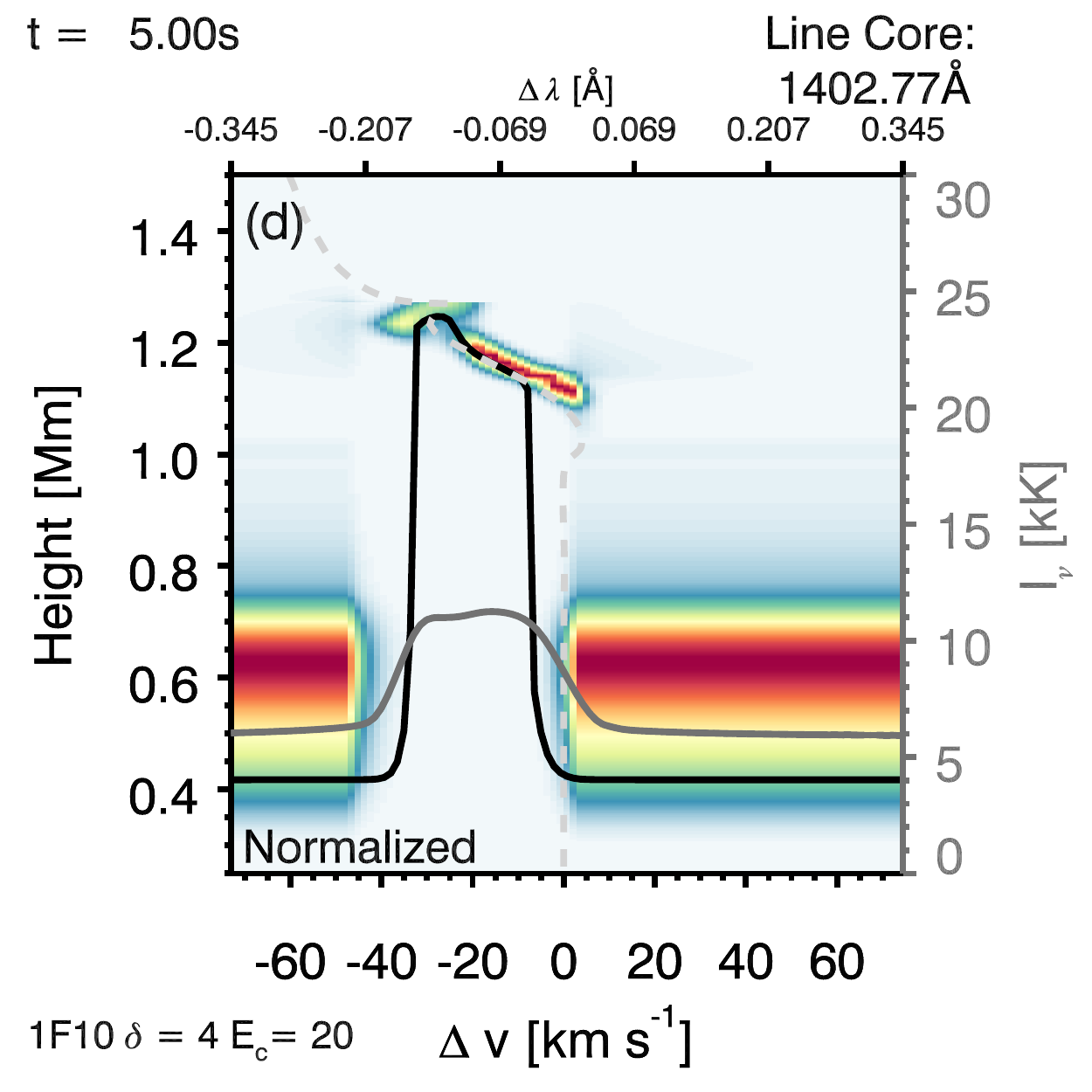}}	
		\subfloat{\includegraphics[width = 0.33\textwidth, clip = true, trim = 0cm 0cm 0cm 0cm]{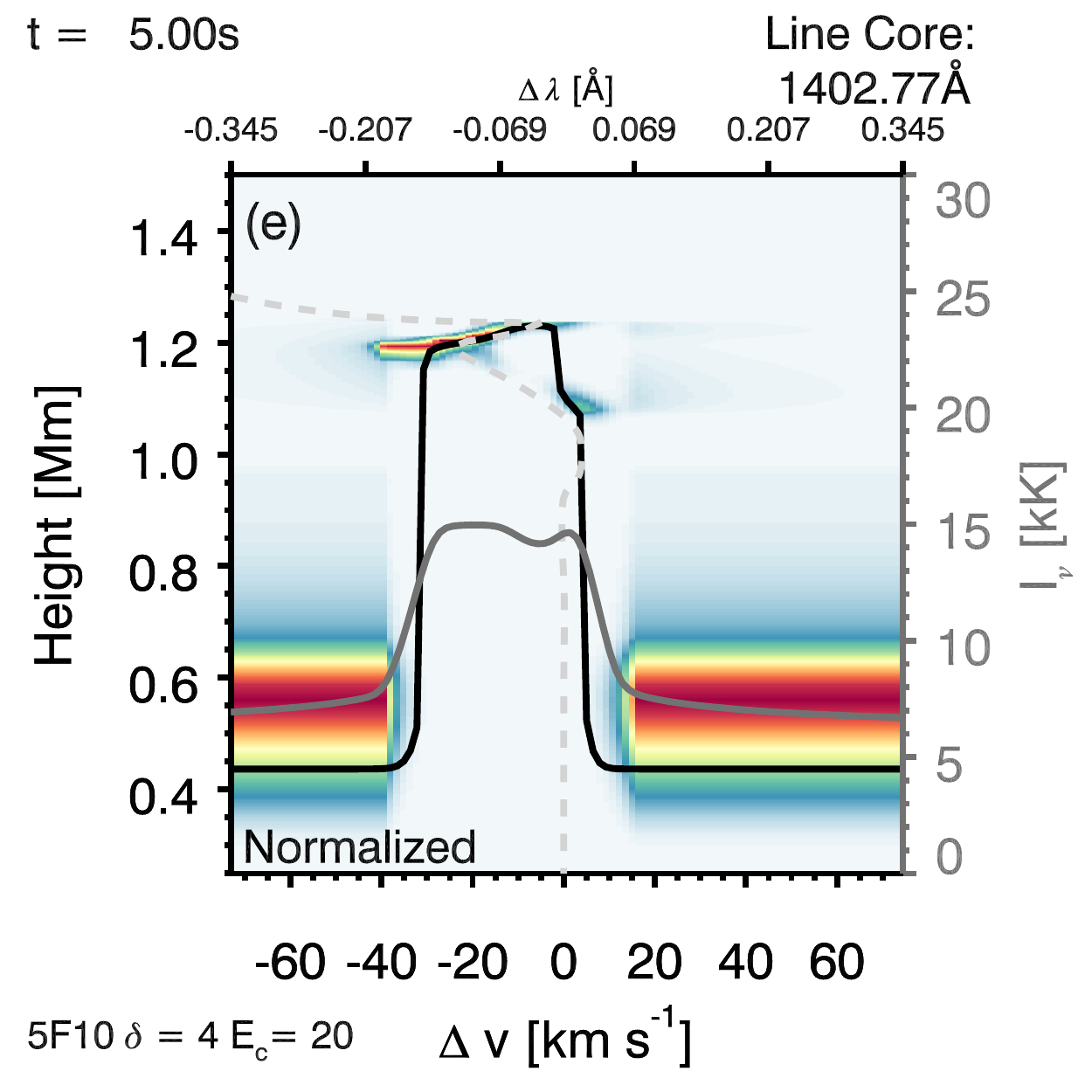}}
		\subfloat{\includegraphics[width = 0.33\textwidth, clip = true, trim = 0cm 0cm 0cm 0cm]{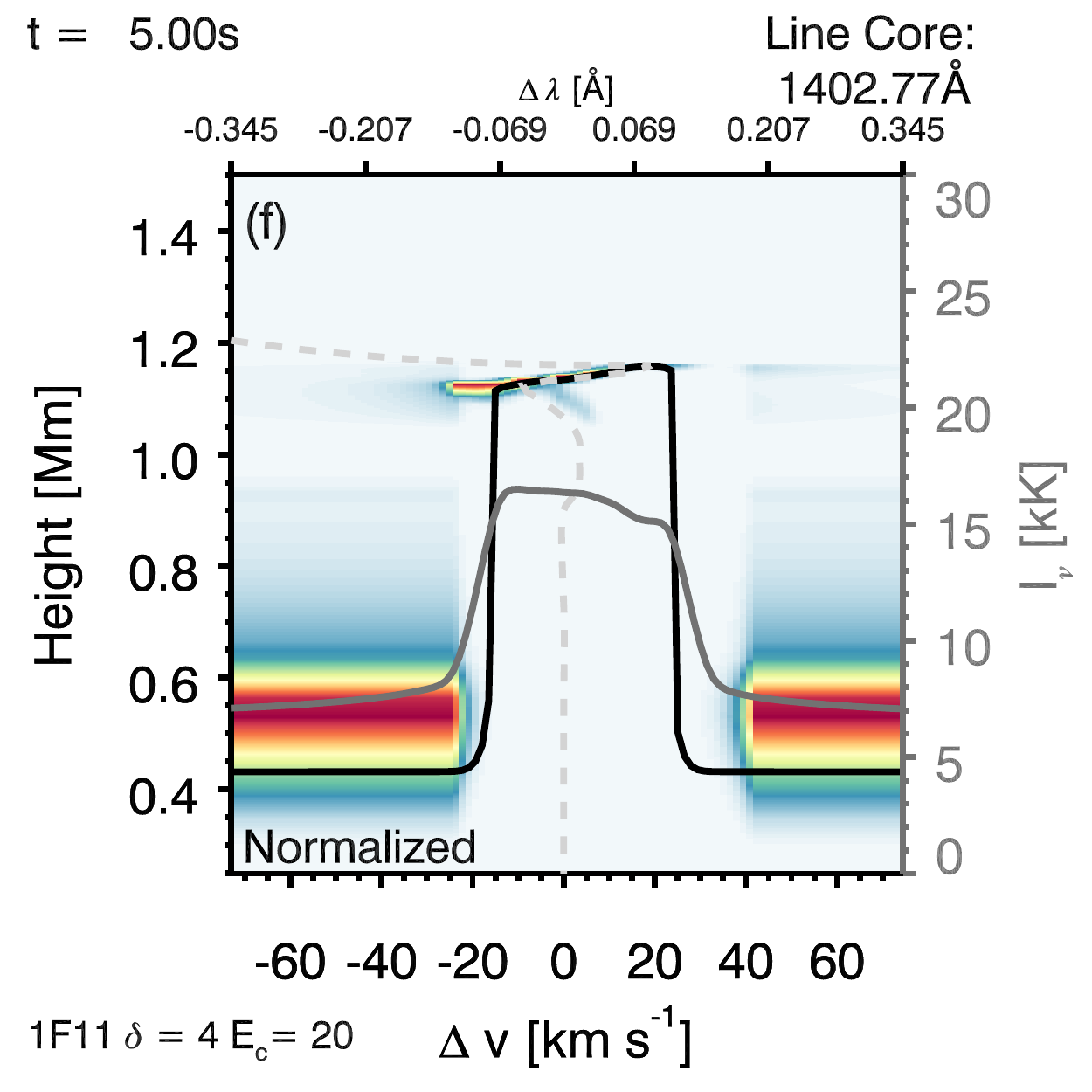}}	
	         }
	\caption{\textsl{Panels illustrating the formation of the Si~\textsc{iv} 1402.77\AA\ at $t=5$~s in six \texttt{MS\_RADYN} simulations with varying injected energy flux. The spectral index and low energy cutoff of the electron beam are $\delta=4$ and $E_{c} = 20$~keV, with the injected energy flux ranging from panel (a)-(f): $F = [5\mathrm{F}8, 1\mathrm{F}9, 5\mathrm{F}9, 1\mathrm{F}10, 5\mathrm{F}10, 1\mathrm{F}11]$~erg~cm$^{-2}$~s$^{-1}$. The background image is the contribution function to the emergent intensity defined on height (left axis), Doppler velocity (bottom axis), and wavelength scales (top scale), where red is more intense. The contribution function has been normalised within each wavelength bin to bring out details in the line wings. On each panel is the atmospheric velocity as a function of height (light grey dashed line), the height at which optical depth unity ($\tau_{\lambda} =1$) is reached (black solid line), and the line profile in units of radiation temperature (right axis; dark grey solid line).}}
	\label{fig:contfns_1}
\end{figure*}

Optical depth effects were present for all simulations in which the injected energy flux was equal to $5\mathrm{F}10$ or greater. For smaller magnitudes of injected energy flux the presence of optical depth effects depends on the other parameters of the non-thermal electron beams. Sufficient energy must be deposited in the upper chromosphere to result in an extended high temperature region over which the Si~\textsc{iv} population becomes enhanced.

The location of energy deposition is a function of both the power-law (spectral) index $\delta$ and the low-energy cutoff of the distribution $E_{c}$. A larger value of $\delta$ means that there are relatively more lower energy electrons than higher energy electrons (if all parameters are fixed), which are thermalised at lower column depths. The distribution contains non-thermal electrons with energies as low as $E_{c}$ so that a smaller value of $E_{c}$ for a fixed energy flux means that more power is carried by the lower energy electrons, resulting in greater heating at low column depth. We show in Table~\ref{tab:opticallythickorthin} in which simulations optical depth effects were present. Note that the length of time over which optical depth effects were present varied in each simulation (though for the high energy simulations they typically lasted for the entire duration of the heating phase), so we just indicate in Table~\ref{tab:opticallythickorthin} if they were present at some point during the flare. If charge exchange processes are excluded then there is less Si~\textsc{iv} present at lower temperatures and consequently a smaller opacity. Table~\ref{tab:opticallythickorthin_nochex} indicates that by excluding charge exchange Si~\textsc{iv} formation is affected by opacity in fewer cases.

\begin{figure*}
	\centering 
	\hbox{
		\subfloat{\includegraphics[width = 0.33\textwidth, clip = true, trim = 0cm 0cm 0cm 0cm]{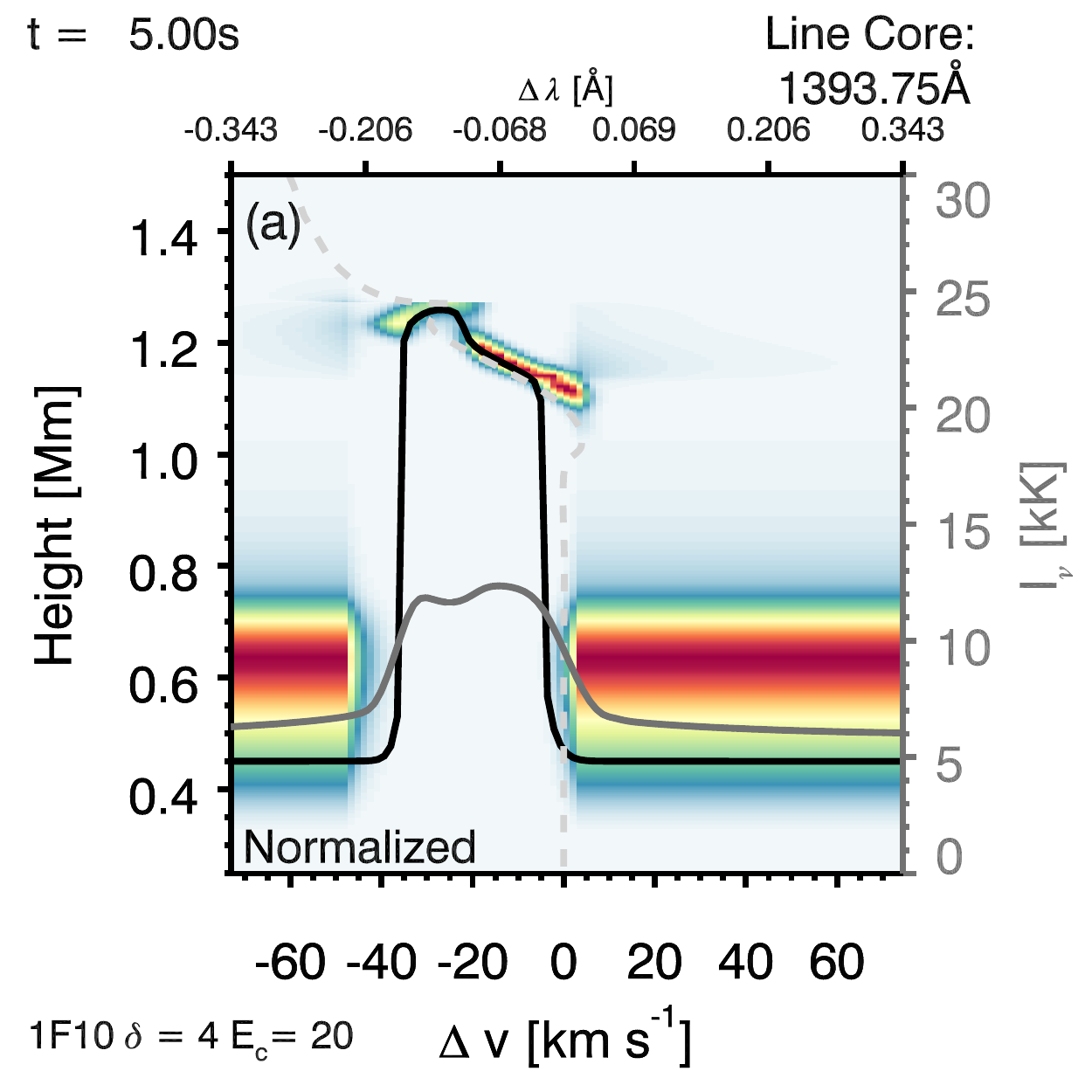}}	
		\subfloat{\includegraphics[width = 0.33\textwidth, clip = true, trim = 0cm 0cm 0cm 0cm]{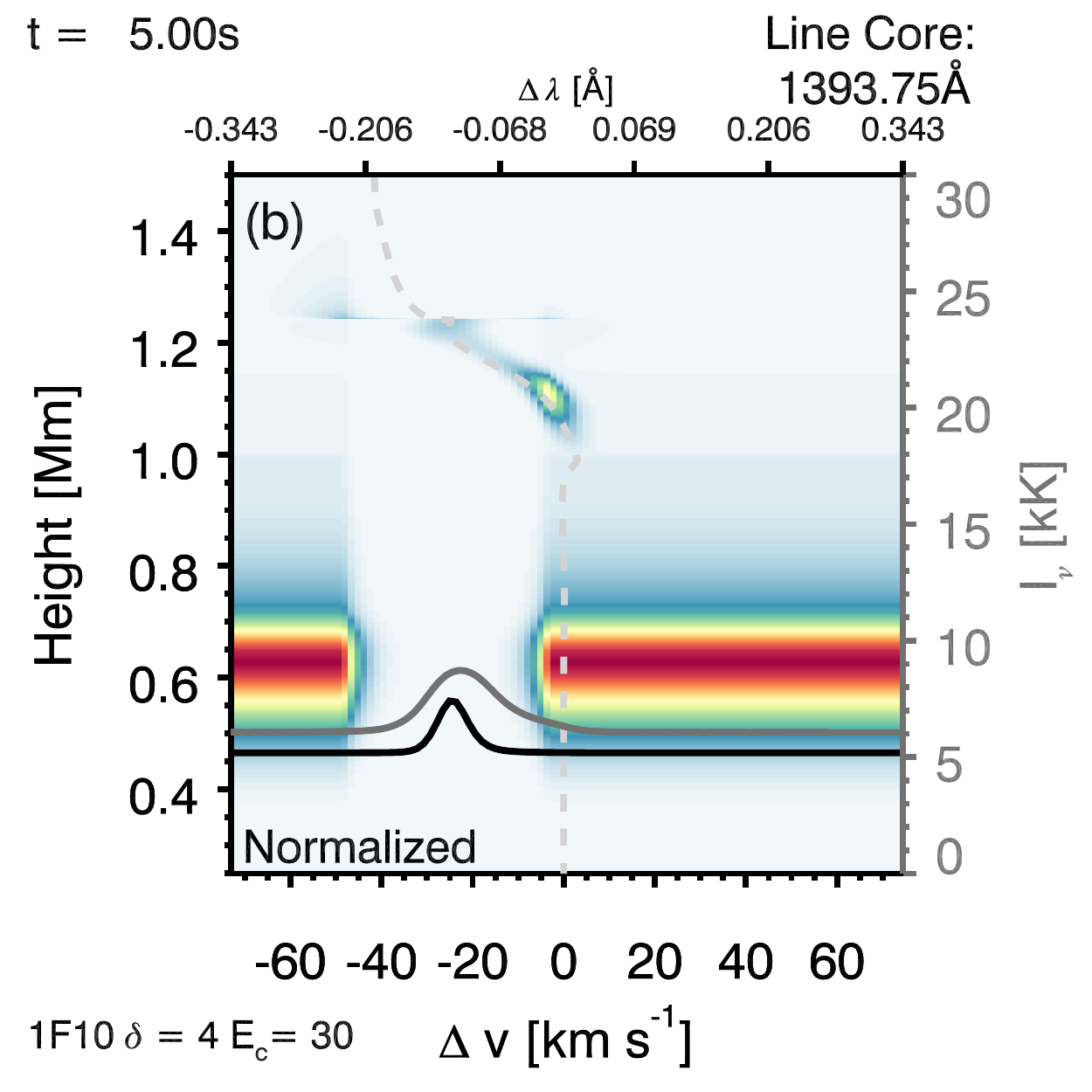}}
		\subfloat{\includegraphics[width = 0.33\textwidth, clip = true, trim = 0cm 0cm 0cm 0cm]{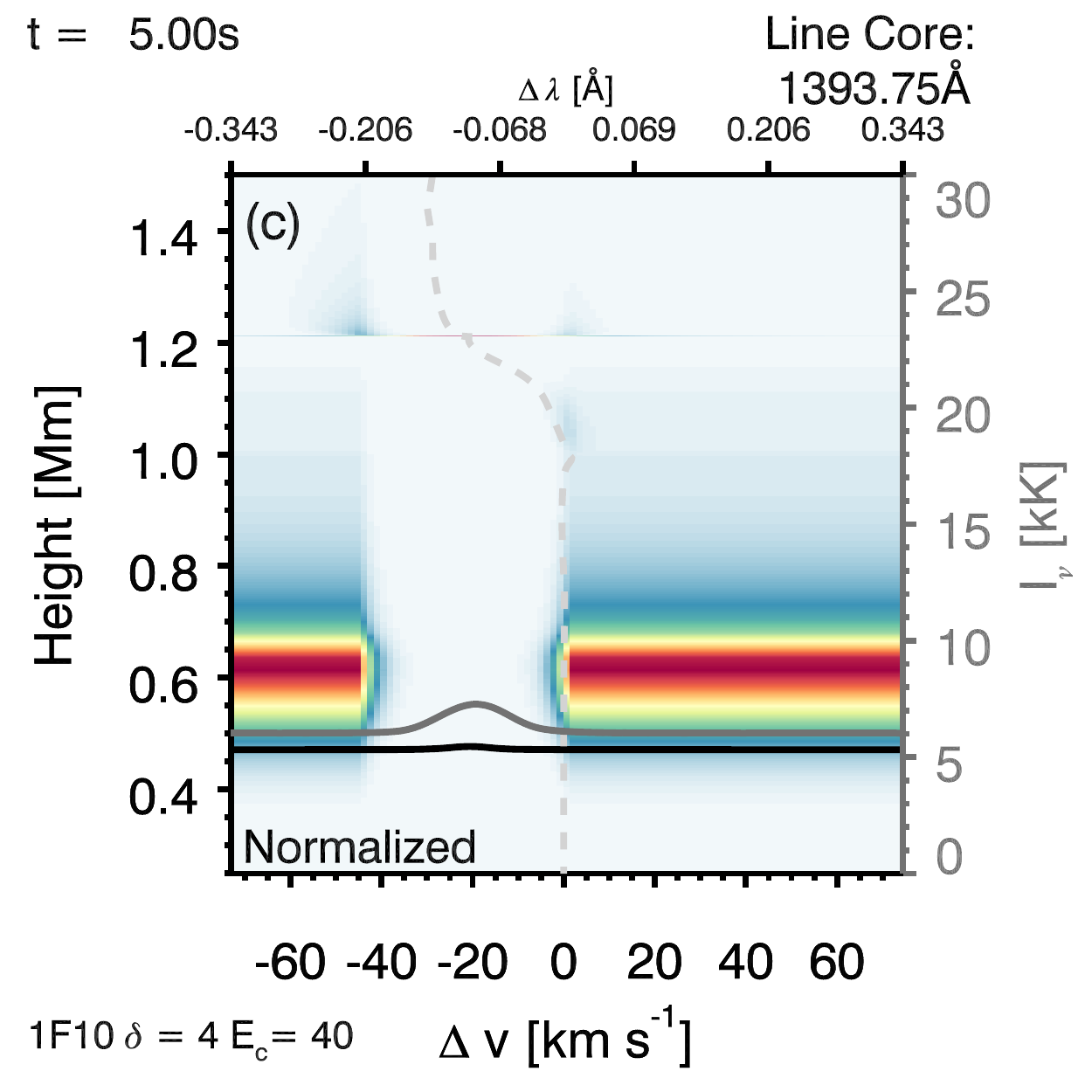}}	
	         }
	\hbox{
		\subfloat{\includegraphics[width = 0.33\textwidth, clip = true, trim = 0cm 0cm 0cm 0cm]{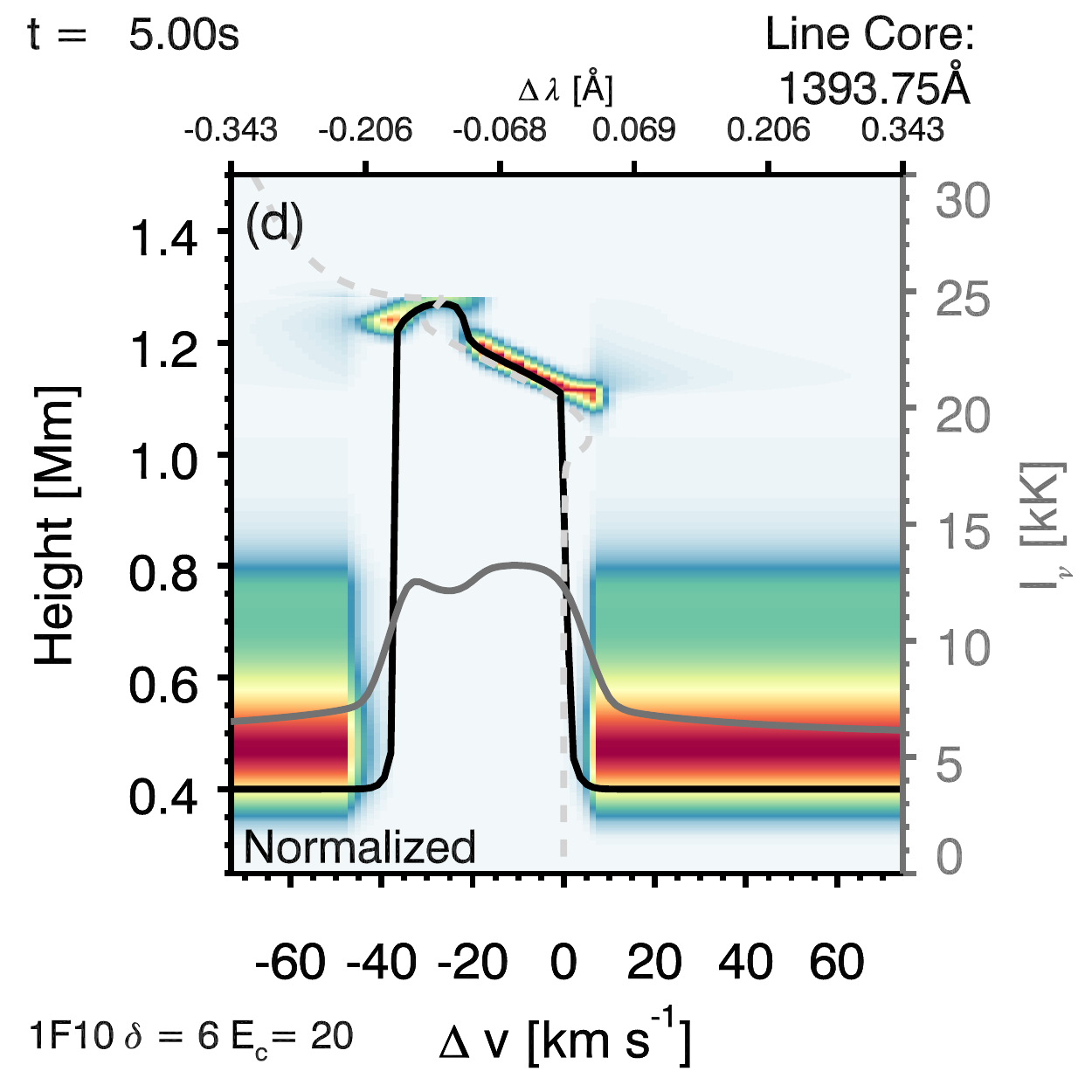}}	
		\subfloat{\includegraphics[width = 0.33\textwidth, clip = true, trim = 0cm 0cm 0cm 0cm]{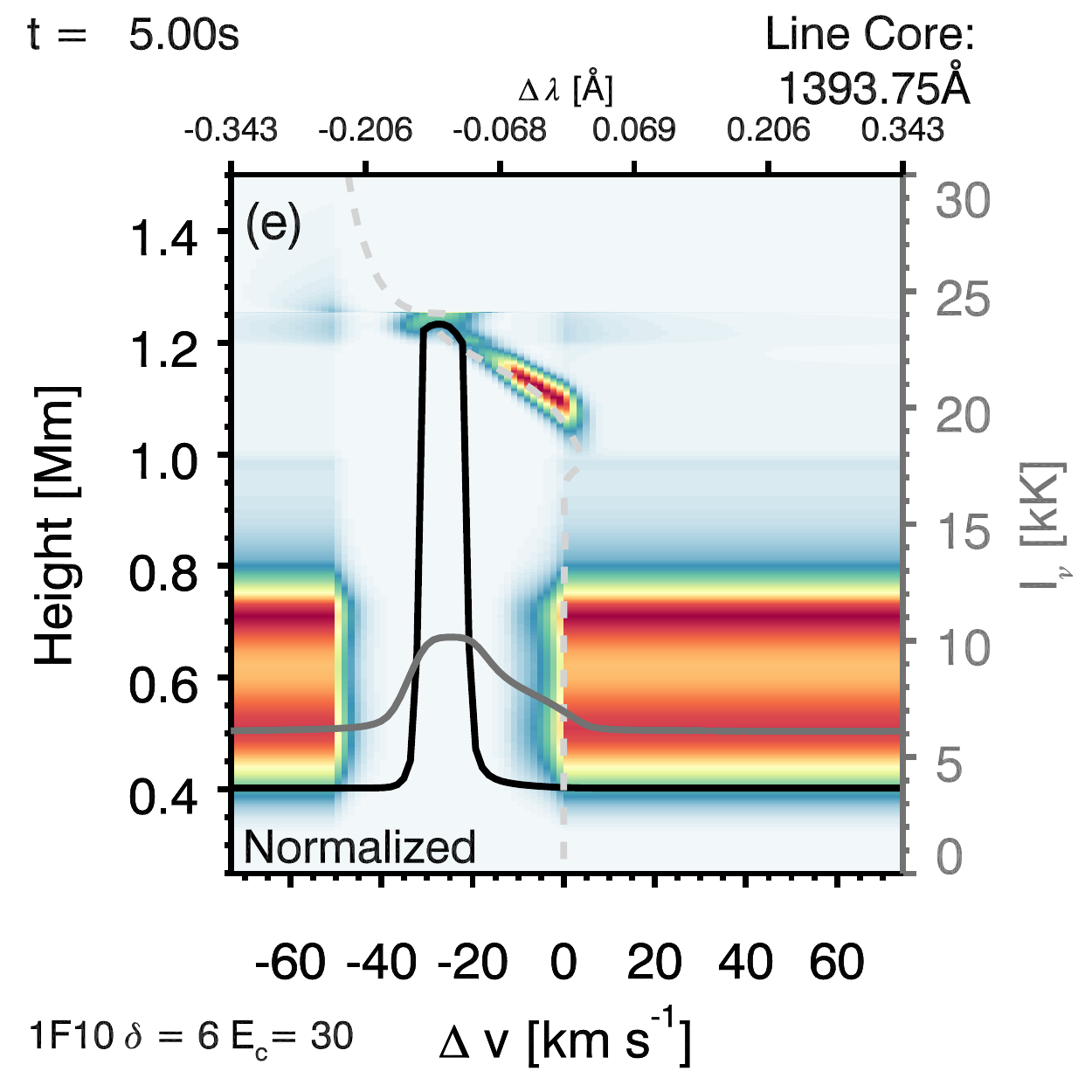}}
		\subfloat{\includegraphics[width = 0.33\textwidth, clip = true, trim = 0cm 0cm 0cm 0cm]{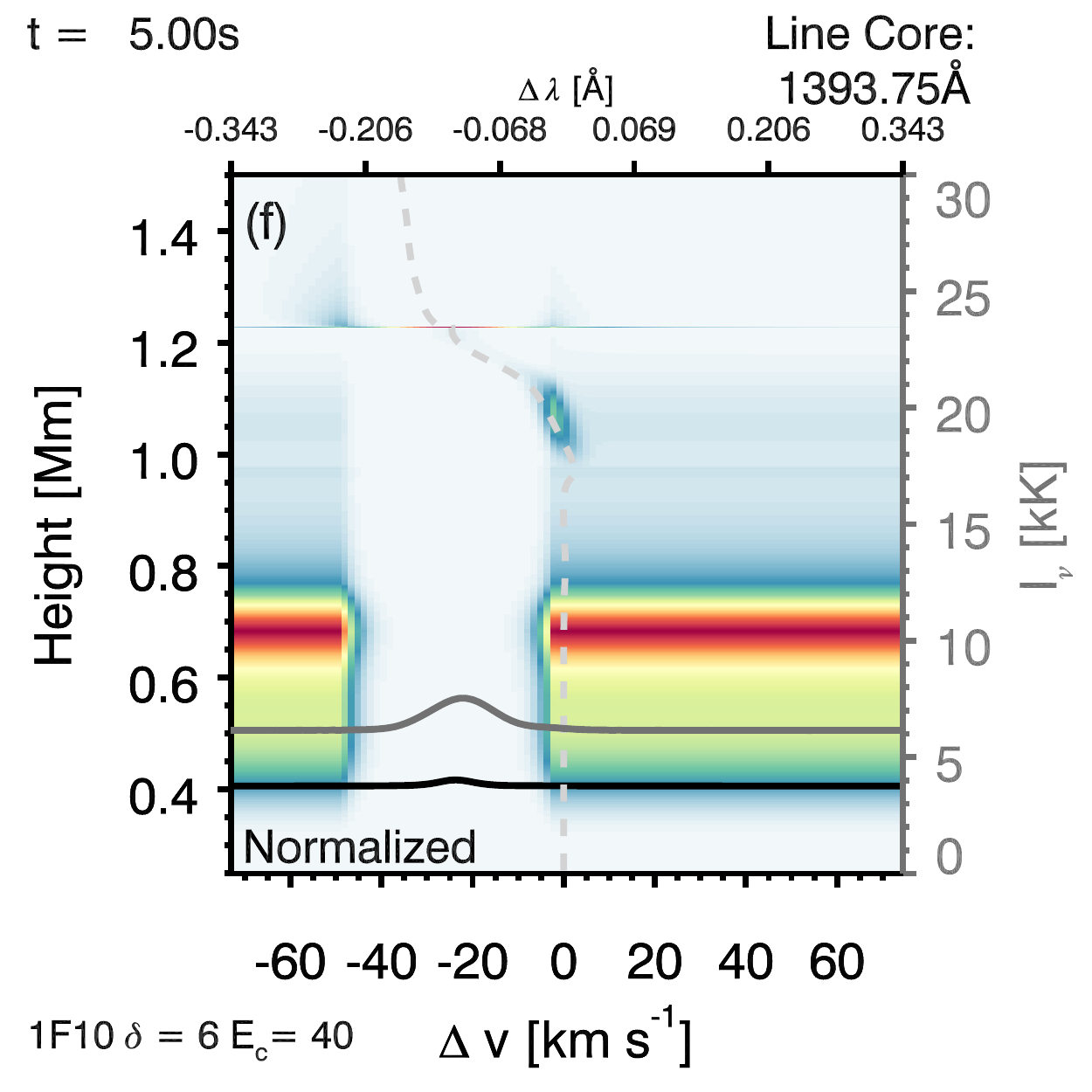}}	
	         }
	\caption{\textsl{Panels illustrating the formation of the Si~\textsc{iv} 1393.75\AA\ at $t=3$~s in six \texttt{MS\_RADYN} simulations with varying constant injected energy flux of $1\mathrm{F}10$~erg~cm$^{-2}$~s$^{-1}$. The spectral index of the top row is $\delta=4$, and the bottom row is $\delta=6$. For each row the low energy cutoff varies from left to right with values of $E_c = [20,30,40]$~keV. The panels are as described in Figure~\ref{fig:contfns_1}.}}
	\label{fig:contfns_2}
\end{figure*}

To illustrate the formation of the line in various simulations we show examples of the contribution function to the emergent intensity in Figures~\ref{fig:contfns_1} \& \ref{fig:contfns_2}. Each of the panels indicates the rest wavelength of the line core in the upper right corner, the time in the simulation in the upper left corner, and the injected electron beam parameters in the lower left corner. The background image in colour is the contribution function, where red is more intense and blue/white is less intense. We have normalised the contribution function within each wavelength bin in order to bring out weaker details in the line wings, which would be washed out by the much more intense line core. The line core formation region is typically very narrow compared to the line wings. Both the core and wings can be combinations of emission formed near the $\tau_{\lambda} = 1$ layer and higher lying optically thin regions. The contribution functions are shown as functions of height (left axis) and Doppler shift (bottom axis), with the corresponding wavelength shift from line center on the top axis. On the same scale the height at which $\tau_{\lambda} = 1$ is shown as a black solid line. The atmospheric velocity as a function of height is shown as a light-grey dashed line. Upflows and blueshifts are negative, downflows and redshifts are positive. Finally, the line profiles in units of radiation temperature (right axis) are shown as dark-grey solid lines. 

Figure~\ref{fig:contfns_1} shows the effect of varying the injected energy flux on the formation of the $\lambda=1402.77$~\AA\ line. Each of the simulated values of injected energy flux for fixed $\delta = 4$ and $E_{c} = 20$~keV are shown at $t=5$~s. For the weakest flare simulations ($5\mathrm{F}8$ \& $1\mathrm{F}9$) the point of optical depth unity is not reached until deep into the atmosphere at $z\approx0.4$~Mm. The line forms in a vanishingly thin layer of the transition region, around $z\approx1.2$~Mm. The lines can be considered to form under optically thin conditions. In panel (b) there is a Doppler shift of the line since it forms in an upflowing region moving at $v\approx10-15$~km~s$^{-1}$. 

In the higher energy simulations opacity effects are present, to varying degrees. Panel (c) shows the $5\mathrm{F}9$ simulation, where the core of the line reaches optical depth unity at a much higher altitude than in the weakest simulations, around $z\approx1.2$~Mm, where that part of the line forms. The line wings are again optically thin. So, the line core is optically thick and blue-shifted but there is a red wing asymmetry caused by an optically thin component that forms over an extended height range along a velocity gradient. If the energy is increased, panel (d), then the contribution function appears similar, but now the range of wavelengths that form under optically thick conditions has increased so that the line core is wider. In this case the velocity gradient produces an optically thick component redward of the line core, increasing the width of the red emission peak relative to the blue. In the strongest flares, (e,f) the dynamics of the atmosphere are more dramatic such that the velocity structure is more complex. Here there are sharper gradients, and strong condensations begin to form. Multiple optically thick components begin to form at different heights and the profiles can appear more complicated.

Figure~\ref{fig:contfns_2} shows the effects of varying the other beam parameters, which effectively is changing the location in the atmosphere at which the energy is injected. The resulting atmospheric structure can vary sufficiently given the location of energy injection so that Si~\textsc{iv} formation is very different even for a fixed amount of injected energy (in these cases we show $1\mathrm{F}10$). The top row panels have $\delta = 4$, and $E_{c} = 20,30,40$~keV (panels a-c respectively) and the bottom row panels have $\delta = 6$, and $E_{c} = 20,30,40$~keV (panels d-f respectively). It is immediately clear that the cases of larger values of $E_{c}$ are more likely to lead to optically thin emission, whereas depositing more energy at higher altitudes can lead to optical depth effects. Similarly, comparing panels (b) and (e) shows that a harder spectrum ($\delta = 4$; meaning that more energy is carried by deeply penetrating electrons) has a similar effect. 

	\subsubsection{Plasma Properties at the Height of Peak Emission}\label{sec:plasmaprops}

During the flares, opacity effects become important for the resonance line photons when the column density of Si~\textsc{iv} through which they have to travel becomes sufficiently large. However, it is not possible to obtain this quantity before running a simulation, nor is it easy to obtain this from observations. It is instructive to turn to another metric of the plasma that might serve as a proxy, and inform under what conditions optical depth becomes significant. {If the Si \textsc{iv} ground state population is sufficiently high then the atmosphere may be optically thick to those wavelengths. In relative terms, this can happen if the Si~\textsc{iv} populations are modest over a large geometric region, or if they are very high over a short geometric region. Integrating the mass density over the rough formation temperature range of Si~\textsc{iv} can provide a proxy of the amount of Si~\textsc{iv} a photon must travel through, and a threshold of at what point the atmosphere becomes opaque to the Si~\textsc{iv} resonance lines. That is, the column mass over the formation region of Si~\textsc{iv} must be at a certain value.}

The temperature range at which we evaluated this threshold was $40$~kK $\leq T \leq 100$~kK. The cumulative sum  $m_{\mathrm{col},} = \sum_{z = z_{100\mathrm{kK}}}^{z_{40\mathrm{kK}}}\rho(z) \delta z$, starting at the height at which $T = 100$~kK, $z_{100\mathrm{kK}}$, until the height at which $T = 40$~kK, $z_{40\mathrm{kK}}$ (where $\rho$ is the mass density). These temperatures represent \textsl{typical} formation temperature ranges, but we note that during the flare the lines can occasionally form outwith this range. For this initial study, though, we deemed it a suitable choice to provide insight to the atmospheric properties where Si~\textsc{iv} forms. Since the temperature structure can be complex with local minima and maxima we selected the minimum height at which $T \geq 40$~kK and the maximum height at which $T \leq 100$~kK. The mass density was then interpolated to a much finer grid with 1000 cells, and the sum performed. 

\begin{figure*}[ht]
	\centering 
	\hspace{0.001in}
	\vspace{-0.2in}
	\hbox{
		\subfloat{\includegraphics[width = 0.7\textwidth, clip = true, trim = 0cm 0cm 1cm 0cm]{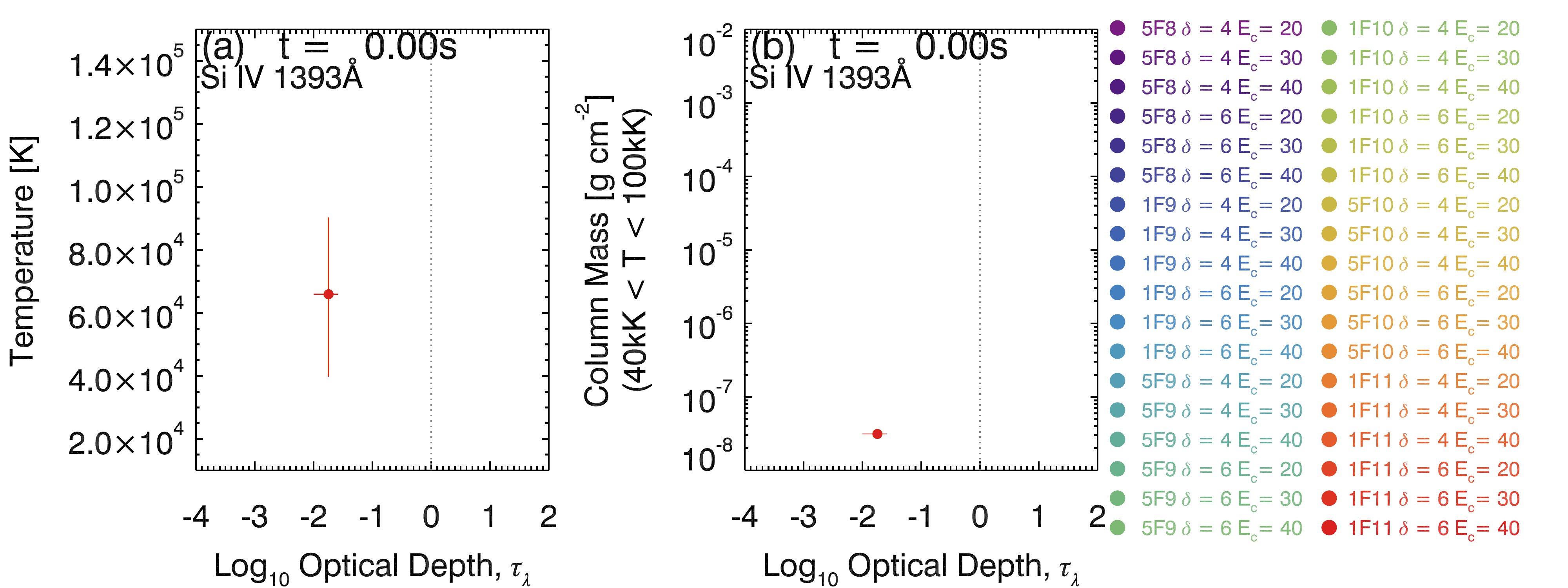}}
		}
	\vspace{-0.2in}
	\hspace{0.001in}
	\hbox{
		\subfloat{\includegraphics[width = 0.7\textwidth, clip = true, trim = 0cm 0cm 1cm 0cm]{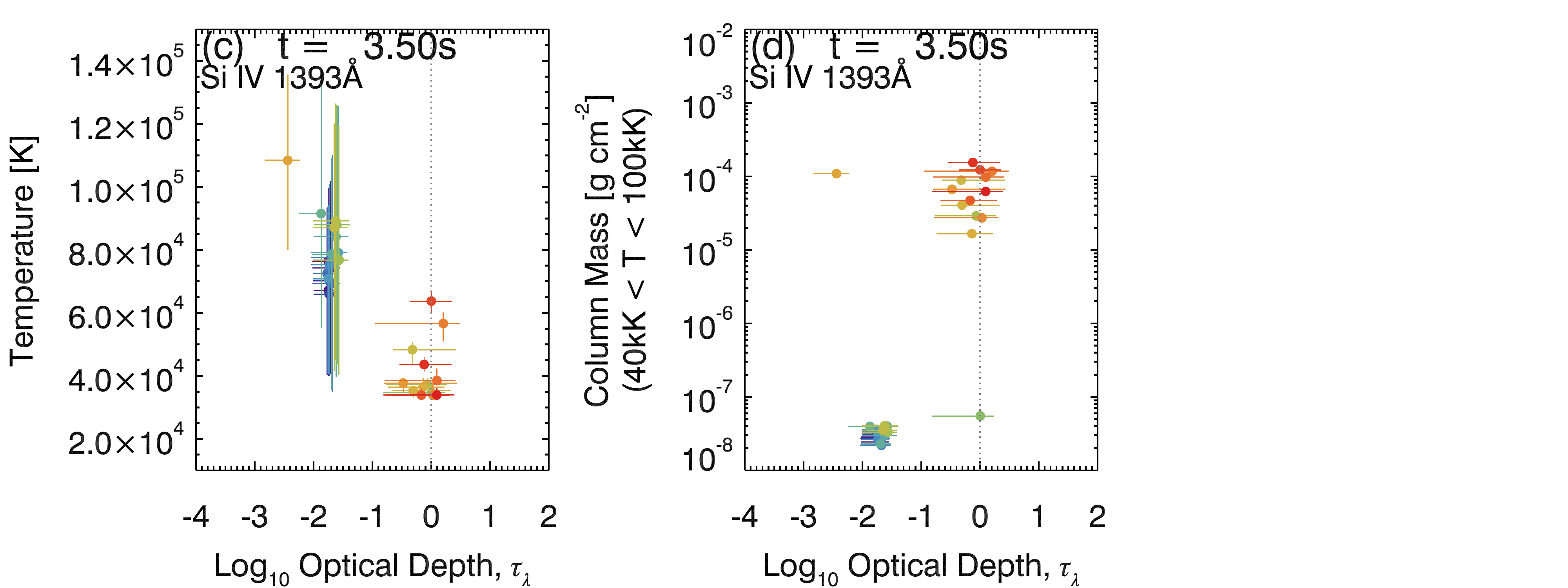}}
		}
	\hspace{0.001in}	
	\vspace{-0.2in}
	\hbox{
		\subfloat{\includegraphics[width = 0.7\textwidth, clip = true, trim = 0cm 0cm 1cm 0cm]{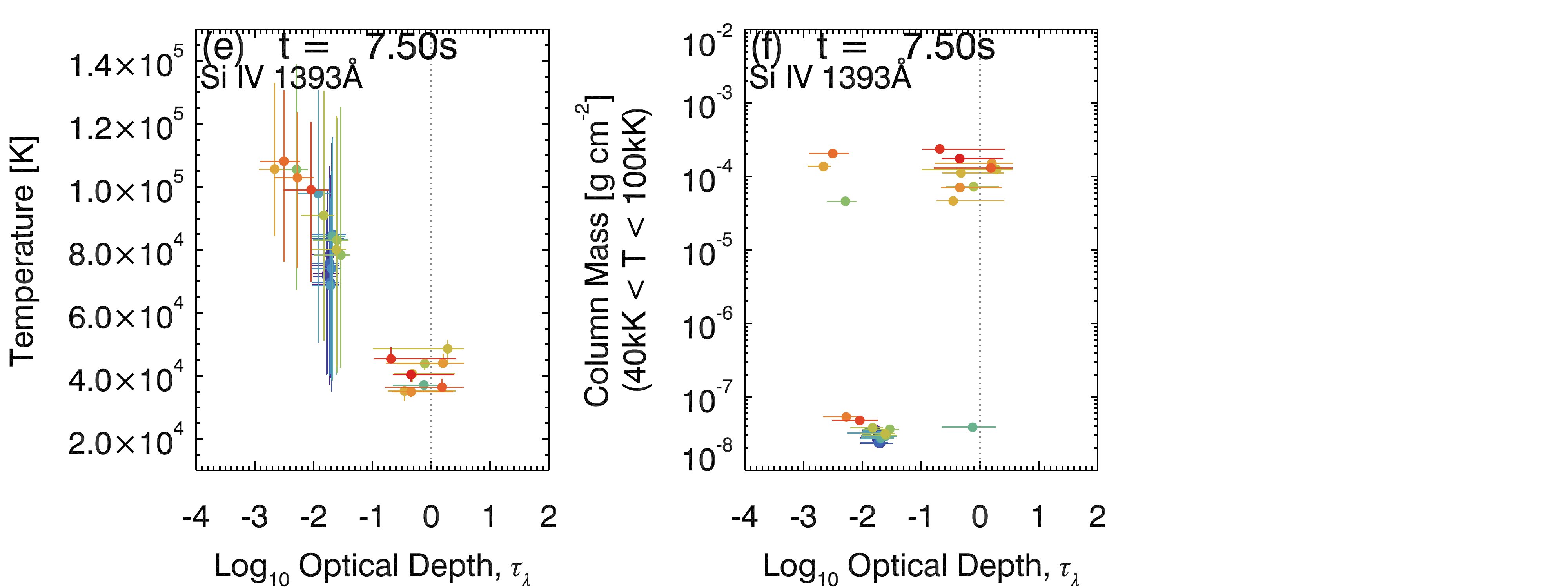}}
		}	
	\hspace{0.001in}
	\hbox{
		\subfloat{\includegraphics[width = 0.7\textwidth, clip = true, trim = 0cm 0cm 1cm 0cm]{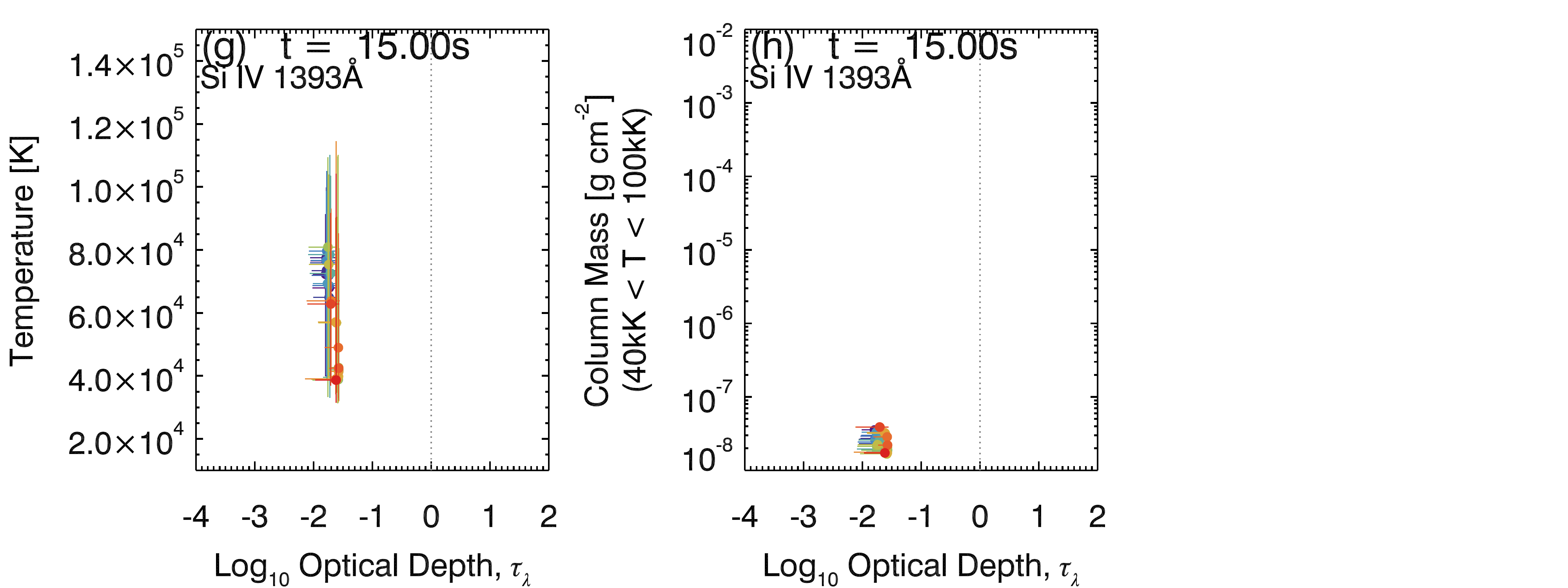}}
		}
		\caption{\textsl{The plasma properties at $t = 0, 3.5, 7.5, 15$~s for the $1393.75$~\AA\ line, from the \texttt{MS\_RADYN} simulations. (a) The temperature at the altitude at which the the contribution function to the emergent intensity at peak line intensity is maximum, as a function of optical depth at that altitude (at the wavelength of line peak intensity). Error bars show temperature and optical depth at heights at which the contribution function reaches FWHM. (b) The column mass, $m_{\mathrm{col}}$, between $40$~kK $\leq T \leq 100$~kK as a function of optical depth at the formation height. In both panels the colours refer to different non-thermal electron distribution parameters as indicated in the annotation. The dotted line shows $\tau_{\lambda} = 1$.}}
	\label{fig:temp_and_mcoldepth}
\end{figure*}

Comparing $m_{\mathrm{col}}$ to the optical depth can allow a rough threshold to be identified, above which the lines are likely to be optically thick. For simulations that produce a sufficiently high $m_{\mathrm{col}}$ the full NLTE radiation transfer should be solved for Si~\textsc{iv}, and for simulations in which $m_{\mathrm{col}}$ is low enough, a process similar to \texttt{Model B} can be used (with the caveat that the intensities will be different by some amount due to exclusion of charge exchange). Observationally this is not a straightforward metric to measure, but we can use our model grids to determine if there is a parameter range for the non-thermal electron distribution that typically results in the atmosphere exceeding this threshold. Hard X-ray observations from the RHESSI \citep{2002SoPh..210....3L} or FERMI \citep{2009ApJ...702..791M} satellites, or from other observatories, can be used to derive parameters for the non-thermal electron distribution, which can be compared to our study, providing some guidance on the liklihood that opacity effects are present. 

For each snapshot output from \texttt{MS\_RADYN} the wavelength at which the intensity of the resonance lines peaked was selected. The temperature and optical depth at the height of the peak of the contribution function for that wavelength was recorded. The heights corresponding to the full width at half maximum (FWHM) of the contribution function, and the temperatures and optical depths at those heights were also recorded. Choosing the peak line intensity helped to avoid the requirement of tracking the line core position (which becomes ambiguous for optically thick lines anyway) in atmospheres with strong flows. In the optically thin cases the line core and position of peak line intensity are typically the same. In the optically thick cases with self-absorption features, the peak line intensity is a component of the line that forms somewhat below the core.

Figure~\ref{fig:temp_and_mcoldepth} shows the gas temperature as a function of the optical depth at the formation height of the line peak at $t=0$~s in panel (a), and $m_{\mathrm{col}}$ also as a function of optical depth in panel (b), at $t=0$~s. The error bars represent the same values at the heights of the FWHM of the contribution function. Figure~\ref{fig:temp_and_mcoldepth}(c,d,e,f,g,h), show the same properties at $t=3.5$~s, $t=7.5$~s \& $t=15$~s respectively, illustrating the progression during the flare simulations. These figures show results derived from the $\lambda = 1393.75$~\AA\ line only. The $\lambda = 1402.77$~\AA\ line produced similar results, but with optical depths that were sometimes smaller than the $\lambda = 1393.75$~\AA\ line. 

At $t=0$~s the temperature of line peak formation was $\mathrm{log}~T\approx 4.82$ ($T\approx65.9$~kK). The optical depth at the formation height was $\tau_{\lambda} = 0.018$. For reference, excluding charge change meant that the line peak formed at $\mathrm{log}~T\approx 4.87$ ($T\approx74.2$~kK), with an optical depth $\tau_{\lambda} = 0.014$.

From these figures it is clear that in many of the simulations the flare environment is such that opacity effects are significant for the Si~\textsc{iv} resonance line profiles, forming in an optically thick regime (with $\tau_{\lambda} \gtrsim0.5$).  Overall, two populations exist, one where opacity effects are important and one in which they are not. In the latter, the lines generally form in hotter plasma, with $T\approx60-100$~kK, with values of $m_{\mathrm{col}}\approx10^{-8}$~g~cm$^{-2}$. Here any increase of the Si~\textsc{iv} fraction is small so that the opacity change (if any) is also small. The plasma is optically thin to the resonance lines. 

In the other population the lines form at somewhat cooler temperatures, with $T\approx30-60$~kK, and over larger column masses on the order $m_{\mathrm{col}} \gtrsim 5\times10^{-6}$~g~cm$^{-2}$ to $m_{\mathrm{col}} \sim 5\times10^{-4}$~g~cm$^{-2}$ so that the opacity increase is significant. Over time as $m_{\mathrm{col}}$ increased there was a transition from $\tau_\lambda << 1$ to $\tau_\lambda \sim 1$, with a rough threshold of $m_{\mathrm{col}}\approx5\times10^{-6}$~g~cm$^{-2}$ required for the plasma to be optically thick to the Si~\textsc{iv} resonance lines. 

The dynamics of the atmosphere are time-dependent, so that some of the simulations take time to reach the $m_{\mathrm{col}}$ threshold and transition from being optically thin to optically thick. The atmosphere can become compressed, reducing the geometric size of the region exhibiting temperature and electron density enhancements. The path length through which photons must pass is consequently reduced. This process that occurs more rapidly when heating is concentrated higher in the atmosphere. The resonance lines can transition back to being optically thin before the end of the heating phase, for example $1\mathrm{F11}\delta6\mathrm{E_{c}}20$ in Figure~\ref{fig:temp_and_mcoldepth}(e,f).  After the heating phase the temperature and electron density both quickly decrease and in all simulations the lines (while still at a greater intensity that their quiet-Sun values) are optically thin.

Our use of $m_{\mathrm{col}}$ between $40$~kK $\leq T \leq 100$~kK as a proxy of optical depth is, of course, not a perfect rule and formation temperatures can fall outside of the bounds we have selected. This leads to outliers from the general trend. Outliers are also possible due to the selection of the peak intensity, as other line components (e.g. a centrally reversed line core) may be optically thick even if the peak intensity is not. However, we believe that this is a robust enough measure for this initial study, providing a proxy for when opacity effects are substantial in many of the simulations. 

\subsection{Formation Height Relative to other IRIS Spectral lines}\label{sec:iris_comparison}
Using $1\mathrm{F11}\delta6\mathrm{E_{c}}20$ as a case study we compare the height of formation of the Si~\textsc{iv} emission to the formation height of other strong spectral lines observed by IRIS in flares. Following the approach described in \cite{2016ApJ...827..101K} \& \cite{GSK_Thesis}, snapshots of the \texttt{RADYN} atmospheres were used as input to the \texttt{RH} radiation transfer code \citep{2001ApJ...557..389U} to obtain synthetic spectra of the Mg~\textsc{ii} k line, the Mg~\textsc{ii} 2791\AA\ line, the C~\textsc{II} 1334\AA\ and the O~\textsc{i} 1356\AA\ line. The \texttt{RH} code was required here because these ions are not simulated by \texttt{RADYN}, and the Mg~\textsc{ii} k line requires a PRD treatment. A caveat here is that \texttt{RH} assumes statistical equilibrium, which is partly mitigated by using the non-equilibrium electron density from \texttt{RADYN}. Future efforts will attempt to simulate these lines using \texttt{MS\_RADYN}, investigating non-equilbrium effects. We use a modified version of \texttt{RH} in which the hydrogen populations as input from \texttt{RADYN} (and thus were the non-equilibrium populations) were held constant while still solving the radiation transfer. Populations of the other elements were allowed to iterate. This means that the computation of the hydrogen opacity (a significant source of background opacity), and the effects of charge exchange, that are important for O~\textsc{i} 1356\AA\ formation \citep{2015ApJ...813...34L}, used the more accurate hydrogen populations from \texttt{RADYN}. 

\begin{figure}
	\centering
		\includegraphics[width = 0.5\textwidth, clip = true, trim = 0.cm 0.cm 0.0cm 0.cm]{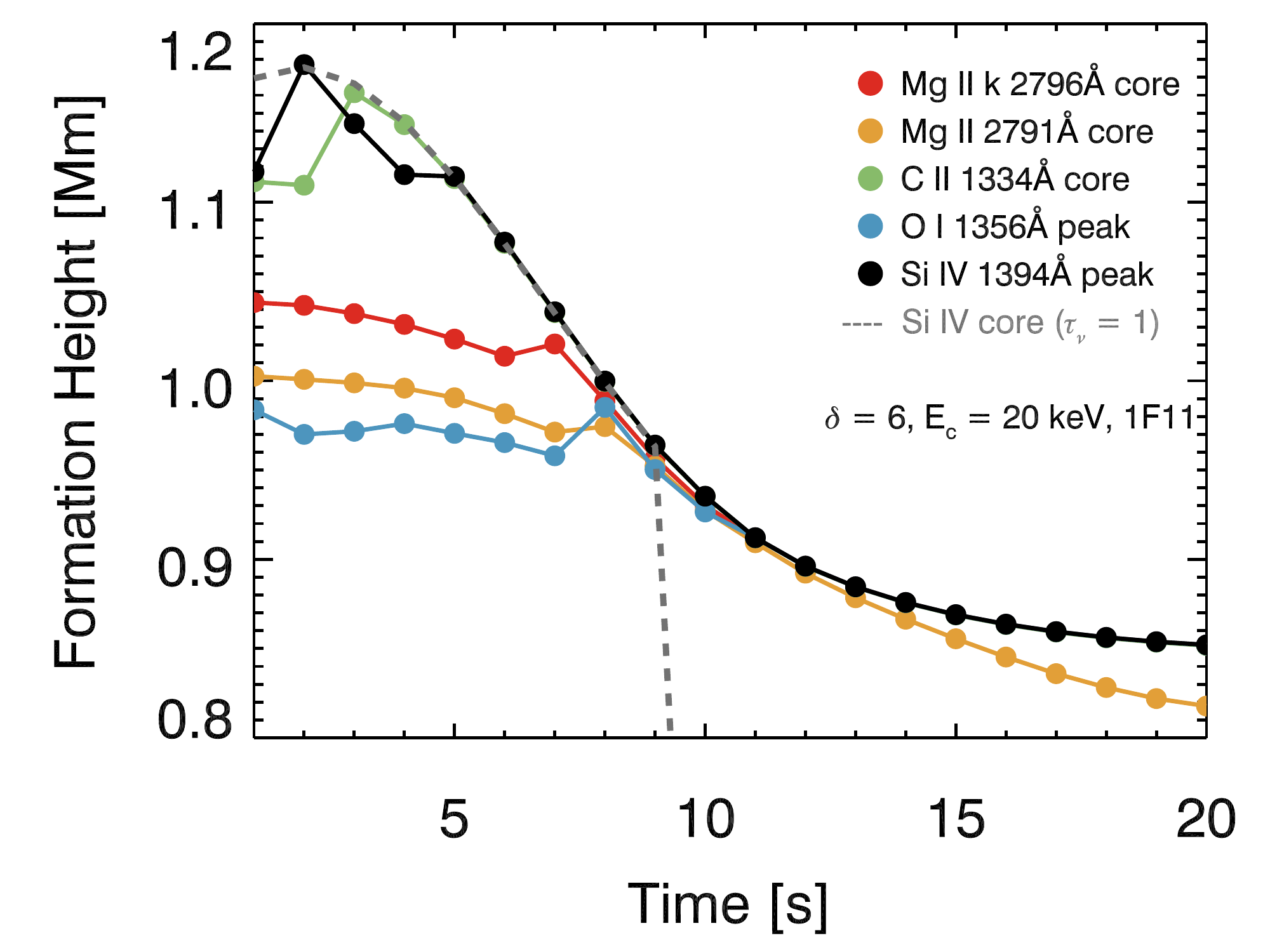}	
          	\caption{\textsl{Formation heights of the Si~\textsc{iv} 1393.75\AA\ core (grey dashed line), Si~\textsc{iv} 1393.75\AA\ emission peak (black circles), C~\textsc{ii} 1334\AA\ core (green circles), Mg~\textsc{ii} k core (red circles), Mg~\textsc{ii} 2791\AA\ core (orange circles), and O~\textsc{i} 1356\AA\ emission peak (blue circles) as functions of time in the $1\mathrm{F11}\delta6\mathrm{E_{c}}20$ simulation. Note that in the gradual phase most lines (apart from Mg~\textsc{ii} 2791\AA) form in a very narrow region (underneath the black symbols).}}
	\label{fig:formheights_comp}
\end{figure}

For optically thick lines (the Mg~\textsc{ii} k, Mg~\textsc{ii} 2791\AA\ and C~\textsc{ii} 1334\AA\ lines) we define the formation height of the line core to be the maximum height of the $\tau_{\lambda} = 1$ surface. These heights are shown in Figure~\ref{fig:formheights_comp}. The O~\textsc{i} 1356\AA\ line is optically thin throughout the flare, and so we instead show the height at which the contribution function to the emergent intensity is largest, for the wavelength corresponding to the peak line intensity. 

Since the Si~\textsc{iv} 1393.75\AA\ line transitions between optically thick and optically thin we show both the height at which $\tau_{\lambda} = 1$  is maximised, and the height at which the contribution function to the emergent intensity is largest, for the wavelength corresponding to the peak line intensity. The former is shown as a dashed grey line, and the latter as black circles. When the line is optically thick the $\tau_{\lambda} = 1$ represents the line core, and the latter metric represents the formation height of the emission peak. When the line forms in optically thin conditions the $\tau_{\lambda} = 1$ surface does not hold any special meaning as the intensity originates far from this height, but we include it in Figure~\ref{fig:formheights_comp} to illustrate the transition. These two measures both show similar heights during the heating phase when the line is optically thick, particularly when the chromosphere compresses so that the formation height difference of the line core and emission peak is small. Shortly before $t=10$~s the height of the $\tau_{\lambda} = 1$ surface drops, whereas the line continues to form in the upper chromosphere / TR. At these times the $\tau_{\lambda} = 1$ no longer represents the line core, and the line is considered optically thin. 

There is a stratification in the heights at which the various lines form, that varies with time in the flare. Initially the core of the Si~\textsc{iv} $1393.75$\AA\ line forms highest in altitude, followed by the C~\textsc{ii} resonance lines, the emission peak of the Si~\textsc{iv} $1393.75$\AA\ line (recall that the line has a self-absorption feature that is the core), the Mg~\textsc{ii} resonance lines, the Mg~\textsc{ii} subordinate lines, and finally the O~\textsc{i} $1356$\AA\ line. After several seconds mass motions, and the changing temperature and density structure, raise the height of the C~\textsc{ii} resonance lines and reduces the height difference between the Si~\textsc{iv} resonance line core and the emission peak. Towards the end of the heating phase the compression of the chromosphere results in the lines forming in a very narrow region, that persists into the cooling phase. The evolution of the stratification of line components will be examined in more detail, in order to identify potential diagnostics such as temperature and velocity gradients.

\section{Summary and Conclusions}\label{sec:summary}

The flare response of the Si~\textsc{iv} resonance lines were computed under the assumption that they are formed in optically thin conditions using atomic data from the CHIANTI database (\texttt{Model B}), and using full NLTE, non-equilibrium radiation transfer via the minority species version of the \texttt{RADYN} radiation hydrodynamics code (\texttt{MS\_RADYN}). In both models we used 36 flare simulations computed via \texttt{RADYN}, that covered a wide range of parameter space of non-thermal electron beams. The initial results from this model grid were presented, illustrating the changes in intensities, shapes, and ratios, as well as some basic formation properties. The Si~\textsc{iv} lines were also placed in context with other strong lines observed by the IRIS spacecraft in a case study of one simulation. 

Generally, weaker flares resulted in reasonably similar Si~\textsc{iv} line emission when using either \texttt{Model B} or \texttt{MS\_RADYN}. In stronger flares the lines differed, sometimes significantly, in terms of intensity, the presence of self-absorption features, asymmetries, Doppler motions, and widths. The intensity ratio of the resonance lines when computed by \texttt{MS\_RADYN} deviate from the optically thin limit of two, in the stronger flares. 

Our results suggest that in flaring conditions, or possibly during smaller scale but intense heating events, the populations of Si~\textsc{iv} can become substantially enhanced over a large enough geometric region so as to result in the upper chromosphere and lower transition region becoming (semi-) opaque to Si~\textsc{iv} resonance line wavelengths. It was found that all flares with injected energy flux $>5\mathrm{F}10$ resulted in optically thick Si~\textsc{iv} emission. Weaker flares also resulted in optical depth effects, depending on the other electron beam parameters. {Empirically, optical depth effects occured when the column mass was sufficiently enhanced, with a rough threshold column mass between $40<T<100$~kK of $m_{\mathrm{col}} \approx 5\times10^{-6}$~g~cm$^{-2}$.} 

The formation temperature in optically thin conditions was around $T\approx60-100$~kK, and in optically thick conditions was somewhat lower, around $T\approx30-60$~kK. There were exceptions to this rule, where the Si~\textsc{iv} resonance line cores formed in significantly higher temperatures $T>100$~kK. When this occurred the line cores were mostly optically thin, and formed in a geometrically narrow region of significantly enhanced electron density (in excess of a few $\times10^{13-14}$~cm$^{-3}$). 

Flares that deposit energy at higher altitudes than those in our simulations might also produce optically thick Si~\textsc{iv} emission, even if the energy flux is relatively low. Longer flare loops might produce optically thick emission even with the lower range of injected flux, due to the larger column mass over which the Si~\textsc{iv} populations could be enhanced.

A large scale survey of Si~\textsc{iv} emission during flares of various sizes, measuring intensities, line ratios and noting the existence or lack of any self-absorption features, should be undertaken. If observational evidence points towards optically thin emission then we must address why modelling predicts the opposite. This is also relevant for the higher-than-observed intensities of other chromospheric synthetic spectral lines, such as Mg~\textsc{ii} h \& k lines, and C~\textsc{ii} resonance lines. As such we also recommend adding to any observational effort the ratios of all the strong resonance lines observed by IRIS, which can be compared to simulations.  

In these simulations the Si~\textsc{iv} resonance lines show opacity effects during the heating phase, and quickly revert to forming in optically thin conditions in the cooling phase. This will likely make detecting these effects on Si~\textsc{iv} difficult in observations without adequate temporal cadence. Even IRIS flare observations with cadences of $t=5-10$~s would only provide one or two profiles, from a particular spatial location during the heating phase, that exhibit the effects described here (assuming an energy deposition time of $t\sim10$~s as in our simulations). Subsequent observations from the same spatial location would be much weaker. Multi-threaded flare simulations \citep[e.g.][]{2016ApJ...827..145R} have shown that sustained heating of various threads into one atmospheric element can explain certain observations of Doppler motions. In such a scenario optical depth effects may be observational for longer as different threads are activated. The Si~\textsc{iv} lines would also likely be weaker as only a fraction of threads would be activated at the same time, with emergent intensity from that pixel averaged over threads that are in their quiescent, heating, and cooling phases.

Finally, it is worth noting that the rapid cooling following cessation of flare energy input is likely to be, in part, due to an overestimation of thermal conduction. It was recently demonstrated \citep{2016ApJ...824...78B,2016ApJ...833...76B, 2017ApJ...835..262B,2018ApJ...852..127B,emslieandbian18} that non-local effects and turbulence can lead to a thermal conduction that is significantly smaller than that predicted by classical Spitzer conductivity (even with saturation to the electron free-streaming limit as currently implemented in \texttt{RADYN}). We plan to implement the suppression of thermal conduction following \cite{emslieandbian18} in a future work. This may extend the time period over which the Si~\textsc{iv} resonance line exhibit optical depth effects. 

Based on these simulations we urge caution when analysing observations of the Si~\textsc{iv} resonance lines, as techniques that assume optically thin conditions may not be valid. Further, producing synthetic emission from simulated, or semi-empirical, flare atmospheres using a purely optically thin assumption, without NLTE radiation transfer, may not be appropriate. Modelling should include the effects of charge exchange which is an important process in the formation of Silicon ions in the solar atmosphere.  \\

\textsc{Acknowledgments:} \small{}
GSK was funded by an appointment to the NASA Postdoctoral Program at Goddard Space Flight Center, administered by USRA through a contract with NASA.  MC acknowledges support from he Research Council of Norway through its Centres of Excellence scheme, project number 262622, and through grants of computing time from the Programme for Supercomputing. JCA and AND acknowledge support from the IRIS mission. IRIS is a NASA small explorer mission developed and operated by LMSAL with mission operations executed at NASA Ames Research Center and major contributions to downlink communications funded by ESA and the Norwegian Space Centre. PRY acknowledges support from NASA grant NNX15AF25G. We appreciate the open data policies of the Opacity Project, and the NIST \& CHIANTI atomic database. The authors thank ISSI \& ISSI-BJ for the support to the teams: ``Diagnosing heating mechanisms in solar flares through spectroscopic observations of flare ribbons'' and ``New Diagnostics of Particle Acceleration in Solar Coronal Nanoflares from Chromospheric Observations and Modeling''. Finally, we would like to thank the anonymous referee who's comments helped to improve this work. 

\bibliographystyle{aasjournal}
\bibliography{Kerr_etal_SiIV_OptThickSims_2018}

\appendix

\section{Silicon Model Atom Details}\label{sec:detailedatom}
Here we present details regarding the construction of our model silicon atom.  \\

{Oscillator strengths, $f_{ij}$, were taken from the CHIANTI database.} Radiative damping coefficients were obtained by summing the Einstein $A_{ji}$ coefficients associated with upper level $j$, so that the lifetime of level $j$ was $t_{j} = 1/\sum A_{ji}$. Direct collisional ionisation rate coefficients, and collisional excitation rate coefficients were taken from CHIANTI. 

The abundance used was  $A_{\mathrm{Si}} = 7.51$ \citep{2009ARA&A..47..481A}, defined on the usual logarithmic scale, with $A_{\mathrm{H}} = 12$. This is assumed to be constant through the loop, and in time. We note that recent work by \cite{2015ApJ...802....5O} and \cite{2016ApJ...817...46M} argues in favour of using coronal abundances for silicon, based on the findings that first ionisation potential (FIP) effects enrich the abundance of low-FIP elements in the upper atmosphere \citep[e.g][]{1992PhyS...46..202F,2004ApJ...614.1063L}. However, \cite{2016ApJ...824...56W} recently reported that in impulsive heating events low-FIP elements have a composition similar to the photosphere. Since we are simulating flares, and given general ambiguity in elemental abundances, we elect to use the photospheric value \citep{2009ARA&A..47..481A}, but provide in Appendix~\ref{sec:coronalabundance} an illustration of the impact on our results of using the coronal abundance.  

Photoionisation cross sections from the TOPbase database\footnote{\url{http://cdsweb.u-strasbg.fr/topbase/topbase.html}} \citep{1993A&A...275L...5C}, remapped to the wavelengths in the \texttt{RADYN} background opacity package, were used for the 33 bound-free transitions. To convert the TOPbase transitions from energy in Rydbergs units to wavelengths in angstroms we used the value of the Rydberg constant for Silicon, $R_{\mathrm{Si}} = 109735.1723$~cm$^{-1}$ ($R_{\mathrm{Si}}=R_{\infty}/(1+m_{e}/M_{\mathrm{Si}})$, for electron mass $m_e$, atomic mass $M_{\mathrm{Si}}$, and Rydberg constant for infinite mass $R_{\infty}$), and corrected the threshold energies according to the NIST database \citep{NIST_ASD}, in the expression:
	\begin{equation}\label{eq:wavel}
		\lambda [\AA]= \frac{1\times10^8}{R_{\mathrm{Si}}(E[\mathrm{Ryd}]|_{\mathrm{TOPbase}}+ \Delta E[\mathrm{Ryd}])},
	\end{equation}
\noindent where $\Delta E$ is the difference in NIST threshold energy and the TOPbase value. As noted in \cite{2013ApJ...772...89L}, TOPbase provides photoionisation cross sections that extend below the threshold energy, which we discard here. 

Recipes from \cite{1985A&AS...60..425A} were used for charge exchange (also known as charge transfer) with H~\textsc{i}, H~\textsc{ii}, He~\textsc{i}, \& He~\textsc{ii}.

For the 33 detailed bound-free transitions the TOPbase cross sections were used for radiative recombinations (the inverse of photoionisations). Only considering radiative recombinations to the levels treated in detail can risk underestimating the total recombination rate. To account for these additional radiative recombinations, and to include dielectronic recombinations, we used rate coefficients from the CHIANTI database in the following way. First, we computed the total radiative recombination rate coefficient to each charge state as a function of temperature at $t=0$~s in a \texttt{MS\_RADYN} test simulation. Then we computed the difference between the total recombination rate coefficients to each charge state from CHIANTI (radiative plus dielectronic) and the radiative recombinations treated in detail by \texttt{MS\_RADYN}. This excess was then included as an additional recombination term in our model atom for flare simulations, {taking place to the ground terms and excited states of the relevant charge state, according to their statistical weights.}

Note that we do not include suppression of dielectronic recombination, a process that occurs at high densities \citep[e.g.][and references therein]{summers_RAL_1974,2013ApJ...768...82N,2018ApJ...855...15Y,2018ApJ...857....5Y}. It has been shown that density sensitive dielectronic recombination can lead to a higher Si~\textsc{iv} fraction at lower temperatures, and a higher peak fraction, compared to the zero-density limit assumed by the CHIANTI rates \cite[e.g.][]{2016A&A...594A..64P}. \texttt{RADYN} has an 
implementation of density effects according to \citet{summers_RAL_1974}, see, \textit{e.g.,} \citet{2015ApJ...811...80R}, but future iterations of this work will modify \texttt{RADYN} to also accept dielectronic rate coeffcients as functions of density, with density dependent rate coefficients from the ADAS\footnote{\url{http://open.adas.ac.uk}} atomic database included in place of those from CHIANTI. For now, we focus on the effects of photoionisations, charge exchange and opacity, but include an illustration of how suppression of dielectronic recombination might affect our results in Appendix~\ref{sec:drsuppression}, by performing an experiment using the suppression factors of \cite{2013ApJ...768...82N} as implemented by \cite{2018ApJ...855...15Y}. 

The silicon data that we used from CHIANTI was from CHIANTI V8.0.7. The sources of those data were:  \cite{2007A&A...466..771D,2008ApJS..179..534T,2012A&A...537A..40A,1989MNRAS.241..209D,2006A&A...447.1165A,2007A&A...474.1051A,2009A&A...500.1263L,2006ApJS..167..334B,2004A&A...426..699Z}.

\section{Using Coronal Abundances}\label{sec:coronalabundance}
In some recent work \citep[e.g][]{2015ApJ...802....5O,2016ApJ...817...46M} is has been argued that it is appropriate to use an abundance larger than the photospheric value for Silicon and other low-First Ionisation Potential (low-FIP) elements \citep{2004ApJ...614.1063L}, for example the coronal abundance from \cite{1992PhyS...46..202F}.  \cite{2016ApJ...824...56W} demonstrate that it is appropriate to use photospheric abundances for impulsive heating events, so we contend that our use of the \cite{2009ARA&A..47..481A} photospheric abundance is valid, particularly given the general ambiguity of elemental abundances in flares. Nevertheless it is useful to provide an illustration of the impact on our results of instead using coronal abundances. 
\begin{figure*}[ht]
	\centering
	\hbox{
	\hspace{1.125in}
	\subfloat{\includegraphics[width = 0.225\textwidth, clip = true, trim = 0cm 0cm 0cm 0cm]{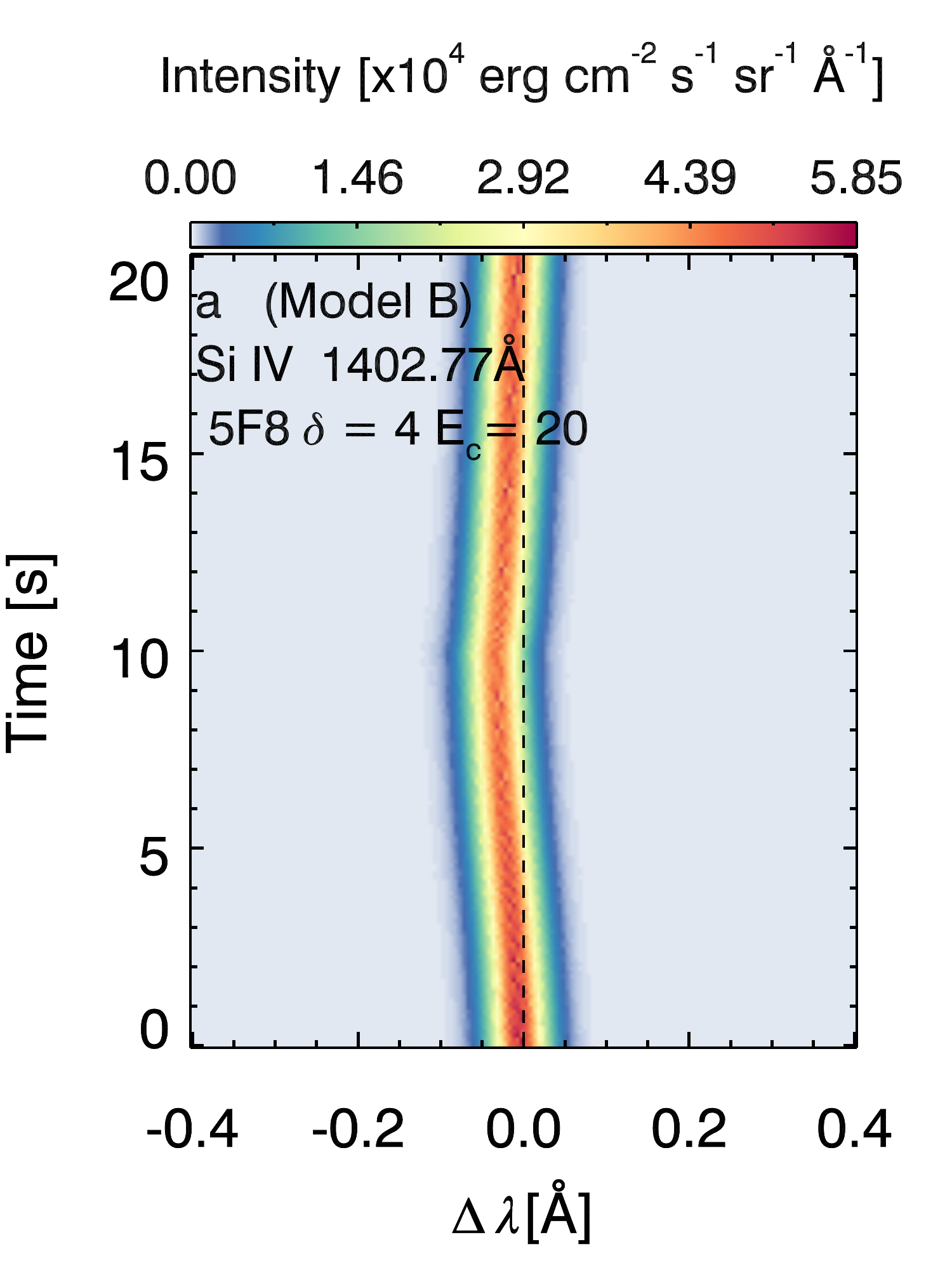}}	
	\subfloat{\includegraphics[width = 0.225\textwidth, clip = true, trim = 0cm 0cm 0cm 0cm]{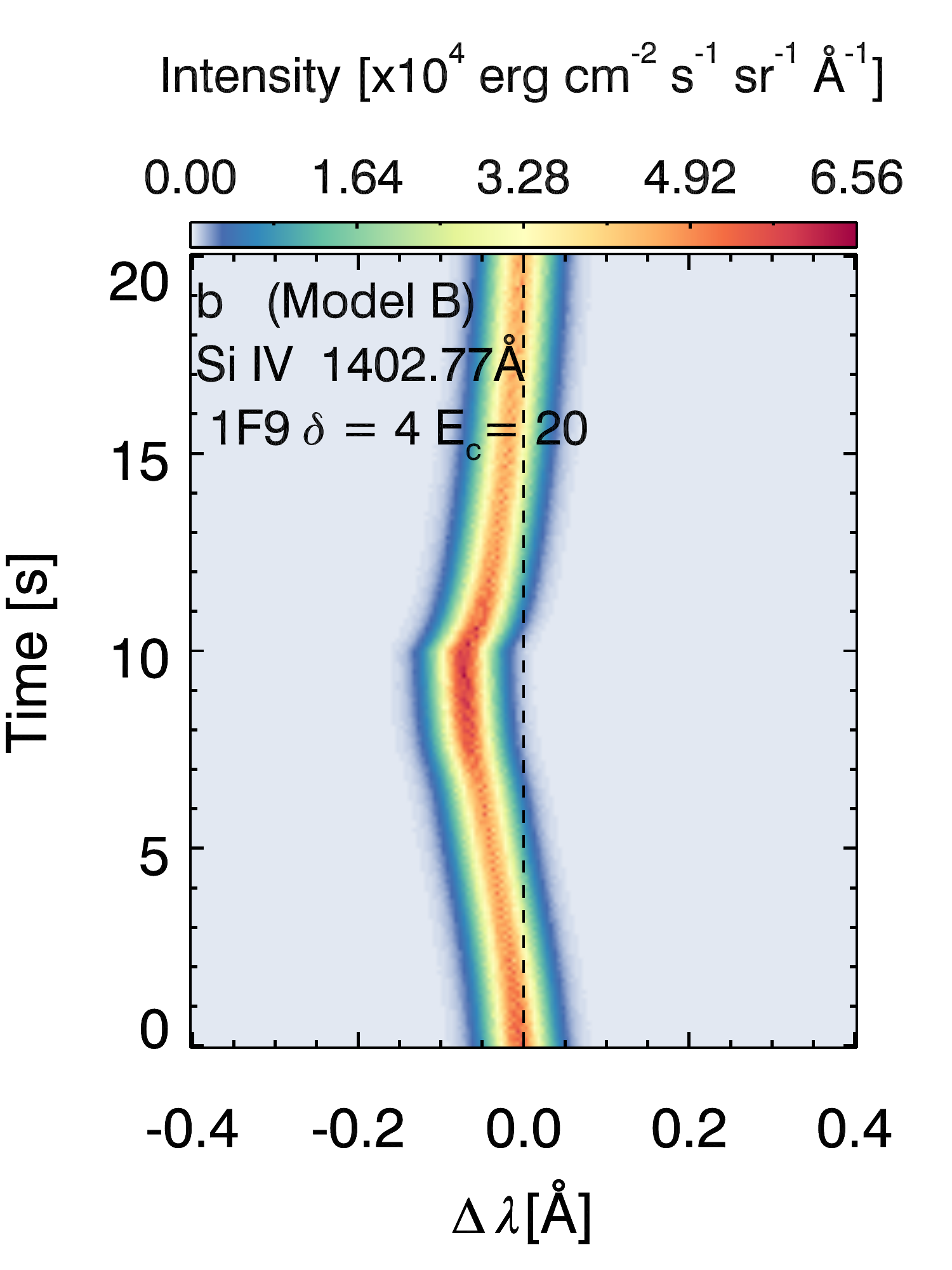}}
	\subfloat{\includegraphics[width = 0.225\textwidth, clip = true, trim = 0cm 0cm 0cm 0cm]{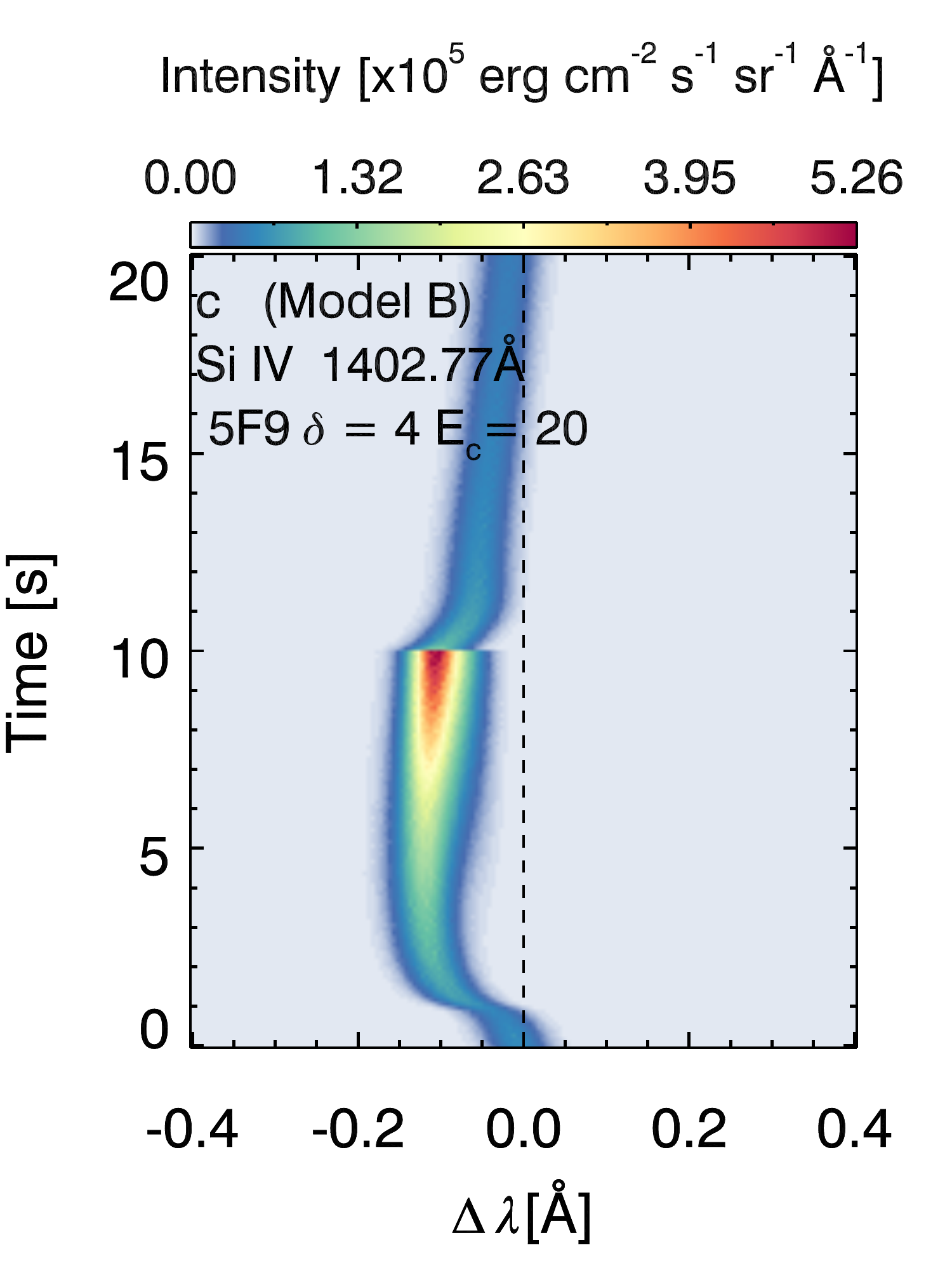}}	
	         }
	 \vspace{-0.2in}
	 \hbox{
	 \hspace{1.125in}
	\subfloat{\includegraphics[width = 0.225\textwidth, clip = true, trim = 0cm 0cm 0cm 0cm]{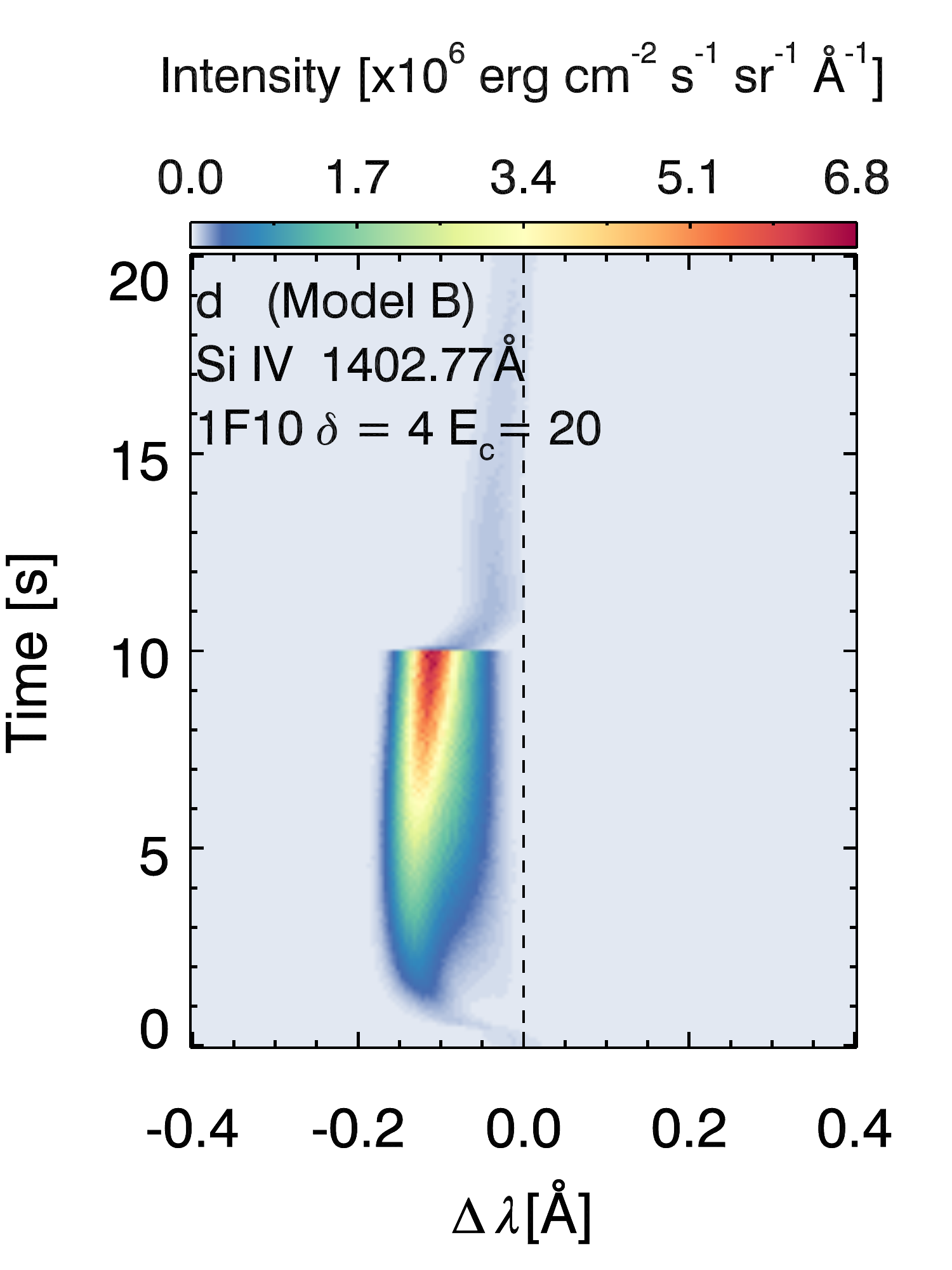}}	
	\subfloat{\includegraphics[width = 0.225\textwidth, clip = true, trim = 0cm 0cm 0cm 0cm]{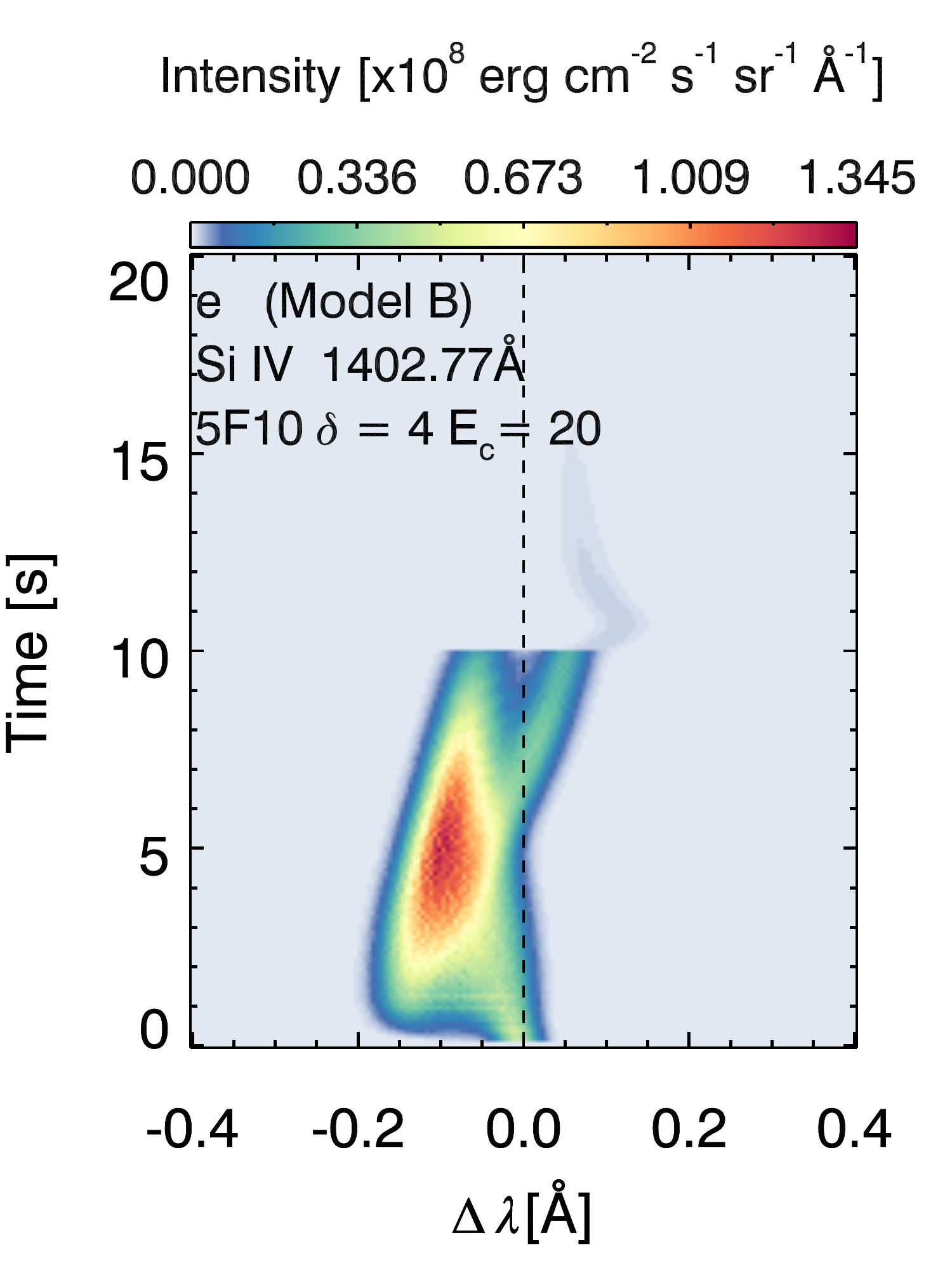}}
	\subfloat{\includegraphics[width = 0.225\textwidth, clip = true, trim = 0cm 0cm 0cm 0cm]{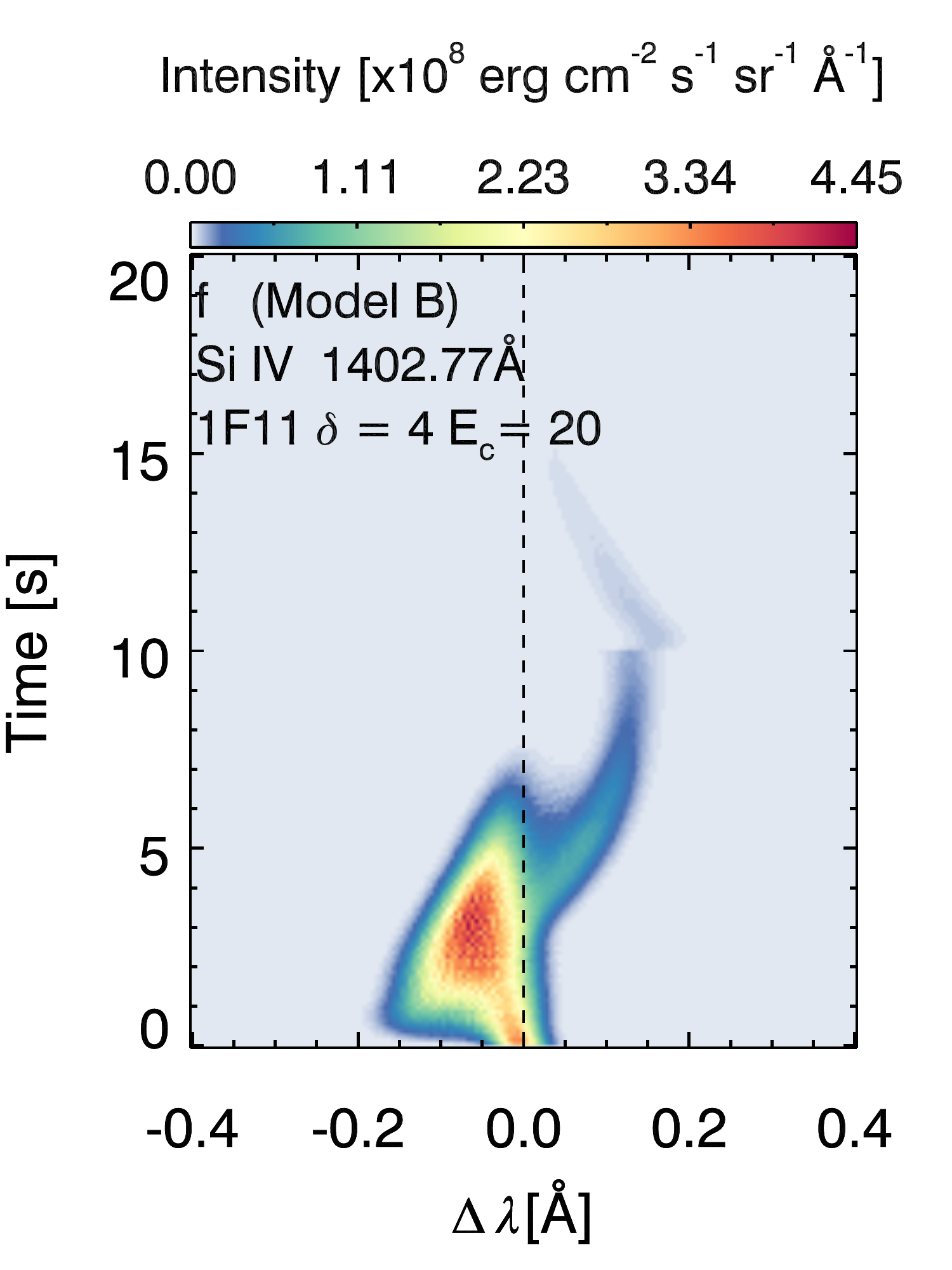}}	
	         }
	\caption{\textsl{Comparing the evolution of the \texttt{Model B} line profiles shown in Figure~\ref{fig:stackplots_1}, to instead using the coronal abundance $A_{\mathrm{Si,Feld}} = 8.10$. Panels and simulations are as described in Figure~\ref{fig:stackplots_1}.}}
	\label{fig:stackplots_coronalabund}
\end{figure*}
\texttt{Model B} results were all recomputed using silicon coronal abundances from \cite{1992PhyS...46..202F}, $A_{\mathrm{Si,Feld}} = 8.10$. Since \texttt{MS\_RADYN} is more computationally demanding we instead ran some test cases using the coronal abundance: $5\mathrm{F}10\delta6E_{\mathrm{c}}20$, and $1\mathrm{F}10\delta4E_{\mathrm{c}}30$. 

\begin{figure*}
	\centering
		\includegraphics[width = 0.85\textwidth, clip = true, trim = 0.cm 0.cm 0.0cm 0.cm]{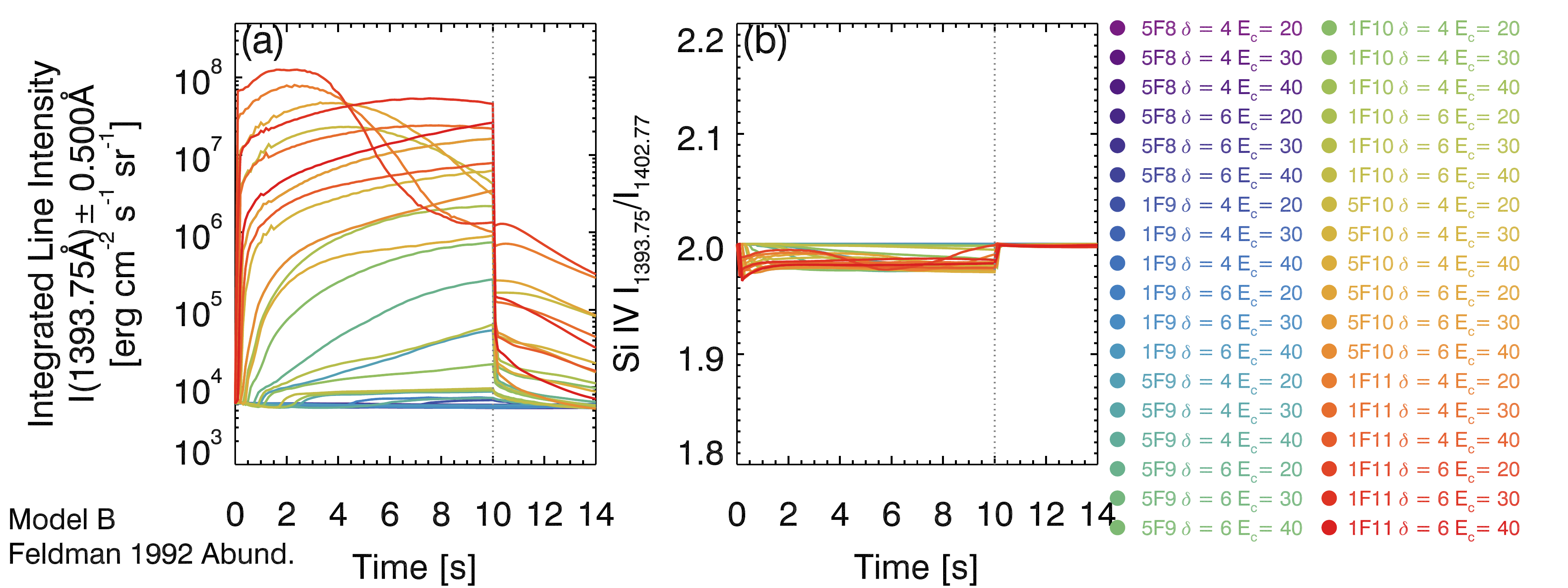}	
          	\caption{\textsl{As in Figure~\ref{fig:lcurves_radyn} but showing results as computed using the optically thin assumption via \texttt{Model B} with coronal abundance $A_{\mathrm{Si,Feld}} = 8.10$.}}
	\label{fig:lcurves_feld}
\end{figure*}

Figure~\ref{fig:stackplots_coronalabund} shows the same \texttt{Model B} simulations presented in Figure~\ref{fig:stackplots_1}, but using $A_{\mathrm{Si,Feld}}$. The line shapes, Doppler motions and general behaviour is preserved, but the lines are more intense. This can also be seen by comparing lightcurves using $A_{\mathrm{Si,Feld}}$, Figure~\ref{fig:lcurves_feld}, and the photospheric value, Figure~\ref{fig:lcurves_chianti}. Again, while temporal behaviour is preserved, the coronal abundances result in much more intense emission. The line ratios are unchanged. 

Results from \texttt{MS\_RADYN} were similarly affected, with the line intensity becoming larger when using the coronal abundance values. The larger abundance also results in the $t=0$~s atmosphere being somewhat more optically thick (so that the line forms closer to $\tau_{\nu} \approx 0.1$), and resulted in the $1\mathrm{F}10\delta4E_{\mathrm{c}}30$ simulation exhibiting optical depth effects in a narrow region of the line core.

\section{Suppression of Dielectronic Recombination}\label{sec:drsuppression}
At high densities, dielectronic recombination rates can be suppressed over the zero-density limits typically used (e.g. the rates in CHIANTI), due to collisions de-populating the higher lying levels through which dielectronic recombination occurs. This is discussed in detail by, for example, \cite{summers_RAL_1974}, \cite{2013ApJ...768...82N}, \cite{2015ApJ...808..116J}, and \cite{2018ApJ...855...15Y}. 

\begin{figure*}[h]
	\centering
		\includegraphics[width = 0.4\textwidth, clip = true, trim = 0.cm 0.cm 0.0cm 0.cm]{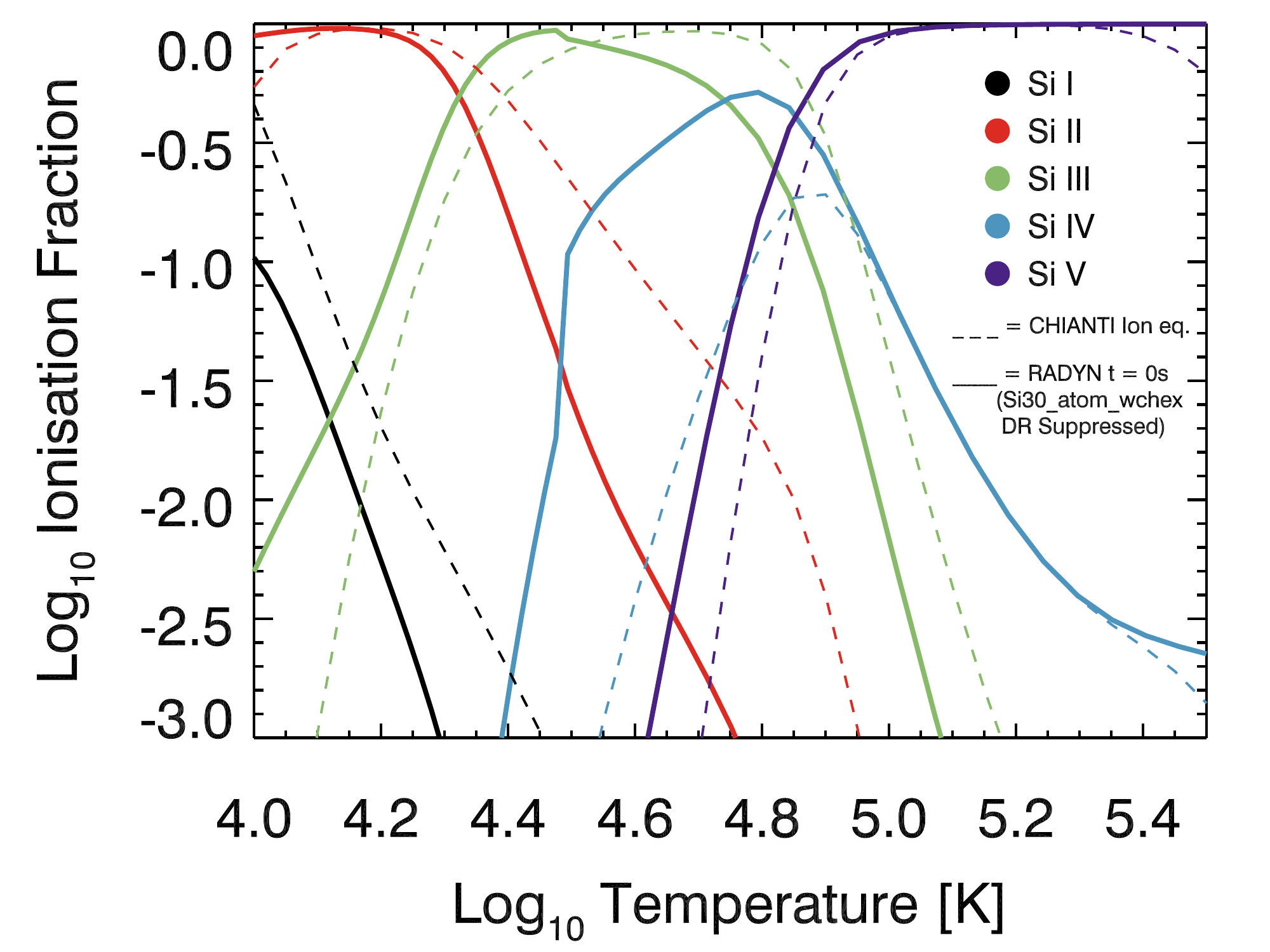}	
          	\caption{\textsl{As in Figure~\ref{fig:si_ionfracs} but showing the effects of including suppression of dielectronic recombination with the \texttt{Si30\_atom\_wchex} atom.}}
	\label{fig:drsuppress_ionfrac}
\end{figure*}

\begin{figure*}[h]
	\centering
	\hbox{
	\hspace{2in}
	\subfloat{\includegraphics[width = 0.225\textwidth, clip = true, trim = 0cm 0cm 0cm 0cm]{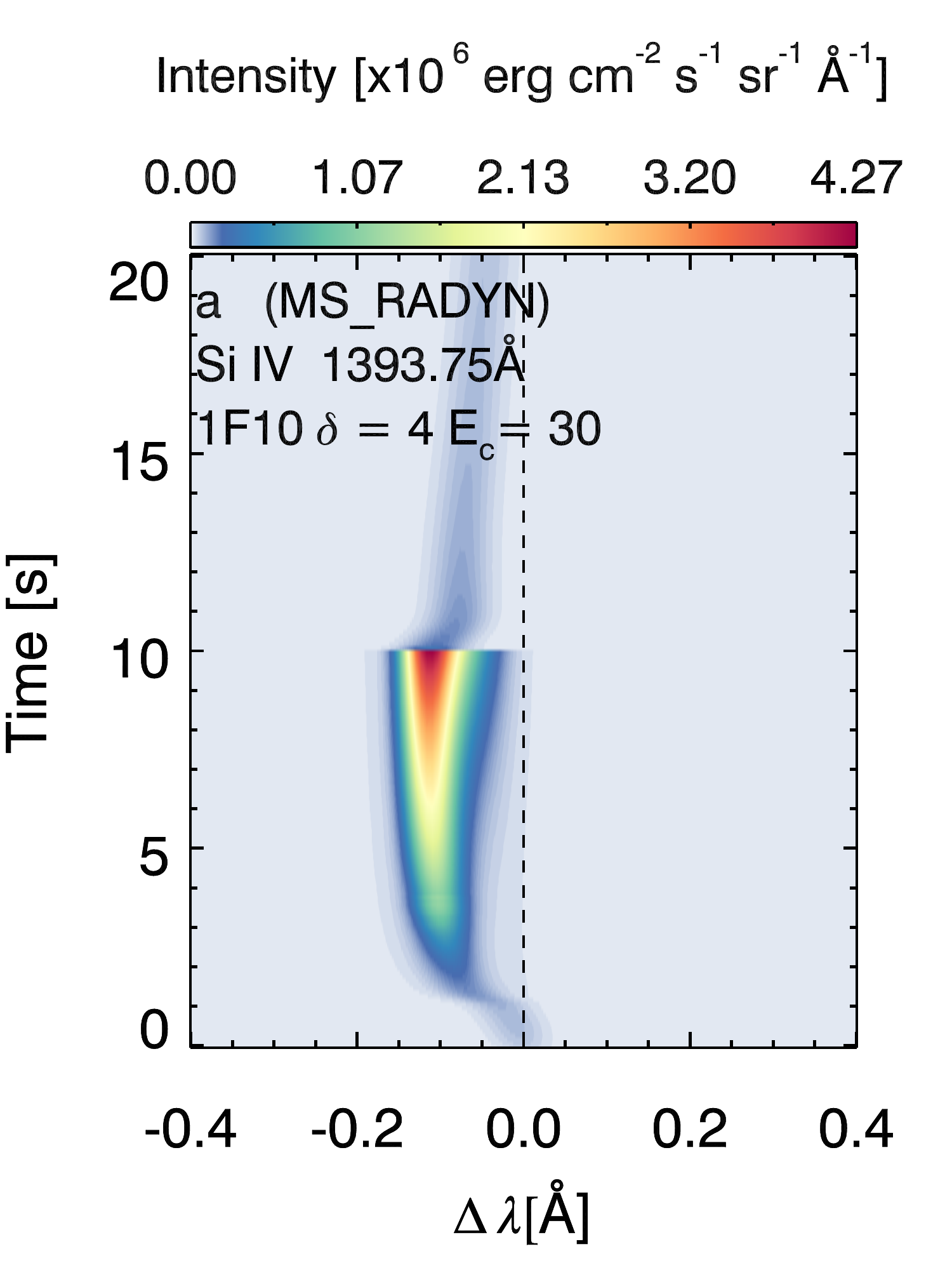}}	
	\subfloat{\includegraphics[width = 0.225\textwidth, clip = true, trim = 0cm 0cm 0cm 0cm]{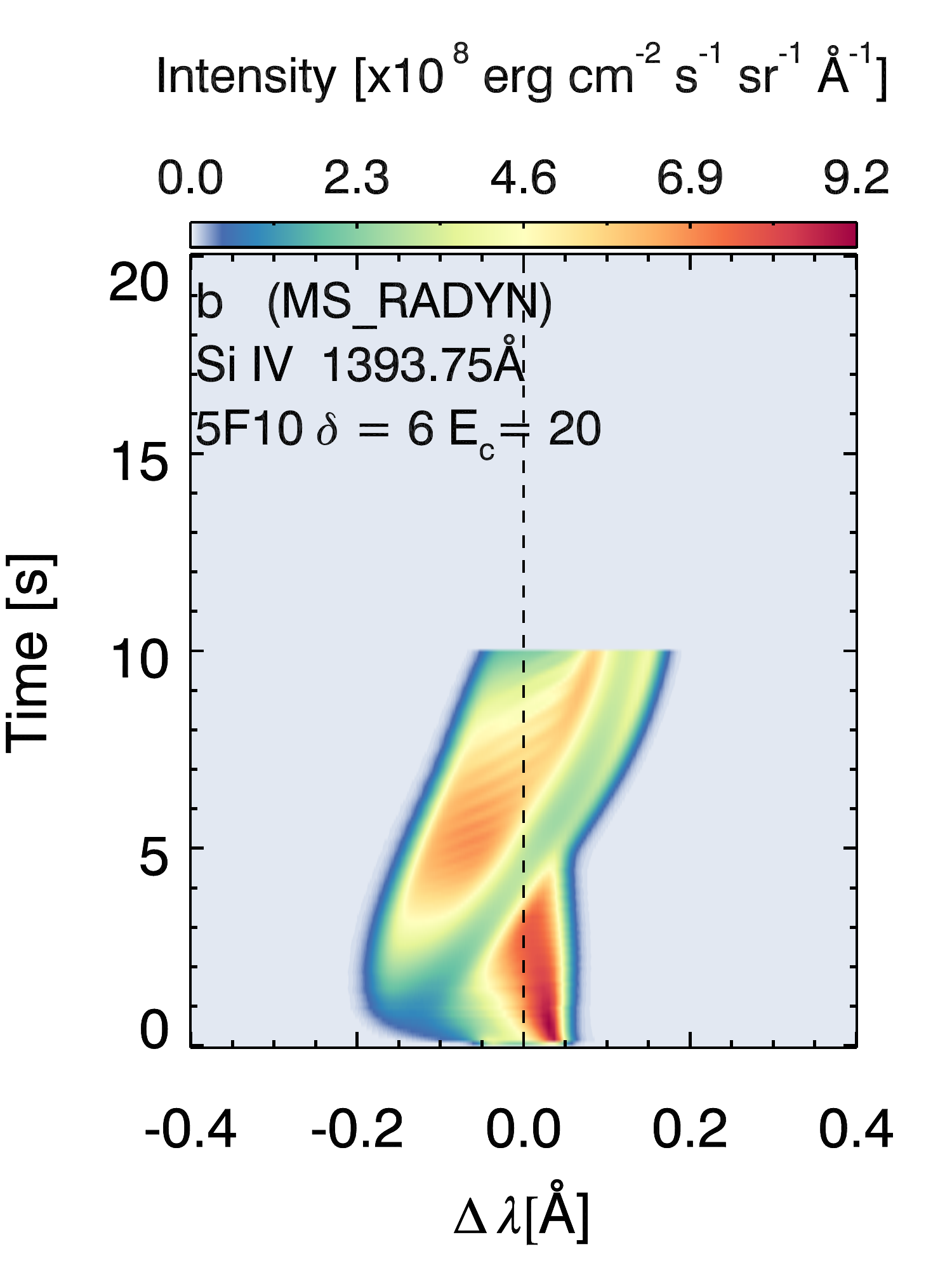}}
	         }
	 \vspace{-0.2in}
	 \hbox{
	 \hspace{2in}
	\subfloat{\includegraphics[width = 0.225\textwidth, clip = true, trim = 0cm 0cm 0cm 0cm]{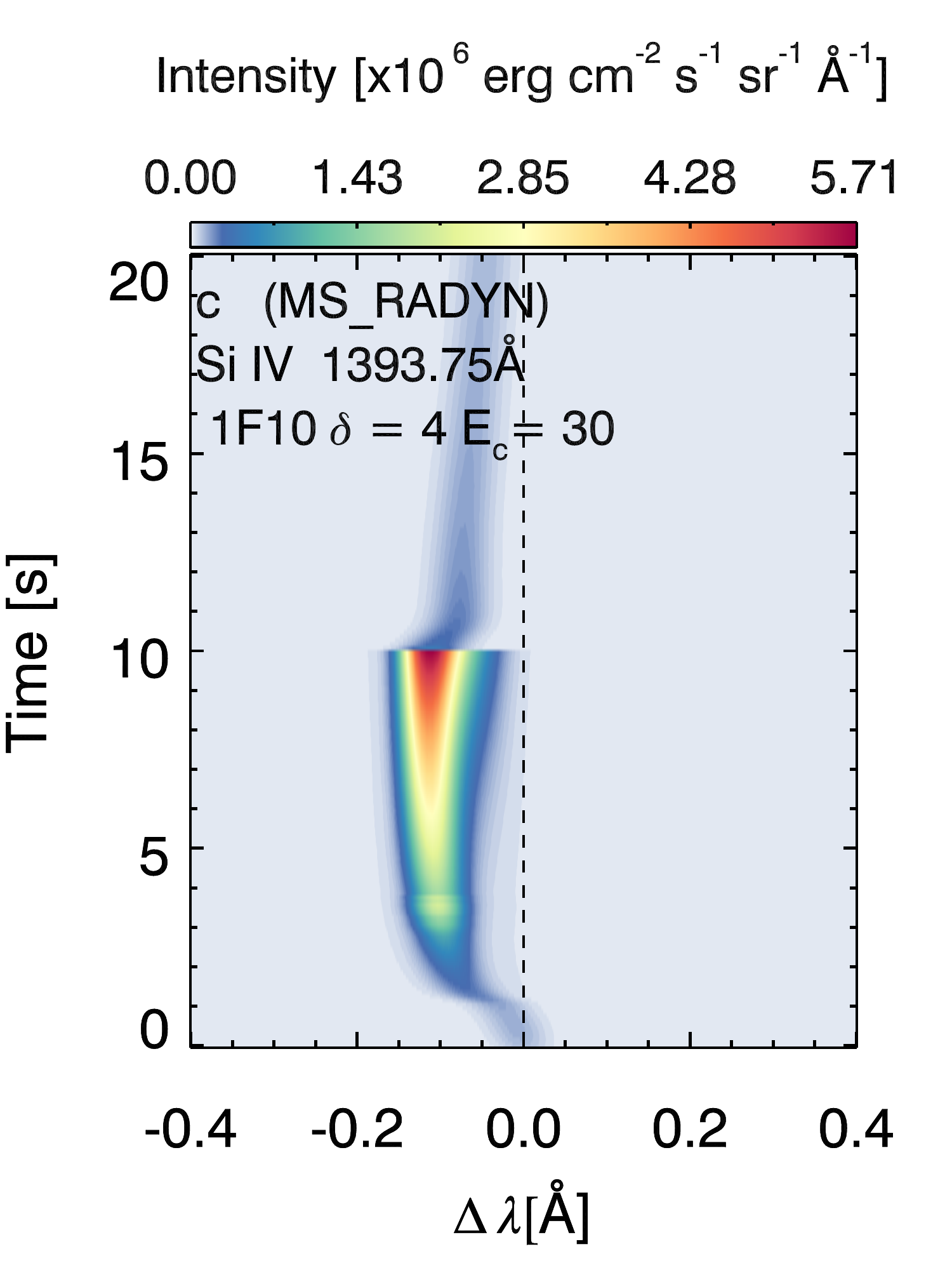}}	
	\subfloat{\includegraphics[width = 0.225\textwidth, clip = true, trim = 0cm 0cm 0cm 0cm]{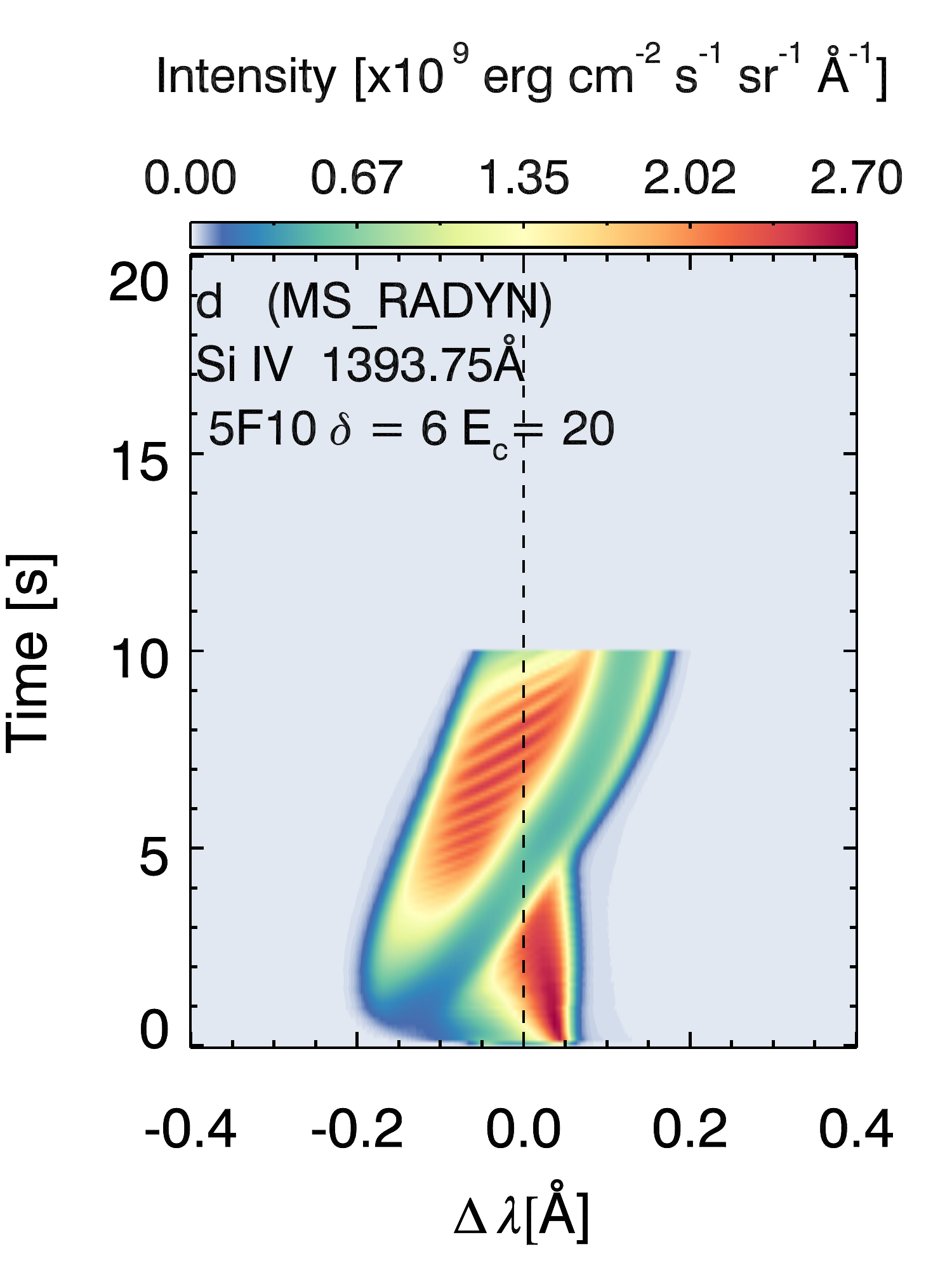}}
	         }
	\caption{\textsl{Comparing Si~\textsc{iv} 1393\AA\ line emission without suppression of dielectronic recombination (a,b) and with suppression (c,d) for two test cases of \texttt{MS\_RADYN} using \texttt{Si30\_atom\_wchex}. Panels are as described in Figure~\ref{fig:stackplots_1}.}}
	\label{fig:drsuppress_stacks}
\end{figure*}

While some atomic databases such as ADAS do contain density sensitive dielectronic recombination rate coefficients, implementing these in \texttt{RADYN} or \texttt{MS\_RADYN} requires modifications to the ways in which the codes deal with atomic data. This will be the focus of future research, but here we demonstrate the potential impact of suppression of dielectronic recombination on our results. 

We use an approximation similar to that of \cite{2018ApJ...855...15Y}, who used suppression factors based on the work of \cite{2013ApJ...768...82N}. The isoelectronic sequence, charge, electron density, and temperature are used in the expressions of \cite{2013ApJ...768...82N} to derive a suppression factor, which we used to suppress the dielectronic rates from CHIANTI in our model atom \texttt{Si30\_atom\_wchex}. This is somewhat artificial in our flare atmospheres as the density is assumed constant for all locations and times when computing the suppressed rates, but is suitable for illustrative purposes. An electron density of $n_{e} = 1\times10^{12}$~cm$^{-3}$ was chosen. While this is too large for the quiet Sun, and the electron density in the flaring transition region varies with flare strength, the results of \cite{2016A&A...594A..64P} show that the differences between $n_{e} = [1\times10^{11}, 1\times10^{12},1\times10^{13}]$~cm$^{-3}$ are smaller than the jump from the zero density limit that was used in the rest of this study. A mid-range value of $n_{e} = 1\times10^{12}$~cm$^{-3}$ therefore seems like an appropriate choice.

Two \texttt{MS\_RADYN} test cases were run with the suppressed dielectronic rates: one moderate flare ($5\mathrm{F}10\delta6E_{\mathrm{c}}20$), and one weaker flare ($1\mathrm{F}10\delta4E_{\mathrm{c}}30$). The latter simulation was chosen to also determine is using the suppression of dielectronic recombination would result in opacity effects becoming important (they were not when using the zero density limit). The ionisation fractions at $t=0$~s are shown in Figure~\ref{fig:drsuppress_ionfrac}, indicating that with suppressed rates Si~\textsc{iv} forms over a similar temperature range as the zero density limit, but with a peak temperature that has shifted lower to $\mathrm{log}~T = 4.77$ ($T = 59.3$~kK), and optical depth $\tau_{\lambda} \approx 0.03$. The peak of the Si~\textsc{iv} fraction also increased in magnitude (compare to Figure~\ref{fig:si_ionfracs}(a)). This results in an increased intensity of emission during the flares, but the line shapes, and features are largely unchanged. The most intense emission is $\approx1.3\times$ larger in the $1\mathrm{F}10\delta4E_{\mathrm{c}}30$ simulation, and $\approx2.9\times$ larger in the $5\mathrm{F}10\delta6E_{\mathrm{c}}20$ simulation. Figures~\ref{fig:drsuppress_stacks}(a,b,c,d) show wavelength versus time images for the cases without suppression of dielectronic recombination (a,b) and with suppression (c,d). Overall the optical depth effects on the line formation were unchanged.

Future efforts will attempt to include suppression of dielectronic recombination self-consistently in \texttt{RADYN} and \texttt{MS\_RADYN}, using the actual atmospheric density with rate coefficients tabulated as functions of both temperature (as currently implemented) and electron density.

\end{document}